\documentclass[DIV=13,11pt,a4paper,abstract=true,date=false]{scrartcl}
\usepackage[left]{lineno}

\usepackage{amsmath,amsthm,amssymb,amsfonts,mathtools}
\usepackage{url}
\usepackage{todonotes}
\usepackage{algorithm}
\usepackage{algpseudocode}
\usepackage{subcaption}
\usepackage{wasysym}
\usepackage{csquotes}
\usepackage{authblk}
\usepackage{hyperref}
\usepackage
[
	backend=bibtex,
	language=british,
	style=numeric,
	sorting=none,
    sortcites=true,
	sortlocale=en-GB,
    giveninits=true,
    maxnames=3,
    autocite=superscript,
]{biblatex}
\DeclareCiteCommand{\supercite}[\mkbibsuperscript]{%
\iffieldundef{prenote}{}{\BibliographyWarning{Ignoring prenote argument}}%
\iffieldundef{postnote}{}{}}%
{\bibopenbracket\usebibmacro{citeindex}\usebibmacro{cite}\usebibmacro{postnote}\bibclosebracket}{\supercitedelim}{}

\newcommand{\R}{\ensuremath{\mathbb{R}}} % set of real numbers
\newcommand{\N}{\ensuremath{\mathbb{N}}} % set of natural numbers
\newcommand{\dt}{\ensuremath{{\mathrm d}t}} % dt
\newcommand{\ds}{\ensuremath{{\mathrm d}s}} % ds
\newcommand{\xx}{\boldsymbol{x}}
\newcommand{\kk}{\boldsymbol{k}}
\newcommand{\vv}{\boldsymbol{v}}

\newcommand{\EE}{\boldsymbol{E}}
\newcommand{\BB}{\boldsymbol{B}}
\newcommand{\jj}{\boldsymbol{j}}

\newcommand{\jcal}{\hat{\mathcal{J}}}
\newcommand{\jcalh}{\hat{\mathcal{J}}_h}

\newcommand{\Ham}{\ensuremath{\mathcal{H}}} % Torus

\newcommand{\HamE}{\ensuremath{\mathcal{H}_{\EE}}} % Torus
\newcommand{\HamB}{\ensuremath{\mathcal{H}_{\BB}}} % Torus
\newcommand{\Hamf}{\ensuremath{\mathcal{H}_{f}}} % Torus

\newcommand{\JB}{\ensuremath{\mathbb{J}_{\BB}}} % Torus

\newcommand{\dxx}{\ensuremath{{\mathrm d}\xx}} % dx in bold
\newcommand{\dvv}{\ensuremath{{\mathrm d}\vv}} % dv in bold 

\newcommand{\vmin}{v_{\min}}
\newcommand{\vmax}{v_{\max}}
\newcommand{\vth}{v_{\text{th}}}

\newtheorem{theorem}{Theorem}[section]
\newtheorem{lemma}[theorem]{Lemma}

\newtheorem{remark}[theorem]{Remark}

\title{Extending the Numerical Flow Iteration to the multi-species Vlasov--Maxwell system through Hamiltonian Splitting}

\makeatletter
\newcommand*\thanksmark[1]{\textsuperscript{\@arabic{#1}}}
\makeatother

\author{
R.-Paul~Wilhelm\thanksmark{1}$^,$\thanksmark{2},
Fabio~Bacchini\thanksmark{1}$^,$\thanksmark{3},
Sebastian~Sch\"ops\thanksmark{4},
Manuel~Torrilhon\thanksmark{2},
Melina~Merkel\thanksmark{4},
Matthias Kirchhart\thanksmark{2}$^,$\thanksmark{5}
}

\begin{document}
%\linenumbers
\maketitle

% --- affiliations as manual footnotes ---
\begingroup
\renewcommand{\thefootnote}{\arabic{footnote}}%
\footnotetext[1]{Centre for mathematical Plasma Astrophysics (CmPA), KU Leuven, Belgium.}%
\footnotetext[2]{Institute for Applied and Computational Mathematics (ACoM), RWTH Aachen University.}%
\footnotetext[3]{Royal Belgian Institute for Space Aeronomy, Brussels, Belgium}
\footnotetext[4]{Institute for Accelerator Science and Electromagnetic Fields (TEMF), TU Darmstadt.}%
\footnotetext[5]{Nvidia GmbH, Munich.}%

\endgroup

\begin{abstract}
The Numerical Flow Iteration (NuFI) method has recently been proposed as a 
memory-slim while accurate in phase-space method for the electro-static 
Vlasov--Poisson system. It stores the temporal evolution of the electric field, 
instead of the distribution functions, and reconstructs the solution in each 
time step by following the characteristics backwards in time and reconstructing 
the solution from the initial distribution. 
NuFI has been shown to be more accurate than other state-of-the-art 
electro-static Vlasov solvers given the same amount of degrees of freedom. 
In this paper, we build on the Hamiltonian structure of the full 
Vlasov--Maxwell system to extend NuFI to handle electro-magnetic kinetic plasma 
dynamics.
We show that the structure-preserving properties of the NuFI time-stepping are 
preserved when extending to the electro-magnetic case. Furthermore we discuss 
how NuFI can be incorporated into existing Semi-Lagrangian codes as an 
efficient while accurate subcycling technique.
\end{abstract}

\vspace{2pc}

\section{Introduction}

In many applications such as nuclear fusion devices or solar wind it is 
relevant to fundamentally understand how turbulence and instabilities in a 
plasma are triggered and how they evolve over time. 
Some phenomena can already be observed through a fluid-description but when 
plasma becomes rarefied or hot the velocity distribution of the charged 
particles can substantially deviate from equilibrium, which breaks
the fundamental equilibrium-assumption of a fluid-descriptions. In this case 
one has to model the plasma evolution through kinetic theory, i.e., including 
the velocity distributions in the model, which results in the Vlasov equation 
\autocite{intro_plasma_physics_chen}
\begin{equation}
\label{eqn:vlasov}
\partial_t f^\alpha + \vv \cdot \nabla_{\xx} f^\alpha 
+ \tfrac{q_\alpha}{m_\alpha}  \left( \EE + \vv \times \BB \right) \cdot \nabla_{\vv} f^\alpha = 0.
\end{equation}
where $f^\alpha: [0,+\infty) \times \R^d \times \R^d \rightarrow [0,+\infty)$ 
is the probability distribution of the species $\alpha$ in the 
up to six-dimensional phase-space and $\EE : [0,+\infty) \times \R^d 
\rightarrow \R^d$ and $\BB: [0,+\infty) \times \R^d 
\rightarrow \R^d$ are the self-induced 
electro-magnetic fields acting upon the species $\alpha$. 
The Vlasov equation is coupled to the Maxwell's equations (in a vacuum)
\begin{gather}
\label{eqn:Faraday_law}
\nabla \times \EE + \partial_t \BB = 0, \\
\label{eqn:Amperes_circuit_law}
\nabla \times \BB - \partial_t \EE = \jj, \\
\label{eqn:Gauss_law}
\nabla \cdot \EE =  \rho, \\
\label{eqn:Gauss_law_magnetism}
\nabla \cdot \BB = 0. 
\end{gather}

The right hand side of the Maxwell's equations is computed from the phase-space
distribution functions $f^\alpha$:
\begin{gather}
\label{eqn:rho}
\rho(t,\xx) = \sum_{\alpha} \rho^\alpha(t,\xx) 
= \sum_{\alpha} \int_{\R^d} f^\alpha(t,\xx,\vv) \dvv, \\
\label{eqn:j}
\jj(t,\xx) = \sum_{\alpha} \jj^\alpha(t,\xx) 
= \sum_{\alpha} \int_{\R^d} \vv f^\alpha(t,\xx,\vv) \dvv,
\end{gather}

where $\rho$ is the charge and $\jj$ the current density. Note that we assume
charge quasi-neutrality~\autocite{intro_plasma_physics_chen}, i.\,e.,
\begin{equation}
\label{eqn:quasi_neutrality}
\int_{\R^d} \rho \, \dxx = 0.
\end{equation}

With this coupling equations \eqref{eqn:vlasov} -- \eqref{eqn:j}
become the (non-linear) Vlasov--Maxwell system. We use the normalized version 
of the Vlasov--Maxwell system in rationalized Gaussian units following 
Morrison~\autocite{MORRISON1980383}, details can be found in appendix 
\ref{appendix_sec:normalization}.

Solving the Vlasov system poses a number of significant challenges: The 
solution lives in the six-dimensional phase-space inducing the curse of 
dimensionality for methods trying to directly discretize the distribution 
function. Hence to simulate the full system one either has to use low 
resolution or reduce the system size with a number of additional assumptions on 
the physics to reduce the dimension of phase space. 
Additionally, kinetic plasma dynamics are commonly only weakly collisional or, 
in the version we use in this paper \eqref{eqn:vlasov}, entirely collisionless, 
which is a valid assumption in e.g. astrophysical 
settings.~\autocite{multi_scale_solar_wind,
multi_species_kin_instability_nufi, intro_plasma_physics_chen}
Therefore the physics lack a (strong) mechanism to dissipate fine structures in 
the distribution function and, due to the non-linearity of the system, fine 
scale structures--often referred to as filaments--develop and remain in the 
distribution function over long time-scales. The filamentation has to be 
resolved to accurately capture the onset of some instabilities and there is 
some evidence that these filaments play a key role in the heat dissipation in a 
plasma.~\autocite{case_for_electron_astrophysics, multi_scale_solar_wind, lapenta_mercury}
However, resolving these fine scale structures also requires the use of
prohibitively high resolutions. Finally, due to the large mass difference 
between electrons and ions, the dynamics of multi-species plasma systems are 
also inherently multi-scale, requiring even higher resolutions if we 
aim to accurately resolve dynamics over all 
scales.~\autocite{multi_scale_solar_wind}

Numerical solvers for the Vlasov equation can generally be grouped into three 
main classes: Eulerian (or Discrete Velocity), Semi-Lagrangian, and Lagrangian 
schemes.
Both Eulerian and Semi-Lagrangian approaches are grid-based, meaning that the 
distribution function $f$ is represented on a phase-space grid and they
differ primarily in how they advance the solution in time.
In Eulerian schemes, the grid values are updated by evaluating the local, 
temporal rate of change, effectively rewriting the Vlasov equation as a large 
system of ordinary differential equations (ODEs)—one for each degree of freedom 
in the discretization \autocite{ARBER2002339, gene_1, GORLER20117053, JUNO2018110, Mandell_Hakim_Hammett_Francisquez_2020, fijalkow_numerical_1999, filbet_conservative_2001, WEN2025114079, KIECHLE2025113693, KORMANN2025118290}.
Although Eulerian methods are widely used in fluid dynamics, their application 
to kinetic equations such as the Vlasov equation is less common. The Eulerian 
framework offers a general methodology for formulating numerical solvers for 
PDEs, yet it typically neglects the intrinsic structural properties of the 
underlying physical system.

The Semi-Lagrangian method aims to encapsule more structure of the Vlasov 
equation by shifting to the Lagrangian viewpoint for each time step. The 
distribution function is still represented on a phase-space grid, but each grid 
point is traced backwards in time over a short time-interval by integrating the 
characteristic equations effectively interpreting it like a particle for one 
time-step. The new-time solution is then obtained by interpolating from the 
previous time step via evaluating $f$ at the foot of the characteristic.\footnote{The approach described here is a 
backward Semi-Lagrangian (BSL) scheme. Propagating grid points forward in time 
instead leads to a forward Semi-Lagrangian (FSL) variant. In the following, we 
use Semi-Lagrangian to refer specifically to the BSL formulation unless stated 
otherwise.} This procedure enhances the conservation properties compared to 
purely Eulerian methods, particularly, when coupled with a symplectic time 
integrator, and also removes the phase-space CFL time-step restriction that 
limits Eulerian schemes (although in case of the Vlasov--Maxwell system the CFL 
condition from the Maxwell equations remains).~\autocite{cheng_integration_1976}
A popular choice for interpolation is are (cubic) B-splines, which provide an 
effective balance between accuracy and computational 
cost~\autocite{Riishøjgaard1998, kormann_massively_parallel},
however, there are also implementations which instead rely on nodal 
interpolation methods.~\autocite{sldg2015, CROUSEILLES20101927, cottet_etancelin_perignon_picard_2014, COTTET2018362, ROSSMANITH20116203, vlasiator_1, VONALFTHAN201424, muphy_I, ALLMANNRAHN2024109064}
Despite their advantages, grid-based Vlasov solvers face serious challenges.
The curse of dimensionality makes storing and updating a full six-dimensional 
grid extremely costly, especially when adaptive resolution or non-rectangular 
geometries are needed. The large memory footprint also causes significant 
communication overhead, reducing parallel efficiency and scalability on modern 
accelerator architectures.~\autocite{kormann_massively_parallel, EINKEMMER2019937, sl_456d_einkemmmer_moriggl, ScRaEi2024}
Incorporating sparse-grid techniques or adaptive meshes is particularly 
difficult in such high dimensions.~\autocite{sparse_kormann_sonnendruecker, Wei_Cheng_2016}
Moreover, discretization on grids inevitably introduces numerical 
diffusion, which--being an artificial source of dissipation--violates the 
conservative, collisionless nature of the Vlasov system. Although various 
strategies aim to reduce this effect, it remains intrinsic to all grid-based 
methods. \autocite{filbet_conservative_2001, CROUSEILLES20101927}
Finally, complex domain geometries, boundary conditions, and expanding velocity 
supports pose additional difficulties for grid-based 
formulations.~\autocite{besse_semi-lagrangian_2003, Pfau_Kempf_2018, Palmroth2018}

An alternative to grid-based discretizations is to employ particle-based, also 
called Lagrangian, solvers. The most widely used example is the 
Particle-In-Cell (PIC) method, though other particle approaches exist, such as 
Smoothed Particle Hydrodynamics (SPH) and the Reproducing Kernel Hilbert Space 
Particle Method (RKHS-PM).~\autocite{cottet_particle_1984, hockney_computer_2021, birdsall_plasma_1985, WILHELM2023111720}
In PIC the distribution function is represented by a collection of marker 
particles, each carrying a value or weight that is transported along the 
characteristics of the Vlasov equation. Because the particle paths 
coincide with the characteristics, their associated weights remain constant in 
time. For PIC a spatial grid is additionally introduced to evaluate and 
interpolate the electromagnetic fields. The right hand side for the Maxwell 
equations is obtained by evaluating the charge and current densities on this 
grid (called charge deposition). 
The simplest deposition strategy counts the number of particles 
per cell, but more accurate methods assign shape functions to particles, 
enabling higher-order charge deposition. \autocite{HarlowEvans2, cottet_raviart_pic, wang_particle--cell_2011, myers_4th-order_2016, ameres_stochastic_2018}
Although PIC methods still face the curse of dimensionality, they offer several 
advantages over grid-based schemes. Particles naturally follow the plasma flow, 
providing automatic adaptivity to the evolving distribution, which especially in case of strong inhomogeneity can be advantageous. Moreover, since 
the data is unstructured, PIC can more easily handle complex geometries, 
boundary conditions, and distributed parallelization.
The main drawback of particle-based representations, however, is the presence 
of statistical noise. Achieving a given level of accuracy often requires 
significantly more particles than grid points in deterministic solvers. This 
issue becomes particularly severe in low-density regions or when high accuracy 
in the phase-space is required. Most PIC implementations are explicit in time
~\autocite{Kraus_Kormann_Morrison_Sonnendrücker_2017, pic_on_gpu, piclas, DEROUILLAT2018351, 10046112, PARODI2025102244}, i.\,e., where the field and 
particle updates are performed using explicit time integration.
Because plasma dynamics are inherently multi-scale, due to the large mass 
disparity between electrons and ions as well as the system size, explicit PIC 
schemes often demand very fine temporal and spatial resolutions to capture all 
relevant physics. 
To address this, various semi-implicit and fully implicit formulations have 
been developed, which relax the stringent stability and resolution constraints 
of explicit methods. \autocite{BRACKBILL1982271, MASON1981233, Lapenta2006, MARKIDIS20101509, LAPENTA2017349, bacchini_relsim, ARSHAD2025109806, Bacchini_2019, Croonen_2024}

A more detailed overview of numerical solvers for the Vlasov equation can be 
found in e.g. the reviews by Filbet and 
Sonnendr\"ucker~\autocite{review_num_kin_filbet} or, more recently, 
Palmroth et al. \autocite{Palmroth2018}

Recently a new class of schemes was introduced to approximate the flow map of 
the Vlasov equation instead of the distribution function, whereby one tries to
further preserve the Lagrangian structure of the Vlasov system. The two main
approaches in this direction are the Characteristic Mapping Method (CMM) 
introduced by Krah et al.~\cite{Krah2023905} and the Numerical Flow Iteration 
(NuFI) introduced by Kirchhart and Wilhelm~\autocite{kirchhart2023numerical} 
for the electro-static Vlasov--Poisson system. CMM directly stores the flow map 
on a phase-space grid and exploits the sub-group property of the flow to 
compose several sub-maps, which can be stored on a coarser grid than the full 
map or distribution function, and thereby achieve good sub-grid resolution. 
However, CMM does not exactly preserve the conservation properties of the 
Vlasov system as the time stepping is not symplectic.
In contrast to this, NuFI employs an iterative-in-time approximation of the 
characteristics based on symplectic operator splitting for the 
time-integration. This avoids the need of directly discretizing and storing the 
distribution function, only requiring to store the time-evolution of the 
electric potentials, while using the symplectic time-stepping scheme leads to 
preservation of the conservation properties of the analytic solution either 
exactly in the case of the $L^p$-norms and entropy or up to the 
time-discretization error in the case of the total energy.
Note that there are also Semi-Lagrangian schemes which can preserve a subset of 
these conservation properties,~\autocite{LIU2025113840, POLLINGER2023112338, CROUSEILLES20101927}, however, the numerical dissipation introduced by 
grid-based discretization cannot be avoided entirely.
This leads to improved results when comparing to Semi-Lagrangian or PIC schemes 
with comparable resolution and interpolation order. Additionally, because NuFI 
only looks at the characteristics instead of a fixed discretization of the 
distribution function in phase-space, it is possible, on the one hand, to zoom 
arbitrarily far into the distribution function in regions of interest (e.g. for 
post-processing) without having to rerun with higher resolution, and, on the 
other hand, allows easy introduction of adaptivity and handling of boundary 
conditions.~\autocite{multi_species_kin_instability_nufi, Wilhelm_2025} 
Since a wide range of plasma phenomena of interest involve electromagnetic 
rather than electrostatic interactions, the goal of this paper is to extend 
NuFI to the fully electro-magnetic case and check to what extend the 
structure-preservation properties are retained in the electro-magnetic case.

In the following, in section \ref{sec:nufi_electro_static} we will briefly 
introduce the Numerical Flow Iteration for the electro-static case before 
explaining how the extension towards the full Maxwell system works via 
revisiting the Hamiltonian of the Vlasov--Maxwell system as well as the 
associated Hamiltonian splitting in the sections \ref{sec:Hamiltonian_VM} 
and \ref{sec:nufi_vm} as well as \ref{sec:nufi_ham_high_order}. In section \ref{sec:conservation_properties_nufi_theory} we give theoretical proof of the 
structure-preserving properties of NuFI. In the sections
\ref{sec:spline_issues_and_gauss_filter}, \ref{sec:nufi_vm_predictor_corrector} 
as well as \ref{sec:restarted_nufi_vm} we discuss the practical implementation 
and, in particular, also how the computational complexity of the approach can 
be reduced. Finally, in section \ref{sec:num_tests} we look at numerical 
benchmarks to test the extension of NuFI to the Vlasov--Maxwell system.

\section{Hamiltonian structure of the Vlasov system}\label{sec:nufi}

In the following we will first introduce the idea of the Numerical Flow 
Iteration (NuFI) for the simpler electro-static case, before showing how the 
extension to the Vlasov--Maxwell system works. As our focus lies solely on the 
numerical algorithm we will in the following assume that the initial data of 
our Vlasov--Maxwell system is sufficiently smooth that a smooth solution exists 
and is unique. 
To the knowledge of the authors the existence of such solutions is not 
conclusively proven for general initial data, however, we refer the interested 
reader to the paper by Pfaffelmoser~\autocite{pfaffelmoser_global_1992}, where 
the existence of smooth solutions for the electro-static case is proven, as 
well as to the more recent paper by Borrin~\autocite{BORRIN2025190} and the 
references therein, where the existence of solutions is discussed in a more 
general setting.

\subsection{The Numerical Flow Iteration in the electro-static case}
\label{sec:nufi_electro_static}

Before we move on to the fully electro-magnetic case of the Vlasov--Maxwell 
system, we briefly recap the ideas behind the Numerical Flow Iteration in the
simplified electro-static case. 
If $\vert \BB \vert \ll \vert \EE \vert$, we can neglect the influence of the
magnetic field on the overall dynamics of the Vlasov--Maxwell system. 
This is in turn allows us to simplify it to the electro-static Vlasov--Poisson 
system, where the Maxwell equations reduce to the Gauss law, which can be 
reformulated to a simple Poisson equation for the electro-static potential 
\begin{equation}
\label{eqn:Poisson}
-\Delta_{\xx} \varphi = \rho
\end{equation}
from which we get the electric field through
\begin{equation}
\label{eqn:potential}
\EE = -\nabla_{\xx} \varphi.
\end{equation}

The Vlasov--Poisson system is a Hamiltonian system with the Hamiltonian
\begin{equation}
\label{eqn:Hamiltonian_Vlasov_Poisson}
\begin{aligned}
\Ham_{VP} = \underbrace{\vphantom{\sum_{\alpha}}\tfrac{1}{2} \int_{\R^d} \EE^2 \dxx}_{\displaystyle \eqqcolon \HamE} 
+ \underbrace{\tfrac{1}{2} \sum_{\alpha} \int_{\R^d\times \R^d} f^\alpha 
\dxx\dvv}_{\displaystyle \eqqcolon \Hamf}.
\end{aligned}
%\begin{aligned}
%\Ham_{VM} &= \HamE &+& \Hamf \\
%&= \tfrac{1}{2} \int_{\R^d} \EE^2 \dxx 
%&+& \tfrac{1}{2} \sum_{\alpha} \int_{\R^d\times \R^d} f^\alpha \dxx\dvv.
%\end{aligned}
\end{equation}

This allows us to use operator splitting techniques for time-stepping, i.\,e., 
solve the associated characteristic equations through splitting the non-linear
Vlasov equation into two linear advection equations associated to the $\HamE$ 
and $\Hamf$ respectively. 
Using the well-known St\"ormer--Verlet splitting this results in the following 
scheme: Starting from $\hat{\xx}^h_n = \xx$ and $\hat{\vv}^h_n = \vv$ at the 
time-step $t=t_n$ compute
%\begin{gather}
%\label{eqn:nufi_sv_first_half_step}
%\hat{\vv}^{h,\alpha}_{i- 1/2} = \hat{\vv}^{h,\alpha}_i - \tfrac{\Delta t}{2} %\tfrac{q_\alpha}{m_\alpha}   \EE(t_i, \hat{\xx}^{h,\alpha}_i), \\ 
%\label{eqn:nufi_sv_second_half_step}
%\hat{\xx}^{h,\alpha}_{i-1} = \hat{\xx}^{h,\alpha}_i 
%- \hat{\vv}^{h,\alpha}_{i-1/2}, \\
%\label{eqn:nufi_sv_third_half_step}
%\hat{\vv}^{h,\alpha}_{i-1} = \hat{\vv}^{h,\alpha}_{i-1/2} - \tfrac{\Delta t}{2}  \tfrac{q_\alpha}{m_\alpha}  \EE(t_{i-1}, \hat{\xx}^{h,\alpha}_{i-1})
%\end{gather}
\begin{align}
\hat{\vv}^{h,\alpha}_{i-1/2}
&= \hat{\vv}^{h,\alpha}_i
- \tfrac{\Delta t}{2} \tfrac{q_\alpha}{m_\alpha}
\EE(t_i, \hat{\xx}^{h,\alpha}_i)
\label{eqn:nufi_sv_first_half_step}
\\
\hat{\xx}^{h,\alpha}_{i-1}
&= \hat{\xx}^{h,\alpha}_i 
- \hat{\vv}^{h,\alpha}_{i-1/2}
\label{eqn:nufi_sv_second_half_step}
\\
\hat{\vv}^{h,\alpha}_{i-1}
&= \hat{\vv}^{h,\alpha}_{i-1/2}
- \tfrac{\Delta t}{2} \tfrac{q_\alpha}{m_\alpha}
\EE(t_{i-1}, \hat{\xx}^{h,\alpha}_{i-1})
\label{eqn:nufi_sv_third_half_step}
\end{align}
for $i=0,...,n$. 

This operator splitting is widely used in Semi-Lagrangian codes\footnote{ The 
same operator splitting idea is also used in Particle-In-Cell codes where it is 
often referred to as \emph{Leapfrog} time-integration.} to either advance the 
characteristics or trace them backwards in time for 1 time-step after which the
results can be interpolated and stored on a phase-space grid. The idea behind 
NuFI is to avoid the phase-space grid altogether and instead use the same
scheme \eqref{eqn:nufi_sv_first_half_step} -- \eqref{eqn:nufi_sv_third_half_step} to iterate backwards to the initial time,
where one can evaluate the initial data -- often prescribed in a closed, analytic form -- through the following expression
\begin{equation}
\label{eqn:eval_f_through_char_num}
f^\alpha(t,\xx,\vv) = f^\alpha_0(\hat{\xx}^{h,\alpha}_0, 
\hat{\vv}^{h,\alpha}_0) + \mathcal{O}\left(\Delta t^2\right)
\end{equation}

based on the second order accuracy of Störmer-Verlet.

Only knowledge of the evolution in time of the electric field is required to
evaluate any distribution function $f^\alpha$ at any arbitrary position in 
phase-space. Therefore -- using some numerical quadrature rule -- we can 
integrate \eqref{eqn:rho} to obtain the right-hand side of \eqref{eqn:Poisson} 
and combining with a Poisson solver we arrive at a numerical solver for the
non-linear Vlasov--Poisson system. A more in detail derivation can be found in 
the previous works of Wilhelm et al.\autocite{kirchhart2023numerical, multi_species_kin_instability_nufi, Wilhelm_2025}

The advantages of the NuFI approach compared to a classic Semi-Lagrangian 
scheme are two-fold: 

First, instead of storing a 6-dimensional distribution function $f^\alpha$ per 
species, one only has to store the evolution of the 3-dimensional electric 
potential $\varphi$, i.\,e., one 4-dimensional instead of several 6-dimensional 
functions. Additionally, the electric potential is substantially smoother than 
the individual distribution functions as one has to take the 2nd derivative of 
$\varphi$ to obtain $\rho$, which in turn is obtained by integrating over 
$f^\alpha$. Hence we need fewer (spatial) degrees of freedom to accurately 
resolve $\varphi$.

Secondly, because the St\"ormer--Verlet time-integration scheme is symplectic
NuFI preserves the conservation properties of the analytic solution. In fact, 
all $L^p$-norms as well as the entropy are preserved analytically while the 
total energy is preserved up to the discretization error, however, without a 
drift. Therefore NuFI is structure-preserving far beyond what conventional 
Semi-Lagrangian schemes for the Vlasov--Poisson system can achieve and, in 
practice, this manifests through substantially improved accuracy for comparable 
numerical resolutions.~\autocite{kirchhart2023numerical}
However, the main disadvantage of NuFI compared to conventional approaches is 
the increased computational complexity: Because the characteristics are now 
traced until $t=0$ (instead of just 1 time-step) the computational complexity
becomes quadratic (instead of linear) in the total number of time-steps in 
a simulation. The increased complexity makes the pure NuFI approach only 
applicable to short simulation periods.

\subsection{Hamiltonian of the Vlasov--Maxwell system}
\label{sec:Hamiltonian_VM}

The extension of the St\"ormer--Verlet-based NuFI scheme from the electro-
static to the fully electro-magnetic Vlasov--Maxwell system is not straight 
forward, if we want to retain the structure-preserving properties of NuFI, and 
requires us to revisit Hamiltonian structure of the full Vlasov system.
The Vlasov--Maxwell system is also a Hamiltonian system with the Hamiltonian 
being \autocite{MARSDEN1982394,MORRISON1980383,CROUSEILLES2015224}
\begin{equation}
\label{eqn:Hamiltonian_Vlasov_Maxwell}
\begin{aligned}
\Ham_{VM} 
= \underbrace{\vphantom{\sum_{\alpha}}\tfrac{1}{2} \int_{\R^d} \EE^2 \dxx}_{\displaystyle \eqqcolon \HamE} 
+ \underbrace{\vphantom{\sum_{\alpha}}\tfrac{1}{2} \int_{\R^d} \BB^2 \dxx}_{\displaystyle \eqqcolon \HamB} 
+ \underbrace{\tfrac{1}{2} \sum_{\alpha} \int_{\R^d\times \R^d} f^\alpha 
\dxx\dvv}_{\displaystyle \eqqcolon \Hamf}.
%&= \HamE &+& \HamB &+& \Hamf \\
%&= \tfrac{1}{2} \int_{\R^d} \EE^2 \dxx &+& \tfrac{1}{2} \int_{\R^d} \BB^2 \dxx
%&+& \tfrac{1}{2} \sum_{\alpha} \int_{\R^d\times \R^d} f^\alpha \dxx\dvv.
\end{aligned}
\end{equation}

The three terms $\HamE, \HamB$ and $\Hamf$ correspond to the electric,
magnetic and kinetic energy respectively. 
Assuming that the solution of the system exists and is smooth enough, we can 
follow the reasoning of Crouseilles et al. \autocite{CROUSEILLES2015224} to 
derive (linear) partial differential equations associated with each of the 
sub-Hamiltonians. 

\subsubsection{Equations for $\HamE$}

The equations associated to the electric field Hamiltonian $\HamE$ are
\begin{align}
\partial_t f^\alpha + \tfrac{q_\alpha}{m_\alpha}\EE \cdot \nabla_{\vv} f^\alpha = 0,
\label{eqn:HamE_1}
\\
\partial_t \EE = 0, 
\label{eqn:HamE_2} \\
\partial_t \BB + \nabla_{\xx} \times \EE = 0.
\label{eqn:HamE_3}
\end{align}

Provided the initial data $\EE_0, \BB_0$ and $f^\alpha_0$ at time $t = 0$ the 
solution of this linear transport equation can be explicitly stated as
\begin{align}
f^\alpha(t,\xx,\vv) &= f^\alpha_0\left(\xx, \vv - t \tfrac{q_\alpha}{m_\alpha} 
\EE_0\left(\xx\right)\right), 
\label{eqn:sol_HamE_1} \\
\EE(t,\xx) &= \EE_0(\xx), 
\label{eqn:sol_HamE_2} \\
\BB(t,\xx) &= \BB_0(\xx) - t \nabla_{\xx} \times \EE_0(\xx).
\label{eqn:sol_HamE_3}
\end{align}

\subsubsection{Equations for $\HamB$}

The equations associated to the magnetic field Hamiltonian $\HamB$ are
\begin{align}
\partial_t f^\alpha + \tfrac{q_\alpha}{m_\alpha} \left( \vv \times \BB(\xx) \right) \cdot \nabla_{\vv} f^\alpha = 0, \label{eqn:HamB_1} \\
\partial_t \EE - \nabla_{\xx} \times \BB = 0, \label{eqn:HamB_2} \\
\partial_t \BB = 0. \label{eqn:HamB_3}
\end{align}

The solution can be written as 
\begin{align}
f^\alpha(t,\xx,\vv) &= f^\alpha_0\left(\xx, \exp\left(-\tfrac{q_\alpha}{m_\alpha} t \JB \right) \vv  \right), \label{eqn:sol_HamB_1} \\
\EE(t,\xx) &= \EE_0(\xx) + t \nabla_{\xx} \times \BB_0(\xx), \label{eqn:sol_HamB_2} \\
\BB(t,\xx) &= \BB_0(\xx). \label{eqn:sol_HamB_3}
\end{align}

Here we used that the cross-product between two vectors $a,b \in \R^3$ can be rewritten into a vector-matrix product through
\begin{equation}
a \times b = \begin{pmatrix}
a_2 b_3 - a_3 b_2 \\
a_3 b_1 - a_1 b_3 \\
a_1 b_2 - a_2 b_1
\end{pmatrix}
= \underbrace{\begin{pmatrix}
0 & b_3  & -b_2 \\
-b_3 & 0 & b_1 \\
b_2 & -b_1 & 0 
\end{pmatrix}}_{\eqqcolon \mathbb{J}_b} a.
\end{equation}

By $\exp(A)$ we denote the natural matrix exponential for a matrix 
$A \in R^{3\times 3}$.

\subsubsection{Equations for $\Hamf$}

The equations associated to the kinetic energy Hamiltonian $\Hamf$ are
\begin{align}
\partial_t f^\alpha + \vv \nabla_{\xx} f^\alpha = 0, \label{eqn:Hamf_1} \\
\partial_t \EE  + \jj = 0 , \label{eqn:Hamf_2} \\
\partial_t \BB = 0. \label{eqn:Hamf_3}
\end{align}

The solution can be written as 
\begin{align}
f^\alpha(t,\xx,\vv) &= f^\alpha_0\left(\xx - t \vv , \vv  \right), \label{eqn:sol_Hamf_1} \\
\EE(t,\xx) &= \EE_0(\xx) + \int_0^t \sum_{\alpha}\left( \int_{\R^3} \vv f^\alpha_0\left(\xx - s \vv, \vv\right) \dvv \right) \ds , \label{eqn:sol_Hamf_2} \\
\BB(t,\xx) &= \BB_0(\xx). \label{eqn:sol_Hamf_3}
\end{align}

Note that \eqref{eqn:sol_Hamf_2} can be further rewritten to
\begin{equation}
\label{eqn:sol_Hamf_2_alternative}
\begin{aligned}
\EE(t,\xx) &= \EE_0(\xx) + \int_0^t \sum_{\alpha}q_\alpha\left( \int_{\R^3} \vv f^\alpha\left(s, \xx, \vv\right) \dvv \right) \ds \\
&= \EE_0(\xx) + \int_0^t \sum_{\alpha}\jj\left( s,\xx \right) \ds.
\end{aligned}
\end{equation}

\begin{remark}
Note that the Gauss law for the electric \eqref{eqn:Gauss_law} and for the 
magnetic field \eqref{eqn:Gauss_law_magnetism} are preserved by construction 
through the solutions of these sub-systems if the Gauss law was fulfilled at 
$t=0$.\autocite{CROUSEILLES2015224}
\end{remark}

\subsection{Numerical Implementation purely through Hamiltonian splitting  }
\label{sec:nufi_vm}

Because the resulting sub-systems are all linear transport equations they admit
analytic solutions. Therefore, we can use operator-splitting techniques to
construct an approximate solution for the Vlasov--Maxwell 
system. \autocite{hairer2006, CROUSEILLES2015224} 

First we consider the first order in time \emph{Lie splitting} approach 
choosing the operator ordering
\begin{equation}
\label{eqn:chosen_order_Lie}
F^\alpha(t) = \exp(\HamE t) \exp(\HamB t) \exp(\Hamf t)  F^\alpha(0),
\end{equation}
where we denote the solution triplet as $F^\alpha(t) = (f^\alpha, \EE, \BB)$.

In the following we restrict ourselves to the case of periodic boundary 
conditions to simplify the derivation of the scheme. 
Similar to the electro-static scheme we need to evaluate the fields during the 
backtracking procedure, so in space we are using B-splines of order $m \in \N$ 
on a uniform grid to represent the fields.~\autocite{kirchhart2023numerical} 
For the time discretization we fix a time-step $\Delta t$. 

At a time $t = t_n$ we assume that we can evaluate $f^\alpha(t_n,\xx,\vv)$ for 
any $(\xx,\vv) \in \Omega \times \R^d$ and are given $\BB_0 \coloneqq \BB(t_n)$
as well as $\EE_0 \coloneqq \EE(t_n)$ such that they can be pointwise 
evaluated. Then we can derive a numerical scheme by composing the analytic 
solutions of the linear sub-equations for one time step.

Furthermore, we can use the fact that we are in a periodic domain to write 
our field update formulas in Fourier space. To this end we transform the 
field data on the spatial grid $\xx_g$ into Fourier space, i.\,e.,
$\hat{\EE}_0 \coloneqq \hat{\EE}_0(\xx_g)$ and 
$\hat{\BB}_0 \coloneqq \hat{\BB}_0(\xx_g)$.

Corresponding to $\Hamf$ we obtain
\begin{align}
f^\alpha_1(\xx,\vv) &= f^\alpha_0(t_n,\xx-\Delta t \vv,\vv), \label{eqn:Ham_f_1_num} \\
\BB_1(\xx) &= \BB_0(\xx), \label{eqn:Ham_f_2_num} \\
\EE_1(\xx) &= \EE_0(\xx) - \sum_{\alpha}q_\alpha \int_0^{\Delta t} \int_{\R^3} \vv f^\alpha(t_n,\xx - s \vv,\vv) \dvv \ds. \label{eqn:Ham_f_3_pre_num}
\end{align}

We can now rewrite equation \eqref{eqn:Ham_f_3_pre_num} into Fourier space for 
the $\kk$-th Fourier mode:
\begin{equation}
\label{eqn:Ham_f_3_Fourier_transform}
\hat{\EE}_{1,\kk} = \hat{\EE}_{0,\kk} - \sum_{\alpha} q_\alpha
\int_{t_n}^{t_{n+1}} \int_{\R^d} \vv \hat{f}^\alpha(t_n) \exp(-i \vv \kk \Delta t) \dvv \dt,
\end{equation}
where we use that equation \eqref{eqn:Ham_f_1_num} corresponds to 
\begin{equation}
\hat{f}_{1,\kk}^\alpha(\vv) = \hat{f}^\alpha_{\kk}(t_n) \exp(-i \vv \kk \Delta t)
\end{equation}
using the Fourier ansatz
\begin{equation}
f^\alpha(t,x,v) = \sum_{\kk} \hat{f}^\alpha(t,v) \exp(i \kk \cdot \vv).
\end{equation}
This allows to analytically solve the time-integral in equation 
\eqref{eqn:Ham_f_3_Fourier_transform}
\begin{align}
\!\! \int_{t_n}^{t_{n+1}} \!\!\!\! \int_{\R^d} \!\! \vv \hat{f}^\alpha(t_n,\vv) \exp(-i \vv \kk (s- t_n)) \dvv \ds &= \int_{\R^d} \vv \hat{f}^\alpha(t_n,\vv)  \int_{t_n}^{t_{n+1}} \exp(-i \vv \kk (s- t_n)) \dvv \ds \\
&= \int_{\R^d} \vv \hat{f}^\alpha(t_n,\vv)  \frac{\left( \exp(-i \vv \cdot \kk \Delta t) - 1 \right)}{-i \vv \cdot \kk} \dvv.
\end{align}

The singularity is resolvable via
\begin{equation}
\frac{\exp(-i \vv \cdot \kk \Delta t) - 1 }{ \vv \cdot \kk } = 
\frac{\exp(-i \vv \cdot \kk \Delta t) - 1 }{ \vv \cdot \kk 
\tfrac{\Delta t}{\Delta t}} = \Delta t \frac{\exp(-i \gamma)}{\gamma}
 \xlongrightarrow{\gamma \rightarrow 0} \Delta t,
\end{equation}
with $\gamma = \vv \cdot \kk \Delta t$. Thus the analytical solution to the 
integral becomes
\begin{equation}
\begin{aligned}
\jcal^\alpha(\kk) &\coloneqq
\int_{t_n}^{t_{n+1}} \int_{\R^d} \vv \hat{f}^\alpha(t_n,\vv) \exp(-i \vv \kk (s- t_n)) \dvv \ds  \\
&= \int_{\R^d} 
\begin{cases}
\vv \hat{f}^\alpha(t_n,\vv)  \frac{\left( \exp(-i \vv \cdot \kk \Delta t) - 1 \right)}{-i \vv \cdot \kk} &, \vert \vv \vert > 0 \\
\Delta t \vv \hat{f}^\alpha(t_n,\vv) &, \vert \vv \vert = 0
\end{cases}
 \dvv.
 \end{aligned}
\end{equation}

The remaining integral in velocity space is then resolved by a mid-point rule in truncated velocity space:

\begin{equation}
\label{eqn:definition_j_hat}
\jcalh^\alpha(\kk) \coloneqq 
\sum_{i_u=1}^{N^{\alpha}_u} \sum_{i_v=1}^{N^{\alpha}_v} 
\sum_{i_w=1}^{N^{\alpha}_w}
\begin{cases}
\vv^{h,\alpha}_{i_u,i_v,i_w} \hat{f}^\alpha(t_n,\vv^{h,\alpha}_{i_u,i_v,i_w})  \frac{\left( \exp(-i \vv^{h,\alpha}_{i_u,i_v,i_w} \cdot \kk \Delta t) - 1 \right)}{-i \vv^{h,\alpha}_{i_u,i_v,i_w} \cdot \kk} &, \vert \vv^{h,\alpha}_{i_u,i_v,i_w} \vert > \epsilon \\
\Delta t \vv^{h,\alpha}_{i_u,i_v,i_w} \hat{f}^\alpha(t_n,\vv^{h,\alpha}_{i_u,i_v,i_w}) &, \vert \vv^{h,\alpha}_{i_u,i_v,i_w} \vert \le \epsilon
\end{cases} 
\end{equation}
where 
\begin{equation}
\label{eqn:definition_velocity_dof}
\vv^{h,\alpha}_{i_u,i_v,i_w} = \begin{pmatrix}
\vmin^{1,\alpha} + (i_u + \tfrac{1}{2}) \Delta u \\
\vmin^{2,\alpha} + (i_v + \tfrac{1}{2}) \Delta v \\
\vmin^{3,\alpha} + (i_w + \tfrac{1}{2}) \Delta w 
\end{pmatrix}.
\end{equation}
We chose $\vmin^{i,\alpha}$ and $\vmax^{i,\alpha}$ as the cut-off in 
velocity boundaries as well as $N^{\alpha}_u, N^{\alpha}_v$ and $N^{\alpha}_w$ 
as number of quadrature points in velocity space, where we also allow to choose 
different discretization parameters for the different species $\alpha$.
The tolerance parameter $0 < \epsilon \ll 1$ is introduced for numerical 
stability of the formula. 

Finally we then obtain 
\begin{equation}
\jcalh(\kk) \coloneqq \sum_{\alpha} \jcalh(\kk)^\alpha
\end{equation}
and 
\begin{equation}
\hat{\EE}_{1,\kk} \approx \hat{\EE}_{0,\kk} - \jcalh(\kk).
\end{equation}

\begin{remark}
%\begin{itemize}
\begin{enumerate}
\item Because the truncation of the velocity space cannot be avoided one has to 
ensure that the loss of mass is negligible. In practice that means that the 
cut-off velocity for each species is chosen such that the distribution function 
only takes values close to or less than machine precision beyond the cut-off. 
For cases where strong heating is to be expected, i.\,e., the velocity support
significantly expands over time one has to either prescribe appropriate cut-off
velocities initially or, more efficiently, track the velocity support over 
time.~\autocite{Wilhelm_2025}
\item In equation \eqref{eqn:Ham_f_3_pre_num} we can also approximate the 
time-integral through a midpoint rule
\begin{equation}
\label{eqn:Ham_f_3_num}
\EE_1(\xx) = \tilde{\EE}_0(\xx) - \Delta t \sum_{\alpha}q_\alpha \int_{\R^3} \vv \tilde{f}^\alpha_0(\xx - \tfrac{\Delta t }{2} \vv,\vv) \dvv.
\end{equation}

This allows us to also transfer the NuFI approach to non-periodic boundary 
conditions. Note that this approximation is also $\mathcal{O}(\Delta t^2)$, 
i.\,e., second order in time. If we are using a higher order splitting, we can 
also choose a corresponding higher order time-integration scheme.
\item In principle we can also choose a different, e.g. higher order or 
adaptive, integration rule for the current density in this case. We chose 
midpoint, (following the choice from our previous work on the electro-static 
Vlasov equation \autocite{kirchhart2023numerical}) for the following benchmarks 
because for smooth initial conditions $f$ is also smooth and hence the 
mid-point rule converges exponentially fast in theory, while being easy to 
implement.\autocite{trefethen_trapezoidal} However, for complicated
multi-species simulations we suggest using the adaptive third order 
scheme (with adaptive velocity boundary) previously presented by Wilhelm et al. 
for multi-species, electro-static 
turbulence.~\autocite{multi_species_kin_instability_nufi}
%\end{itemize}
\end{enumerate}
\end{remark}

Next we discretize the equations corresponding with $\HamB$:
\begin{align}
f_2^\alpha(\xx,\vv) &= f_1^\alpha(\xx,\exp(-\tfrac{q_\alpha}{m_\alpha}\Delta t J_{\BB_1(\xx)})\vv), \label{eqn:Ham_B_1_pre_num} \\
\BB_2(\xx) &= \BB_1(\xx), \label{eqn:Ham_B_2_pre_num} \\
\EE_2(\xx) &= \EE_1(\xx) + \Delta t \nabla_{\xx} \times \BB_1(\xx). \label{eqn:Ham_B_3_pre_num}
\end{align}

Plugging \eqref{eqn:Ham_f_1_num} -- \eqref{eqn:Ham_f_3_num} into the above we
obtain

\begin{align}
f_2^\alpha(\xx,\vv) &= f_0^\alpha\left(t_n,\xx-\Delta t \left(\exp\left(-\tfrac{q_\alpha}{m_\alpha}\Delta t J_{\BB_0(\xx)}\right)\vv\right),\exp\left(-\tfrac{q_\alpha}{m_\alpha}\Delta t J_{\BB_0(\xx)}\right)\vv\right), \label{eqn:Ham_B_1_num} \\
\BB_2(\xx) &= \tilde{\BB}_0(\xx), \label{eqn:Ham_B_2_num} \\
\EE_2(\xx) &= \EE_0(\xx) - \hat{\jj}(\xx) + \Delta t \nabla_{\xx} \times \tilde{\BB}_0(\xx). \label{eqn:Ham_B_3_num}
\end{align}

In Fourier space this corresponds to
\begin{gather}
\label{eqn:Ham_B_3_num_Fourier}
\hat{\EE}_2(\kk) = \hat{\EE}_0(\kk) - \jcalh + \Delta t \kk \times \hat{\BB}_0(\kk).
\end{gather}

Finally we can also plug this into the analytic solution corresponding to 
$\HamE$ to obtain
\begin{align}
f^\alpha_3(\xx,\vv) &= f^\alpha_2(\xx,\vv - \Delta t \tfrac{q_\alpha}{m_\alpha} \EE_2(\xx)), \label{eqn:Ham_E_pre_num_1} \\
\BB_3(\xx) &= \BB_2(\xx) - \Delta t \nabla_{\xx} \times \EE_2(\xx), \label{eqn:Ham_E_pre_num_2}  \\
\EE_3(\xx) &= \EE_2(\xx). \label{eqn:Ham_E_pre_num_3}
\end{align}

Translating equation \eqref{eqn:Ham_E_pre_num_2} into Fourier space, 
plugging in that $\BB$  did not change during the evolution corresponding 
to $\HamB$ as well as $\Hamf$ and equation \eqref{eqn:Ham_B_3_num_Fourier}
we get
\begin{equation}
\hat{\BB}_3(\kk) = \hat{\BB}_0(\kk) - \Delta t \kk \times (\hat{\EE}_0(\kk) - \jcalh + \Delta t \kk \times \hat{\BB}_0(\kk)).
\end{equation}

So for the fields at time $t = t_{n+1}$ we arrive at
\begin{align}
\hat{\BB}(t_{n+1},\kk) &= \hat{\BB}_0(\kk) - \Delta t \kk 
\times (\hat{\EE}_0(\kk) - \jcalh + \Delta t \kk \times \hat{\BB}_0(\kk)), \label{eqn:B_final_NuFI_Ham_Lie} \\ 
\hat{\EE}(t_{n+1},\kk) &= \hat{\EE}_0(\kk) - \jcalh + \Delta t \kk \times 
\hat{\BB}_0(\kk). \label{eqn:E_final_NuFI_Ham_Lie}
\end{align}

Hence for $f^\alpha(t_{n+1})$ we obtain 
\begin{equation}
\label{eqn:f_final_NuFI_Ham_Lie}
\begin{aligned}
f^\alpha(t_{n+1},\xx,\vv) = f_0^\alpha(t_n,\xx - \Delta t ( \exp(- \Delta t \tfrac{q_\alpha}{m_\alpha} J_{\BB(t_n,\xx)}) 
( \vv - \Delta t \tfrac{q_\alpha}{m_\alpha} \EE(t_{n+1},\xx) ),  \\
\exp(- \Delta t \tfrac{q_\alpha}{m_\alpha} J_{\BB(t_n,\xx)}) ( \vv - \Delta t \tfrac{q_\alpha}{m_\alpha} \EE(t_{n+1},\xx))).
\end{aligned}
\end{equation}

These formulas can now be used to construct a scheme similar to the 
electro-static NuFI. For this we first define the numerical flow approximation given by the advection formulas \eqref{eqn:Ham_f_1_num}, \eqref{eqn:Ham_f_2_num} and \eqref{eqn:Ham_f_3_num}, and denote it for the time-step $n \in \N$ by
\begin{equation}
\label{eqn:definition_one_step_numerical_flow}
f^\alpha(t_{n+1}, \xx, \vv) = f_3^\alpha(\xx,\vv) 
= f^\alpha(\Psi_{t_n}^{t_{n+1},\alpha}(\xx,\vv)).
\end{equation}

Thus if the evolution of the electromagnetic fields $\EE$ and $\BB$ is known 
for $t=0,...,t_{n+1}$ we can evaluate $f^\alpha(t_{n+1})$ at an arbitrary 
position in phase space through
\begin{equation}
\label{eqn:eval_f_through_numerical_flow}
f^\alpha(t_{n+1},\xx,\vv) = f_0^\alpha\circ \Psi_{t_0}^{t_1} \circ \hdots \circ  
\Psi_{t_n}^{t_{n+1}} (\xx,\vv) = f_0^\alpha(\Psi_{t_0}^{t_{n+1}}(\xx,\vv)).
\end{equation}

Note that by construction, because we only need to know $f^\alpha(t_n)$ to 
compute the fields at $t_{n+1}$, we can use $\EE(t_{n+1})$ in formula 
\ref{eqn:definition_one_step_numerical_flow}. Hence we can now formulate the 
full NuFI algorithm for the multi-species Vlasov-Maxwell system using the 
Hamiltonian splitting (NuFI-Ham).

\begin{algorithm}[H]
\begin{algorithmic}
\Function{NuFI}{$f_0^\alpha$, $\BB_0$, $\EE_0$, $N_t$, $\Delta t$, 
$(N_{x_i}^\alpha)_{i,\alpha}$, $(N_{v_i}^\alpha)_{i,\alpha}$, 
$(\vmin^{i,\alpha})_{i,\alpha}$, $(\vmax^{i,\alpha})_{i,\alpha}$
}
  \State Interpolate $\EE_0$ and $\BB_0$ on spatial grid.
  \State Compute Fourier transformations $\hat{\EE}(0)$ and $\hat{\BB}(0)$.
  \State Compute $\jcalh(0)$ on spatial grid from $f_0$ 
  using \eqref{eqn:definition_j_hat}.  
\For{$n = 1,...,N_t$}
  \State Compute $\hat{\BB}(t_n)$ and $\hat{\EE}(t_n)$ using \eqref{eqn:B_final_NuFI_Ham_Lie} and \eqref{eqn:B_final_NuFI_Ham_Lie} from
   $\hat{\BB}(t_{n-1},\kk)$ and $\hat{\EE}(t_{n-1},\kk)$.
   \State Fourier-transform $\hat{\EE}(t_{n})$ and $\hat{\BB}(t_{n})$ back
   into real space and interpolate.
  \State Evaluate $\jcalh$ on spatial grid using \eqref{eqn:eval_f_through_numerical_flow} to evaluate $f(t_n)$ as well as \eqref{eqn:definition_j_hat}.
\EndFor
\EndFunction
\end{algorithmic}
\caption{ \label{alg:pure_nufi_vm}
Numerical Flow Iteration for the Vlasov--Maxwell system (NuFI-Ham).}
\end{algorithm}

Note that the quasi-neutrality condition \eqref{eqn:quasi_neutrality} implies 
through \eqref{eqn:Gauss_law} that the mean of the electric field is 0. 
Practically we have to fix the mean of $E_i$ to 0, as otherwise Amp\`ere's law 
can introduce unphysical drift into the simulation. This can be easily 
implemented during the computation of $\hat{\EE}$, where we just set the 
$\kk = 0$ mode to 0 for each component.

\begin{remark}
To evaluate formula \eqref{eqn:definition_j_hat} we have to compute the 
Fourier transform of $f^\alpha$. Note, however, that in contrast to a 
Semi-Lagrangian approach we want to avoid having to explicitly store the full
distribution function data on a grid. Instead a more efficient approach in 
terms of memory complexity is to iterate over the velocity degrees of freedom,
i.\,e., compute
\begin{equation}
A(\xx_{i_x,i_y,i_z}) = f^\alpha(t,\xx_{i_x,i_y,i_z},\vv)
\end{equation}
for $t=t_n$ and $\vv_{i_u,i_v,i_w}$ fixed, calculate the 
FFT~\autocite{fftw_reference} of $A$ and finally sum over the velocity degrees 
of freedom.
\end{remark}

\subsection{Higher order splitting in time}
\label{sec:nufi_ham_high_order}

The algorithm presented in the previous chapter is first order in time. Higher
orders in time can be achieved by using higher order Strang splitting. Similar
to the electro-static case we can also here -- through clever choice of 
operator ordering -- extend to 2nd order in time without significantly 
increasing the workload:
\begin{equation}
\label{eqn:chosen_order_Strang_2nd_order}
F^\alpha(\Delta t) = \exp(\HamE \tfrac{\Delta t}{2}) \exp(\HamB \tfrac{\Delta t}{2}) \exp(\Hamf \Delta t)  \exp(\HamB \tfrac{\Delta t}{2}) \exp(\HamE \tfrac{\Delta t}{2}) F^\alpha(0).
\end{equation}

The most expensive operation in NuFI is the evaluation of the distribution 
function. By placing the advection with respect to $\Hamf$ in the middle we 
only need to evaluate the distribution function once, i.\,e., the dominating 
discretization cost remains the same as for the Lie splitting.
Implementation-wise the major difference is that $f^\alpha$ is no longer 
evaluated at $t=t_n$ but instead at a time after the half-step advection 
$\exp(\HamB \tfrac{\Delta t}{2}) \exp(\HamE \tfrac{\Delta t}{2})$.

Note that even higher order in time can be achieved through even higher order 
splitting~\autocite{hairer2006}, however, that would actually involve higher 
computational cost as more evaluations of the distributions would be necessary.

\subsection{Conservation properties}
\label{sec:conservation_properties_nufi_theory}

One major advantage of NuFI in the electro-static limit is that due to the
symplectic nature of the time-stepping, the approach is exactly 
conserving.~\autocite{kirchhart2023numerical} In the following, we prove that 
an analogous result also holds for the electro-magnetic case.

\begin{lemma}[Conservation properties]
\label{lemma:conservation}
Let $(f_0, \EE_0, \BB_0)$ be sufficiently smooth such that a unique, classical 
solution to the Vlasov--Maxwell system 
\eqref{eqn:vlasov} -- \eqref{eqn:Gauss_law_magnetism} exists. Then for the 
numerical solution $f_h^\alpha$ produced by algorithm \ref{alg:pure_nufi_vm}
\begin{equation}
\int_{\Omega} \int_{\R^d} g(f^\alpha_h(t_n,\xx,\vv)) \dvv \dxx 
= \int_{\Omega} \int_{\R^d} g(f^\alpha_0(\xx,\vv)) \dvv \dxx 
\end{equation}
holds for any Lebesgue measurable function $g: \R \rightarrow \R$ and any $t_n$.
\end{lemma}
\begin{proof}
If the initial data is smooth enough such that a unique, classical solution 
exists, it coincides with the solution through the method of characteristics. 
In this case the approximate characteristics $\Psi_0^{t_n}$ computed by 
algorithm \ref{alg:pure_nufi_vm} exist and are also unique. It holds further
\begin{equation}
\Psi_0^{t_n}(\xx,\vv) = \Psi_0^{t_1} \circ  \Psi_{t_1}^{t_2} 
\circ ... \Psi_{t_{n-1}}^{t_n} (\xx,\vv).
\end{equation}

Further for any $i \in \{0,...,n\}$ we have
\begin{equation}
\Psi_{t_{i-1}}^{t_i} = \Psi_E \circ \Psi_B \circ \Psi_f,
\end{equation}
where by $\Psi_E$, $\Psi_B$ and $\Psi_f$ we denote the maps corresponding to 
the respective sub-Hamiltonians and fields at $t_{i-1}$. For readability we 
drop the time index for the fields and distribution function. We obtain
\begin{align}
\Psi_f(\xx,\vv) = \begin{pmatrix}
\xx - \Delta t \vv \\ \vv
\end{pmatrix} 
\Rightarrow D_{\xx,\vv} \Psi_f(\xx,\vv) = \begin{pmatrix}
I & - \Delta t I \\ 
0 & I
\end{pmatrix} 
\Rightarrow \text{det}\left( D_{\xx,\vv} \Psi_f(\xx,\vv) \right) = 1,
\end{align}
where by $I$ we denote the identity matrix in $\R^{3\times 3}$.

For 
\begin{equation}
\Psi_B(\xx,\vv) = \begin{pmatrix}
\xx \\ \exp(-\tfrac{q}{m} \Delta t J_B) \vv
\end{pmatrix}
\end{equation}
we use that det$(\exp(A)) = \exp($tr$(A))$ and $J_B$
has trace 0 to arrive at det$\left( D_{\xx,\vv} \Psi_B(\xx,\vv) \right) = 1$.

Analogously to $\Psi_f$, for $\Psi_E$ we compute
\begin{align}
\Psi_E(\xx,\vv) = \begin{pmatrix}
\xx \\ \vv - \Delta t \tfrac{q}{m} \EE
\end{pmatrix} 
\Rightarrow D_{\xx,\vv} \Psi_E(\xx,\vv) = \begin{pmatrix}
I & 0 \\ 
- \Delta t \tfrac{q}{m} D_{\xx}\EE & I
\end{pmatrix} 
\Rightarrow \text{det}\left( D_{\xx,\vv} \Psi_E(\xx,\vv) \right) = 1.
\end{align}

Combining the above and using that det$(A B) =$ det$(A)$det$(B)$ for any 
matrices $A,B\in \R^{3\times 3}$ we arrive at det$(\Psi_{t_{i-1}}^{t_i})=1$.
Consequently also det$(\Psi_{t_{0}}^{t_n})=1$ holds from which concludes the 
proof by using a change of variables.
\end{proof}

Due to the above Lemma we know that all $L^p$-norms as well as the entropy
$\int_{\Omega\times \R^d} f \log(f) \dxx\dvv$ are conserved analytically, 
i.\,e., if one was able to evaluate the involved integrals involved exactly, 
one would see that their values remain constant for all times. Note that this 
result holds independently of the choice of discretization. It is easy to see 
that the same proof also extends to higher order splitting results.

\begin{remark}
The Vlasov--Maxwell system admits only a Poisson bracket with non-canonical 
variables, i.\,e., the system is not a Hamiltonian system in a strict but 
rather non-canonical sense.~\autocite{MORRISON1980383,hairer2006}
We cannot expect exact energy preservation, however, the energy error remains 
bounded over long times, in contrast to methods that do not respect the 
Hamiltonian structure. ~\autocite{morrison2017}
\end{remark}

As in our algorithm \ref{alg:pure_nufi_vm} we use Amp\`ere's law to update 
the electric field it is not trivially clear why the Gauss law should be 
fulfilled. Hence in the following we proof an estimate of the Gauss law error.

\begin{lemma}[Gauss law error]
\label{lemma:gle}
Let $(f_0, \EE_0, \BB_0)$ be sufficiently smooth that a unique, classical 
solution to the Vlasov--Maxwell system 
\eqref{eqn:vlasov} -- \eqref{eqn:Gauss_law_magnetism} exists. 
Then the Gauss law error is bounded from above by the velocity integration 
error for the current density, i.\,e.,
\begin{equation}
\Vert \nabla_{\xx} \cdot \EE_h(t) - \rho_h(t) \Vert_{L^\infty} 
\lesssim \Vert \jcalh - \jcal \Vert_{L^\infty}, 
\end{equation}
where by $\EE_h(t)$ and $\rho_h(t)$ we denote the numerical solution 
by algorithm \ref{alg:pure_nufi_vm}.
\end{lemma}
\begin{proof}
For the subflows associated with $\HamE$ and $\HamB$ we can see,
following the argumentation from the proof of Lemma \ref{lemma:conservation},
that the corresponding transformation $\Psi_E$ and $\Psi_B$ are volume 
preserving in velocity and hence starting from $f_0^\alpha$ we have
\begin{equation}
\int_{\R^3} \sum_{\alpha} f^\alpha(0,\Psi(\xx,\vv)) \dvv 
= \int_{\R^3} \sum_{\alpha} f^\alpha(0,\xx,\vv) \dvv 
\end{equation}
for $\Psi = \Psi_E, \Psi_B$. For $\HamE$ we have $\partial_t \EE = 0$ and for $\HamB$ we have
\begin{equation}
\partial_t \EE(t,\xx) = \nabla_{\xx} \times \BB(t,\xx) 
\Rightarrow \nabla_{\xx} \cdot \left(\partial_t \EE(t,\xx) \right) = 
\nabla_{\xx} \cdot \left( \nabla_{\xx} \times \BB(t,\xx) \right) = 0.
\end{equation}
Therefore the Gauss law is exactly preserved during the subflows associated to 
$\HamE$ and $\HamB$. For the subflow associated to $\Hamf$, assuming exact integration, we have 
\begin{align}
\nabla_{\xx} \cdot \EE(t,\xx) &= \nabla_{\xx} \cdot \EE(0,\xx)
- \sum_{\alpha} q_\alpha \int_0^t \int_{\R^3} \vv \cdot \nabla_{\xx} 
\left( f_0^\alpha(\xx - s \vv,\vv)\right) \dvv \ds \\
&= \sum_{\alpha} q_\alpha \int_{\R^3} f_0^\alpha(\xx,\vv) \dvv
- \sum_{\alpha} q_\alpha \int_0^t \int_{\R^3} \partial_s \left( 
f_0^\alpha(\xx-s\vv,\vv) \right) \dvv \ds \\
&= \sum_{\alpha} q_\alpha \int_{\R^3} f_0^\alpha(\xx,\vv) \dvv
- \sum_{\alpha} q_\alpha \int_{\R^3} \left( f_0^\alpha(\xx-t\vv,\vv)
 - f_0^\alpha(\xx,\vv) \right) \dvv \ds \\
&= \sum_{\alpha} q_\alpha \int_{\R^3} f^\alpha(t,\xx,\vv) \dvv.
\end{align}

In our numerical implementation of $\Hamf$ we cannot calculate the 
involved integrals exactly: Using the transformation to Fourier space we 
evaluate the integration in time over the charge current exactly, but the 
velocity space integral to obtain the charge current itself is only an 
approximation. Hence the error introduced during each time-step in the Gauss 
law is bounded by the velocity integration error made during the calculation 
of $\jcalh$.
\end{proof}

\begin{remark}
Note that if we were to extend algorithm \ref{alg:pure_nufi_vm} to a non-
periodic setting where the exact integration-in-time through the Fourier 
transformation does no longer work, we would additionally introduce a 
time-discretization-error through the involved time-integral.
\end{remark}

\subsection{Spline Representation, Oscillations and Gauss filtering}
\label{sec:spline_issues_and_gauss_filter}

In algorithm \ref{alg:pure_nufi_vm} we are updating our fields in Fourier 
space, but we are not using this spectral representation to evaluate them.
During the NuFI backtracking procedure we have to repeatedly evaluate the 
fields both quickly and accurately. A spectral representation is less suitable 
for this as the evaluation of exponential or trigonometric basis functions is 
an expensive operation. A good compromise both in terms of accuracy and 
speed is a local B-spline representation, i.\,e., after calculating the fields, 
we transform them back to real space and calculate their coefficients in 
B-spline basis, analogous to the electro-static version of 
NuFI.~\autocite{kirchhart2023numerical}

\begin{figure}[h!]
\centering
\begin{subfigure}{0.49\textwidth}
\includegraphics[width=0.99\textwidth]{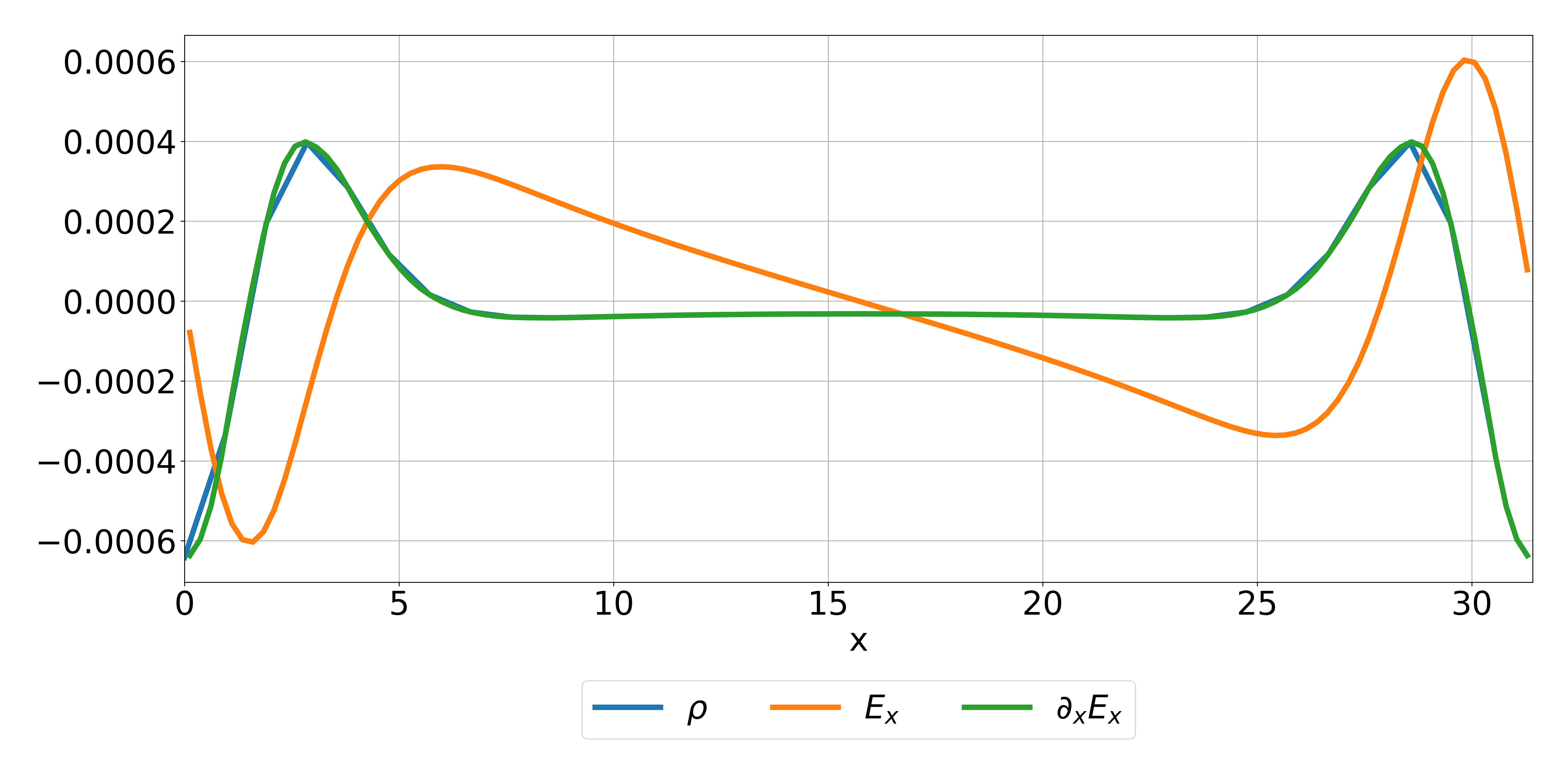}
\subcaption{\label{fig:rho_Ex_dxEx_70}
$\rho, E_x$ and $\partial_x E_x$ at $t=70$.
}
\end{subfigure}
\begin{subfigure}{0.49\textwidth}
\includegraphics[width=0.99\textwidth]{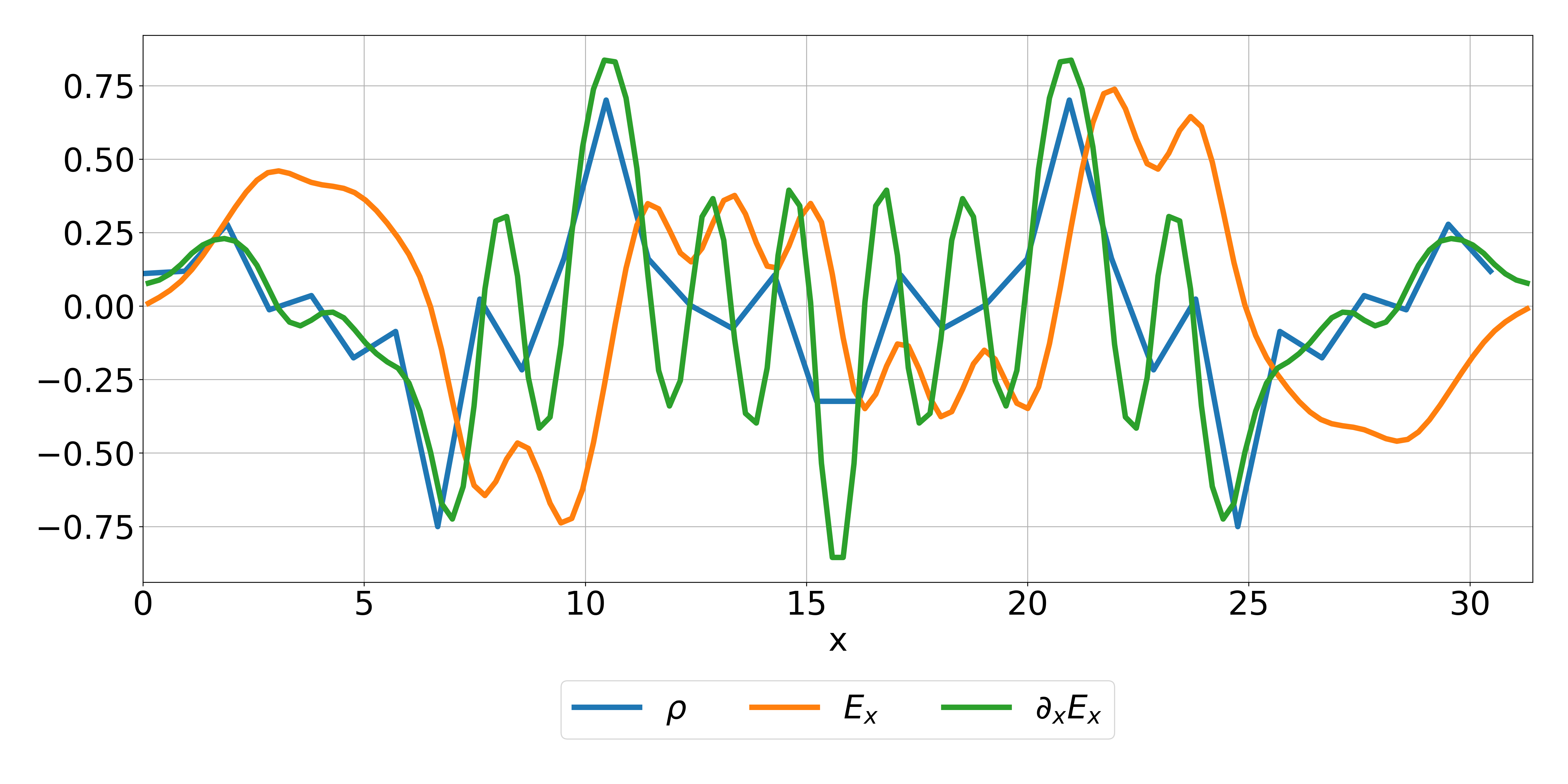}
\subcaption{\label{fig:rho_Ex_dxEx_100}
$\rho, E_x$ and $\partial_x E_x$ at $t=100$.
}
\end{subfigure}
\begin{subfigure}{0.49\textwidth}
\includegraphics[width=0.99\textwidth]{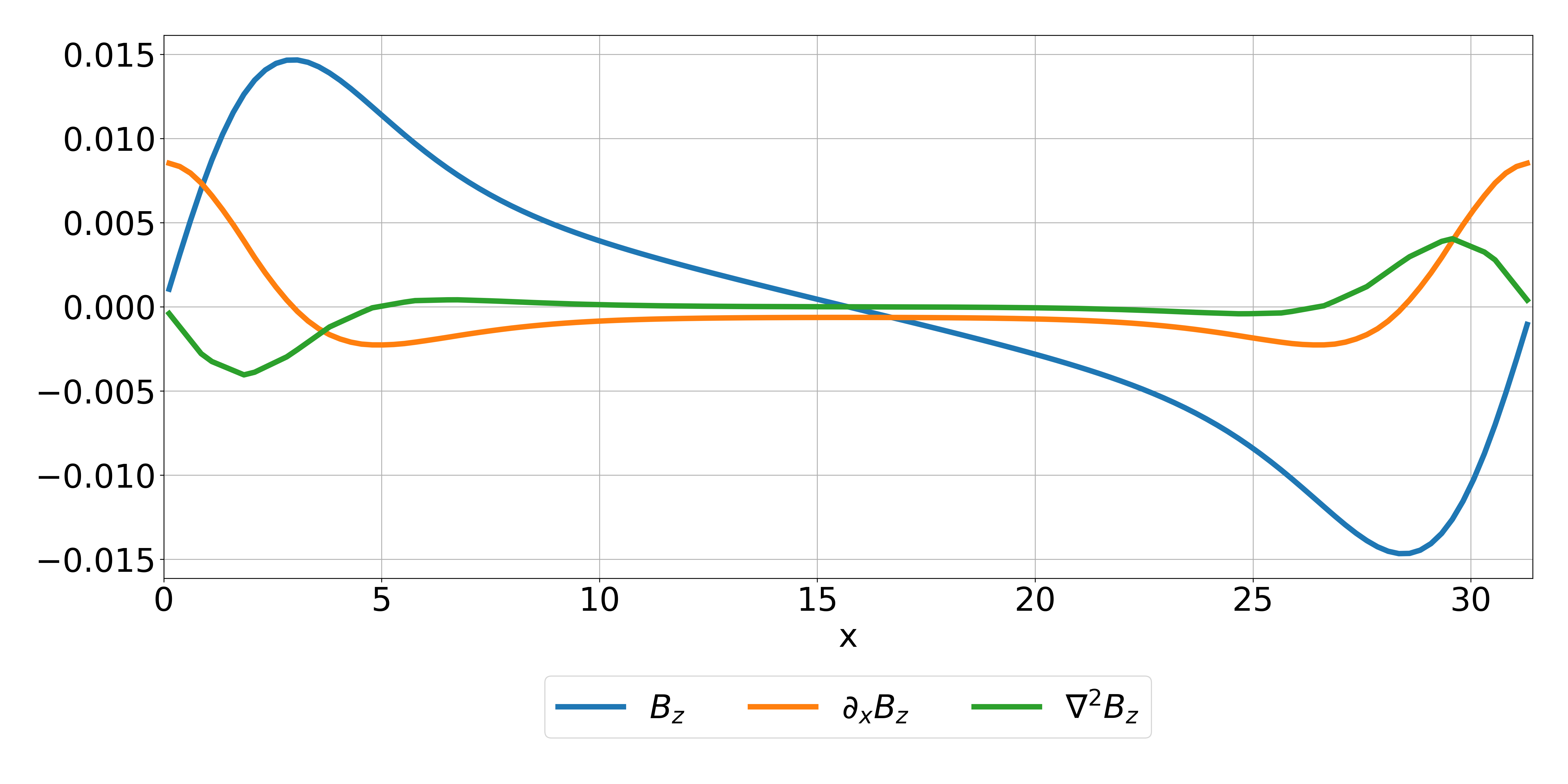}
\subcaption{\label{fig:B_dxBz_dxdxBz_70}
$B_z, \partial_x B_z$ and $\partial_x^2 B_z$ at $t=70$.
}
\end{subfigure}
\begin{subfigure}{0.49\textwidth}
\includegraphics[width=0.99\textwidth]{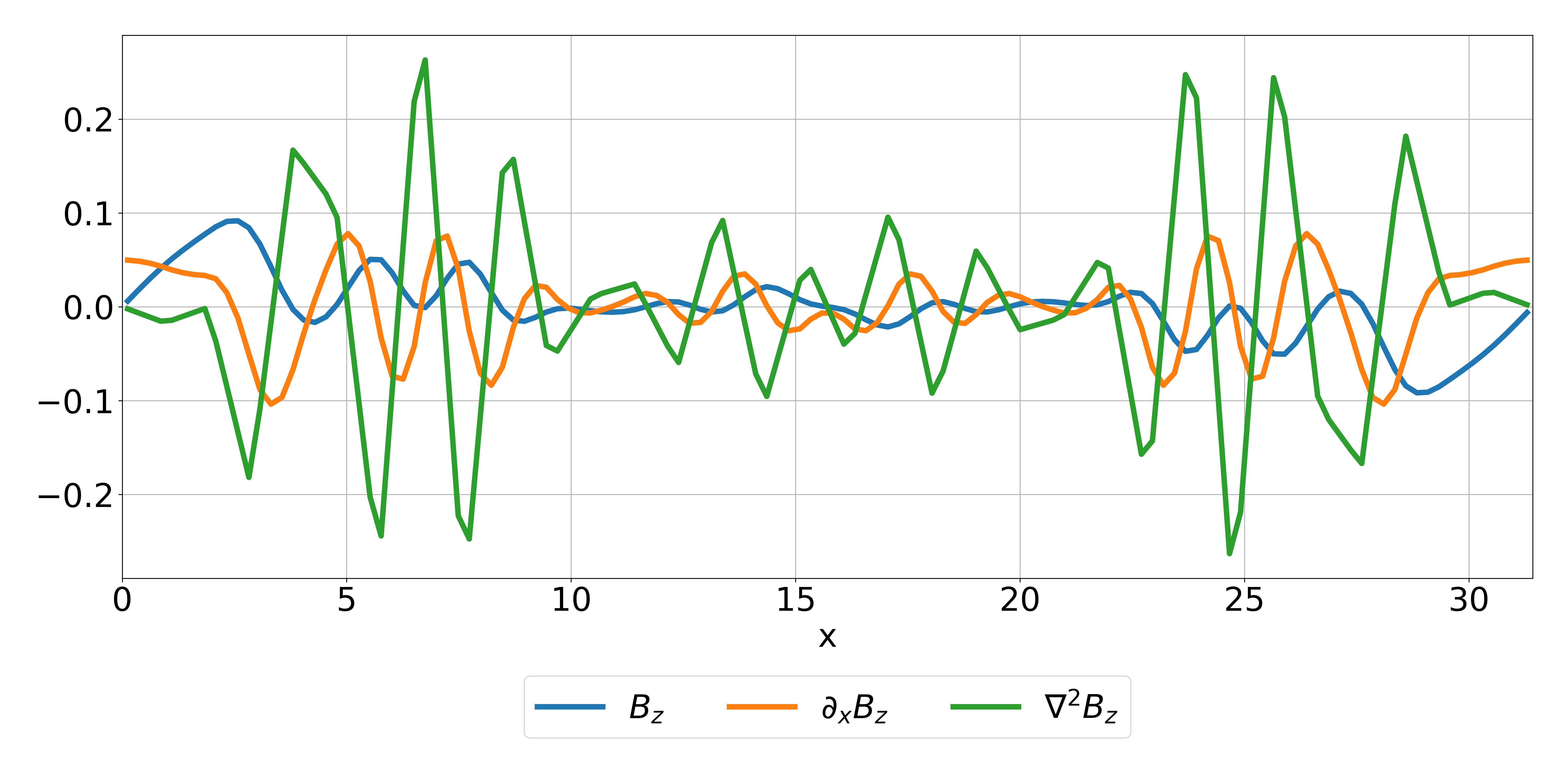}
\subcaption{\label{fig:B_dxBz_dxdxBz_100}
$B_z, \partial_x B_z$ and $\partial_x^2 B_z$ at $t=100$.
}
\end{subfigure}
\caption{\label{fig:rho_E_B}
Fields taken from a (unfiltered) simulation of the 1x2v \emph{streaming Weibel 
instability}, see section \ref{sec:streaming_weibel}, with a phase-space 
resolution of $32^3$. 
At time $t=70$, which is late in the linear growth stage of the instability, 
the fields are still smooth and there is an overall good agreement between 
$\rho$ and $\partial_x E_x$, i.\,e., the Gauss law is well satisfied. Later at 
$t=100$, which is during the non-linear stage of the instability, we observe 
some significant overshoots and oscillations in the spline representation. 
This deteriorates the Gauss law conservation and, more importantly, can also 
lead to unphysical dynamics such as overshooting the correct energy level.
}
\end{figure}

For the Vlasov--Maxwell system we are updating the electric field via  
Amp\`ere's law \eqref{eqn:Amperes_circuit_law}. This is a pure advection 
equation meaning that any oscillations present in the system are propagated 
and there is no mechanism in this update to dampen them. Additionally, because 
of the splitting we are taking higher order derivatives -- 2nd order for Lie
and 4th order for Strang splitting -- which in turn can worsen the problem.
For lower resolution splines we noticed that these oscillations can translate
into overshoots in the spline representation, see figure \ref{fig:rho_E_B}. 
In the worst case these (potentially unphysical) oscillations can lead to 
numerical instability, especially for some more complicated, slowly developing 
kinetic instabilities such as the later discussed \emph{streaming Weibel 
instability}. 

\begin{remark}
Note that this is not a purely NuFI-related problem and is also present to 
varying extent in other schemes. Notably PIC methods are well-known to 
become substantially more stable if one introduces artificial smoothing on 
the electromagnetic fields (or their source terms).~\autocite{CHEN20117018}
On the one hand, for PIC the origin of the oscillations is (statistical) noise, 
while for NuFI, on the other hand, it is rather the high (subgrid) resolution 
in phase-space, capturing small-scale oscillations. Hence both approaches seem 
to suffer from similar numerical stability issues, albeit resulting from very 
different origins.

Notably, in our experiments with the Semi-Lagrangian method described by 
Crouseilles et al.~\autocite{CROUSEILLES2015224} building upon the same 
Hamiltonian splitting, we noticed similar oscillation-related stability issues,
however, to a lesser degree. We suspect that grid-based methods are less prone 
to these oscillation-related issues due to the present numerical diffusion 
through the phase-space grid.

Finally, let us also note that such problems are absent for the electro-static
NuFI approach as the Poisson equation used to calculate the electro-static 
potential effectively acts a smoothing operator. 
\end{remark}

How to best tackle the oscillation-issue is currently part of our on-going 
research and will be presented in more detail in a forthcoming manuscript. 
For the present work we restrict ourselves to two simple algorithmic additions, 
\emph{Gauss cleaning} and \emph{exponential filtering}, which improve stability 
sufficiently well for the test case we discuss later.

\emph{Gauss cleaning} describes the process, where we explicitly enforce the 
Gauss law after each time step, i.\,e., calculate $\rho(t)$ and solve
\begin{equation}
\Delta_{\xx} \varphi_{\text{defect}}(t,\xx) = \nabla_{\xx} \cdot \EE(t,\xx) - \rho(t,\xx)
\end{equation}
for the potential of the Gauss defect. Then we can correct the electric field
via 
\begin{equation}
\EE_{\text{corrected}}(t,\xx) = \EE(t,\xx) 
- \nabla_{\xx} \varphi_{\text{defect}}(t,\xx).
\end{equation}

Note that while Lemma \ref{lemma:gle} gives us a theoretical bound on the Gauss
law error, in practice the error in the electric field is susceptible to errors
in the field representation. Hence, errors induced by the unphysical 
oscillations of the spline-representation of the fields, can manifest in the 
Gauss law error. In our numerical experiments we noticed that Gauss law 
cleaning turns out to be a sufficiently effective mean to suppress such 
oscillations.

In cases where strong oscillations first appear in the magnetic field it can be
beneficial to additionally introduce a filter for fast oscillating modes before
the spline-interpolation step. To suppress high-frequency oscillations, we 
apply a smooth exponential filter directly to the Fourier coefficients. 
For a grid of size $N_x$, we define the maximum resolved wavenumber as 
$k_{\max} = N_x/2$ and introduce the normalized wavenumber
\begin{equation}
\eta = \frac{|k|}{k_{\max}} \in [0,1].
\end{equation}
Each Fourier mode is then multiplied by a damping factor
\begin{equation}
\sigma(\eta) = \exp\!\bigl(-\alpha \, \eta^p\bigr),
\end{equation}
where $\alpha > 0$ controls the overall strength of the filter and $p$ 
determines its order. This filter leaves low-frequency modes essentially 
unchanged while exponentially damping modes close to the Nyquist frequency. 
Note that this is different from the numerical diffusion introduced by 
a phase-space grid, as we are only removing potential numerical artefacts 
below the spatial grid resolution used for the field. This does not reduce
our phase-space resolution.

\subsection{Combing NuFI with a Predictor-Corrector Maxwell update}
\label{sec:nufi_vm_predictor_corrector}

In section \ref{sec:nufi_vm} we discussed using Lie splitting for both the NuFI 
advection for $f^\alpha$ and the electro-magnetic fields. While this NuFI-Ham
approach is structure-preserving, we also see that it entails some potential 
for stability issues especially related to the spline representation and the
high order derivatives that have to be taken when composing the subflows. 
Another alternative to the smoothing approaches discussed in section 
\ref{sec:spline_issues_and_gauss_filter} is to replace the Maxwell update 
arising from the Hamiltonian splitting through another scheme such as the
\emph{predictor-corrector Maxwell solver}~\autocite{birdsall_plasma_1985}, 
which relies on at most first order derivatives.

Following the same recipe as in section \ref{sec:nufi_vm} using the operator order 
\begin{equation}
\label{eqn:Lie_E_B_f}
F^\alpha(t) = \exp(\Hamf) \exp(\HamB) \exp(\HamE) F^\alpha(0),
\end{equation}
we get by applying $\HamE$ to $F^\alpha(0) = (f^\alpha_0,\EE_0,\BB_0)$
\begin{equation}
\label{eqn:eval_f_1_Lie_E_B_f}
f^\alpha_1(\xx,\vv) = f^\alpha_0(\xx, \vv - \Delta t \tfrac{q_\alpha}{m_\alpha} \EE_0(\xx)).
\end{equation}
After applying $\HamB$ we get
\begin{equation}
\label{eqn:eval_f_2_Lie_E_B_f}
f^\alpha_2(\xx,\vv) = f^\alpha_1(\xx, \exp(- \tfrac{q_\alpha}{m_\alpha} 
\Delta t J_{\BB_1(\xx)}) \vv - \Delta t \tfrac{q_\alpha}{m_\alpha} \EE_0(\xx)).
\end{equation}

We have
\begin{equation}
\BB_1(\xx) = \BB_0 - \Delta t \nabla_{\xx} \times \EE_0(\xx)
\end{equation}
and for simplicity of notation we denote
\begin{equation}
\vv^\alpha_2(\xx) = \exp(- \tfrac{q_\alpha}{m_\alpha} \Delta t 
J_{\BB_0 - \Delta t \nabla_{\xx} \times \EE_0(\xx)}) \vv - \Delta t \tfrac{q_\alpha}{m_\alpha} \EE_0(\xx).
\end{equation}

Finally, we can apply $\Hamf$
\begin{equation}
\label{eqn:eval_f_3_Lie_E_B_f}
f^\alpha_3(\xx,\vv) = f^\alpha_2(\xx - \Delta t \vv, v)
\end{equation}
and hence the NuFI-update formula becomes
\begin{equation}
\label{eqn:f_new_lie_split}
f^\alpha(t_{n+1},\xx,\vv) = f^\alpha(t_n, \tilde{\xx}, \vv^\alpha_2(\tilde{\xx})),
\end{equation}
where we denote $\tilde{\xx} = \xx - \Delta \vv$. 

The new NuFI-update formula does not depend on intermediate evaluations of the 
current density, which means that we can now predict $\rho^{n+1}$ as well as 
$\jj^{n+1}$ without knowledge of the electro-magnetic fields at $t^{n+1}$. 
For our implementation we chose the predictor-corrector 
approach~\autocite{birdsall_plasma_1985}: Discretizing $\EE$ and $\BB$ on a 
uniform in space but staggered in time grid we use
\begin{gather}
\label{eqn:E_predictor_corrector}
\EE_{n+1} = \EE_n + \Delta t \nabla_{\xx} \times \BB_{n+1/2} 
- \tfrac{\Delta t}{2} (\jj^n + \jj_{n+1} ), \\
\label{eqn:B_predictor_corrector}
\BB_{n+1/2} = \BB_{n-1/2} - \Delta t \nabla_{\xx} \times \EE_n.
\end{gather}
The thus obtained values are then interpolated on the spatial grid using 
B-Splines. Because we require $\BB^n$ for \eqref{eqn:f_new_lie_split}, we also
compute and interpolate
\begin{equation}
\label{eqn:B_average}
\BB_n = \tfrac{1}{2}(\BB_{n+1/2} + \BB_{n-1/2}).
\end{equation}
Furthermore, if we only prescribe $\BB_0$ then $\BB_{-1/2}$ can be predicted 
via
\begin{equation}
\label{eqn:predict_B_-1/2}
\BB_{-1/2} = \BB_0 + \tfrac{\Delta t}{2} \nabla_{\xx} \times \EE_0.
\end{equation}

This leaves us with the following NuFI-PC algorithm
\begin{algorithm}[H]
\begin{algorithmic}
\Function{NuFI}{$f_0^\alpha$, $\BB_0$, $\EE_0$, $N_t$, $\Delta t$, 
$(N_{x_i}^\alpha)_{i,\alpha}$, $(N_{v_i}^\alpha)_{i,\alpha}$, 
$(\vmin^{i,\alpha})_{i,\alpha}$, $(\vmax^{i,\alpha})_{i,\alpha}$
}
  \State Interpolate $\EE_0$ and $\BB_0$ on spatial grid.
  \State Compute and interpolate $\BB_{-1/2}$ via \eqref{eqn:predict_B_-1/2}.
  \State Compute $\jj_0$ on spatial grid using \eqref{eqn:f_new_lie_split} and 				a midpoint rule.
  \State Compute and interpolate $\BB_{1/2}$ using 		 
         \eqref{eqn:B_predictor_corrector}.
\For{$n = 1,...,N_t$}
   \State Compute $\jj_n$ on spatial grid using \eqref{eqn:f_new_lie_split} 
   			and a midpoint rule.
    \State Compute and interpolate $\EE^n$ using 
           \eqref{eqn:E_predictor_corrector}.
   \State Compute and interpolate $\BB_{n+1/2}$ using 		 
         \eqref{eqn:B_predictor_corrector}.
    \State Compute and interpolate $\BB^n$ using \eqref{eqn:B_average}.
    \State $\jj_{n-1} = \jj_n$.
\EndFor
\EndFunction
\end{algorithmic}
\caption{ \label{alg:nufi_vm_pc}
NuFI for the Vlasov--Maxwell system coupled with Predictor-Corrector. (NuFI-PC)}
\end{algorithm}
\section{Reducing the Computational Complexity of NuFI}\label{sec:restarted_nufi_vm}

As discussed in section \ref{sec:nufi_electro_static}, NuFI has a 
computational complexity of $\mathcal{O}( n_t^2 )$ both in the electro-static 
as well as now in the electro-magnetic case. As we explicitly discretize  
Maxwell's equations in the fully electro-magnetic case, we also introduce a 
CFL-type stability restriction on our approach, which restricts the choice of 
time-step size and requires us to make more steps. Hence the simulation of long 
time periods may become prohibitively expensive with the pure NuFI approach.

The quadratic computational complexity in time can be reduced back to linear by 
restarting NuFI through storing a snapshot of the distribution functions for 
each species each $n_t^r > 0$ time-steps and replacing $f^\alpha_0$ by the 
stored snapshot $f_r$. 
This is similar to a Semi-Lagrangian approach with sub-cycling for all species, 
however, without the need to store the intermediate results, and in fact if we 
let $n_t^r \rightarrow 1$, then this approach becomes equivalent to a classic 
backwards Semi-Lagrangian scheme.~\autocite{multi_species_kin_instability_nufi, wilhelm2025restartingnumericalflowiteration} 

In the prototype implementation used in this work we have implemented cubic 
interpolation on a uniform grid to store the snapshots. A simple implementation 
consists of computing values of the distribution function on a uniform grid in 
phase-space and evaluating in-between points using a cubic stencil (e.g. with a 
Catmull-Rom spline).~\autocite{CATMULL1974317}

While quite simplistic, it is sufficient to resolve the test cases, which we 
are interested in as a proof-of-concept in section \ref{sec:num_tests}. 
Furthermore, this storage option is naturally compatible with low rank 
compression. \autocite{kormann_tensor_train, wilhelm2025restartingnumericalflowiteration}
Additionally, with $n_t^r \rightarrow 1$ this implementation reduces back to 
the classic cubic-spline backwards Semi-Lagrangian approach commonly employed.

Note, however, that this procedure again introduces some amount of numerical 
diffusion back into the scheme and therefore does not preserve neither mass nor 
energy. We will look into this in section \ref{sec:conservation_properties_restarted_nufi}.
Improved restart procedures are under investigation in on-going work. In 
particular, a promising alternative is -- instead of storing the distribution 
function -- to store the flow map directly following the Characteristic Mapping 
Method (CMM) approach.~\autocite{Krah2023905} This approach has been recently 
explored for the electro-static case and is promising due to its additional 
structure-preservation properties in comparison to direct discretization of the 
distribution function.~\autocite{cmm_nufi} 
There is also currently on-going research for structure-preserving sparse~\autocite{POLLINGER2023112338, oblapenko2026}
or low-rank representations~\autocite{EINKEMMER2023112060, guo_mei_2024}, which are also promising directions for NuFI as 
restart techniques.
\section{Numerical Tests}\label{sec:num_tests}

In the following we are testing algorithms \ref{alg:pure_nufi_vm} and
\ref{alg:nufi_vm_pc} for a number of classical benchmarks. 
First we look at \emph{weak Landau damping} to test the consistency of the 
Vlasov--Maxwell solver in the electro-static limit. 
Then we look at a \emph{streaming Weibel instability} as well as a
\emph{filamentation instability} in $1x2v$ and $2x3v$.

The implementation used for the simulations can be found in the NuFI GitHub
repository\footnote{\url{https://github.com/paulwilhelmvlasov/NumericalFlowIteration}} in the branch \emph{tensor\_svd\_approx}. 
The implementation is natively in up to six dimensions and allows for electrons 
as well as ions of arbitrary mass and charge. 

In the following we will also be comparing to an implementation of the 
algorithm suggested by Crouseilles et al.\autocite{CROUSEILLES2015224}. We abbreviate their Semi-Lagrangian approach with Hamiltonian splitting by SL-Ham.

\subsection{Weak Landau Damping}
\label{sec:weak_landau_damping}

Before we move on to show results of fully electro-magnetic simulations, 
we verify that our implementation is consistent with the electrostatic limit. 
To this end we consider the classic \emph{weak Landau damping} benchmark. For 
the initial data we have
\begin{equation}
\label{eqn:f0_weak_landau}
f_0(x,v) = \tfrac{1}{\sqrt{2 \pi}} (1 + \alpha \cos(k x)) 
\exp(-\tfrac{v^2}{2}),
\end{equation}
where $\alpha = 0.01$ and $k = 0.5$. The initial electric field is 
\begin{equation}
\EE_0(x) = \begin{pmatrix}
-\tfrac{\alpha}{k} \sin(kx) \\ 0 \\ 0
\end{pmatrix}
\end{equation}
and $\BB_0 = 0$. The $y$ and $z$ direction are set to be uniform and for the 
$x-v_{x}$ phase-space we choose $[0,\tfrac{2 \pi}{k}] \times [-5,5]$. As base 
resolution for both NuFI-Ham and NuFI-PC we choose $N_x = 32$ and $N_v = 32$.

\begin{figure}[h!]
\centering
\includegraphics[width=0.9\textwidth]{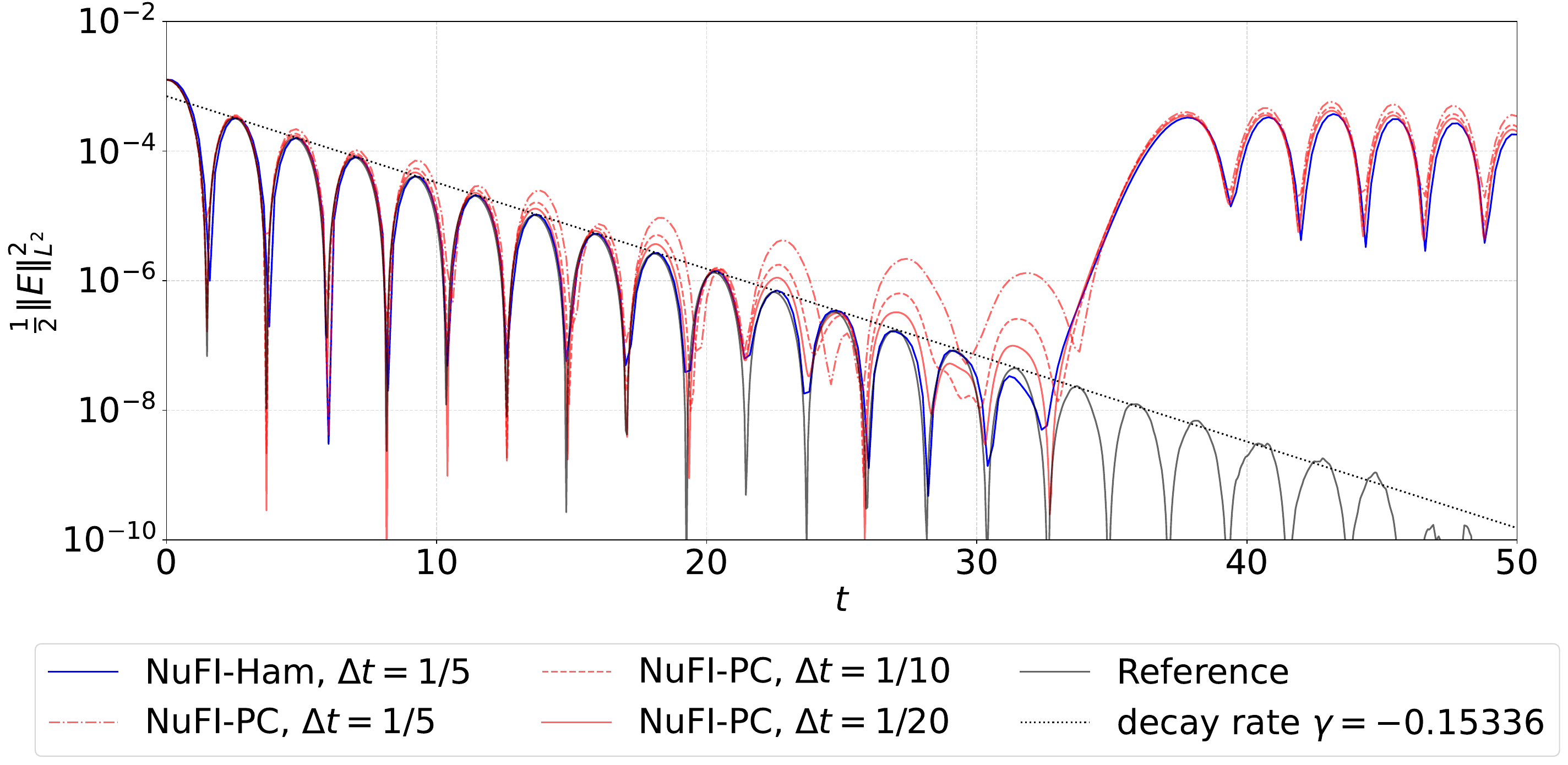}
\caption{\label{fig:weak_landau}
Comparison between NuFI-Ham, NuFI-PC and a reference solution (electro-static 
NuFI) of the \emph{Weak Landau Damping} benchmark. The analytically predicted
decay rate is $\gamma = -0.15336$, which is captured by all approaches, 
however, less precisely by NuFI-PC than NuFI-Ham.
}
\end{figure}

In figure \ref{fig:weak_landau} we show the evolution of the electric energy 
over time. The expected damping rate $\gamma = -0.15336$ of the electric field
is recovered exactly by NuFI-Ham even with a time-step as large as 
$\tfrac{1}{5}$, however, NuFI-PC is less accurate with the large time step
and exhibits some minor energy overshoots after $t \approx 15$, but recovers 
after the recurrence.
Decreasing $\Delta t$ also decreases the strength of the overshoots, which 
suggests that even though both NuFI-Ham and NuFI-PC are technically first order
in time the field solver coming from the Hamiltonian splitting is more accurate 
than the predictor-corrector scheme employed for NuFI-PC.

Both NuFI-Ham and NuFI-PC are consistent with the electro-static limit and
$\BB = 0$ is preserved throughout the entire simulation.

\subsection{Filamentation Instability}
\label{sec:filamentation_instability}

Next we consider a \emph{filamentation instability}, a fully electro-magnetic 
variant of the \emph{two stream instability}. The initial condition are two 
beams in the $v_y$ direction, i.\,e., perpendicular to the dominant spatial 
direction:
\begin{equation}
\label{eqn:f0_filamentation}
f_0(x,u,v) = \tfrac{1}{2} \exp\left( -\tfrac{v^2}{2\vth^2}\right) 
\left(  \exp\left( -\tfrac{(u-v_b)^2}{2\vth^2}\right)
+ \exp\left( -\tfrac{(u+v_b)^2}{2\vth^2}\right) \right),
\end{equation}

where the thermal velocity is uniformly set to $\vth = 0.1$ and the beam
velocity to $v_b = 0.4$. The instability is triggered through an initial  
magnetic field
\begin{equation}
\BB_0(x) = \begin{pmatrix}
0 \\ 0 \\ \beta \cos(k x)
\end{pmatrix},
\end{equation}
with perturbation strength $\beta = 10^{-3}$ and mode $k = 2$. The spatial 
domain is $[0,\tfrac{2\pi}{k}]$ and the velocity domain is truncated to 
$[-1,1] \times [-1.2,1.2]$. 

As baseline resolution in phase-space we use $32^3$ for all approaches. In time
we discretize with Lie splitting and $\Delta t = \tfrac{1}{20}$ for NuFI-PC as
well as SL-Ham. For NuFI-Ham with Lie-splitting we had to take 
$\Delta t = \tfrac{1}{30}$\footnote{Preliminary numerical experiments show that 
we otherwise are too close to the stability limit of the Fourier-based Maxwell 
solver.}. Additionally, we also consider NuFI-PC with a phase-space resolution 
of $64^3$.

\begin{figure}[h!]
\centering
\includegraphics[width=0.79\textwidth]{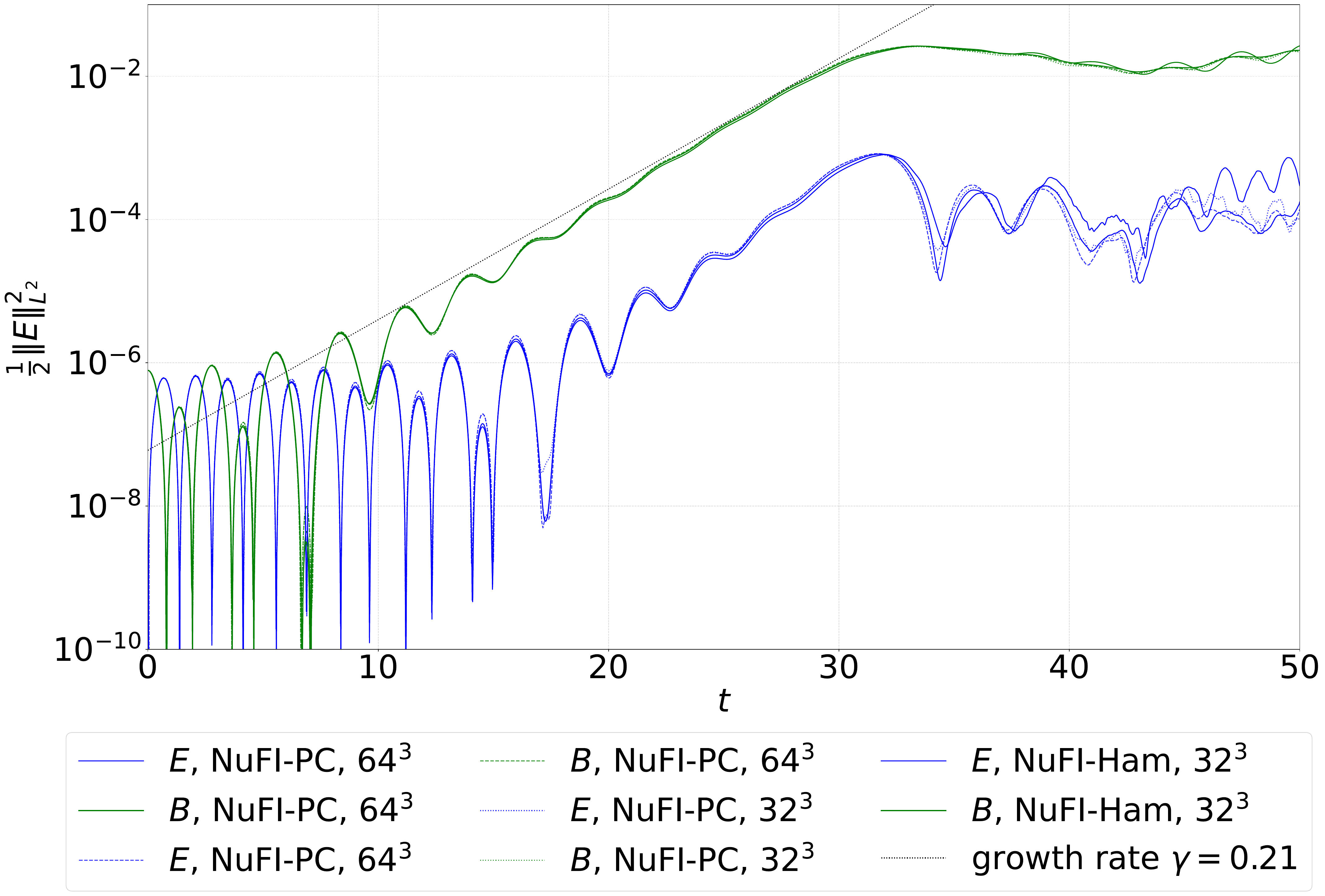} 
\caption{\label{fig:energies_filamentation_instability_nufi_ham_vs_pc}
Evolution of electric and magnetic energy over time for a simulation of the 
\emph{Filamentation instability} compared between NuFI-Ham and NuFI-PC without 
restarts.
}
\end{figure}

\begin{figure}[h!]
\centering
\begin{subfigure}{0.325\textwidth}
\includegraphics[width=0.99\textwidth]{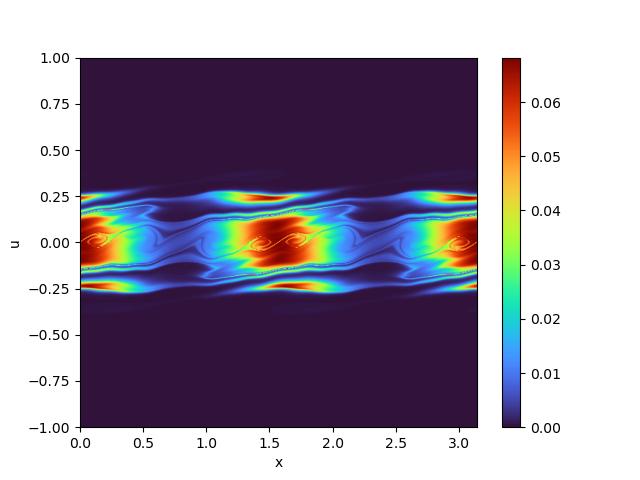}
\subcaption{\label{fig:filamentation_instability_f_x_vx} 
NuFI-Ham $f(50,x,u,0)$
}
\end{subfigure}
\begin{subfigure}{0.325\textwidth}
\includegraphics[width=0.99\textwidth]{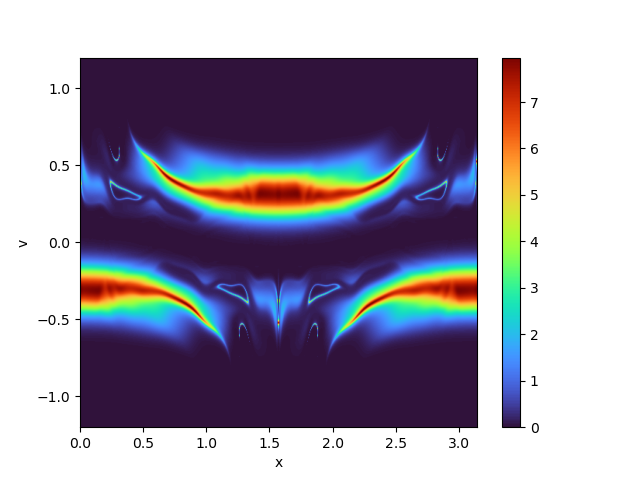}
\subcaption{\label{fig:filamentation_instability_f_x_vy} 
NuFI-Ham $f(50,x,0,v)$
}
\end{subfigure}
\begin{subfigure}{0.325\textwidth}
\includegraphics[width=0.99\textwidth]{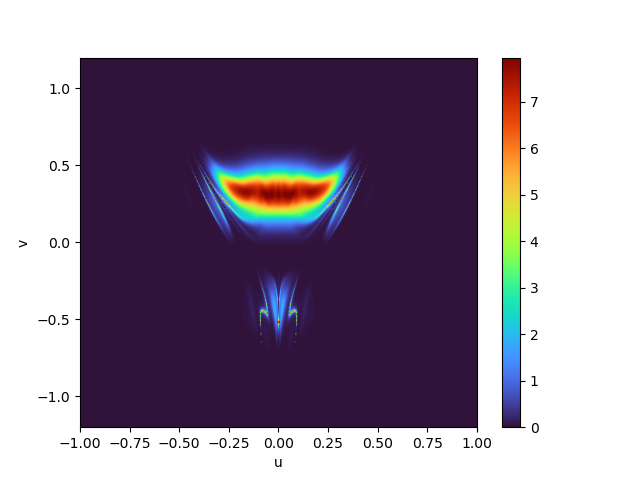}
\subcaption{\label{fig:filamentation_instability_f_vx_vy} 
NuFI-Ham $f(50,\tfrac{L_x}{2},u,v)$
}
\end{subfigure}
\begin{subfigure}{0.325\textwidth}
\includegraphics[width=0.95\textwidth]{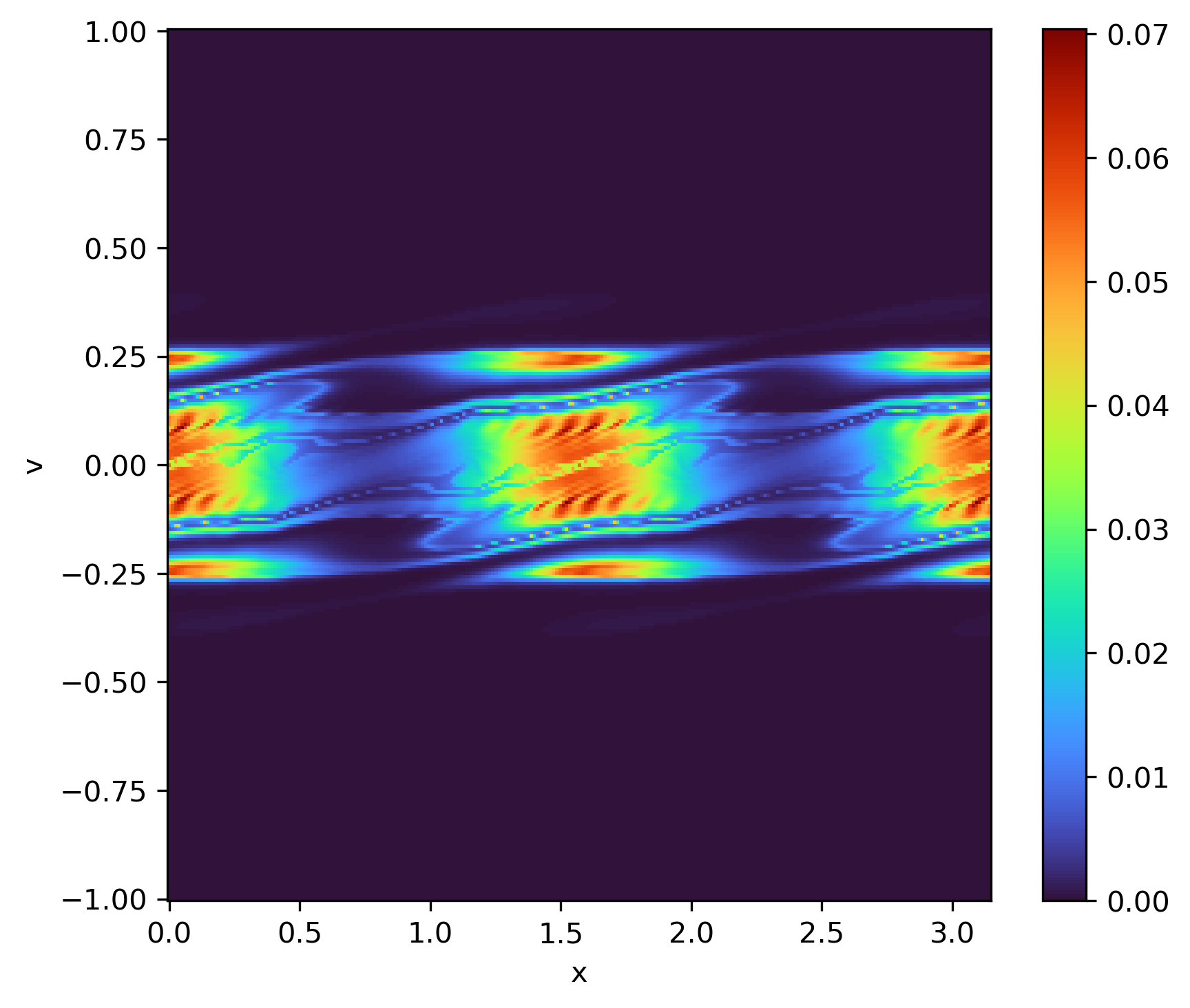}
\subcaption{\label{fig:filamentation_instability_f_x_vx} 
NuFI-PC $f(50,x,u,0)$
}
\end{subfigure}
\begin{subfigure}{0.325\textwidth}
\includegraphics[width=0.91\textwidth]{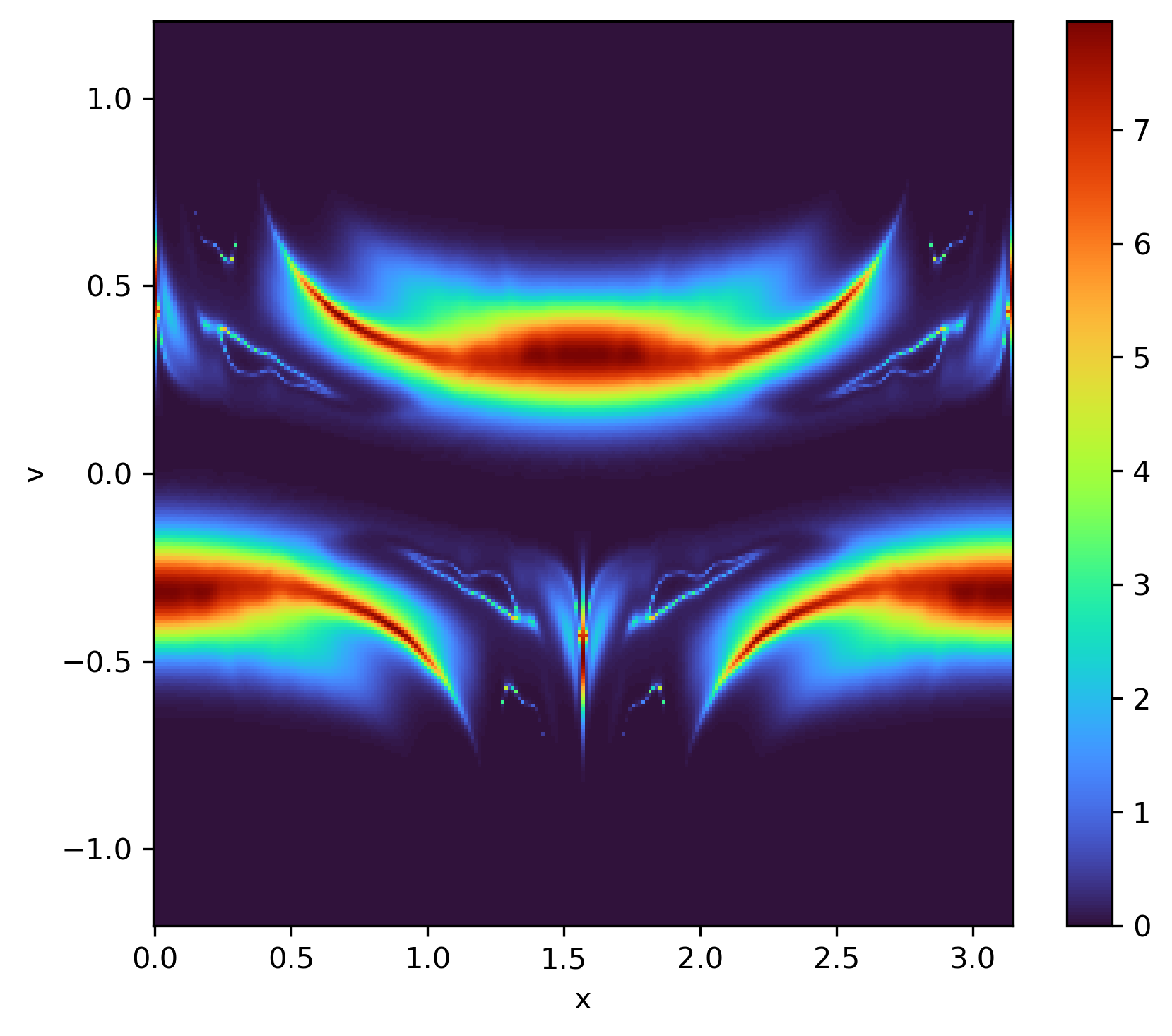}
\subcaption{\label{fig:filamentation_instability_f_x_vy} 
NuFI-PC $f(50,x,0,v)$
}
\end{subfigure}
\begin{subfigure}{0.325\textwidth}
\includegraphics[width=0.91\textwidth]{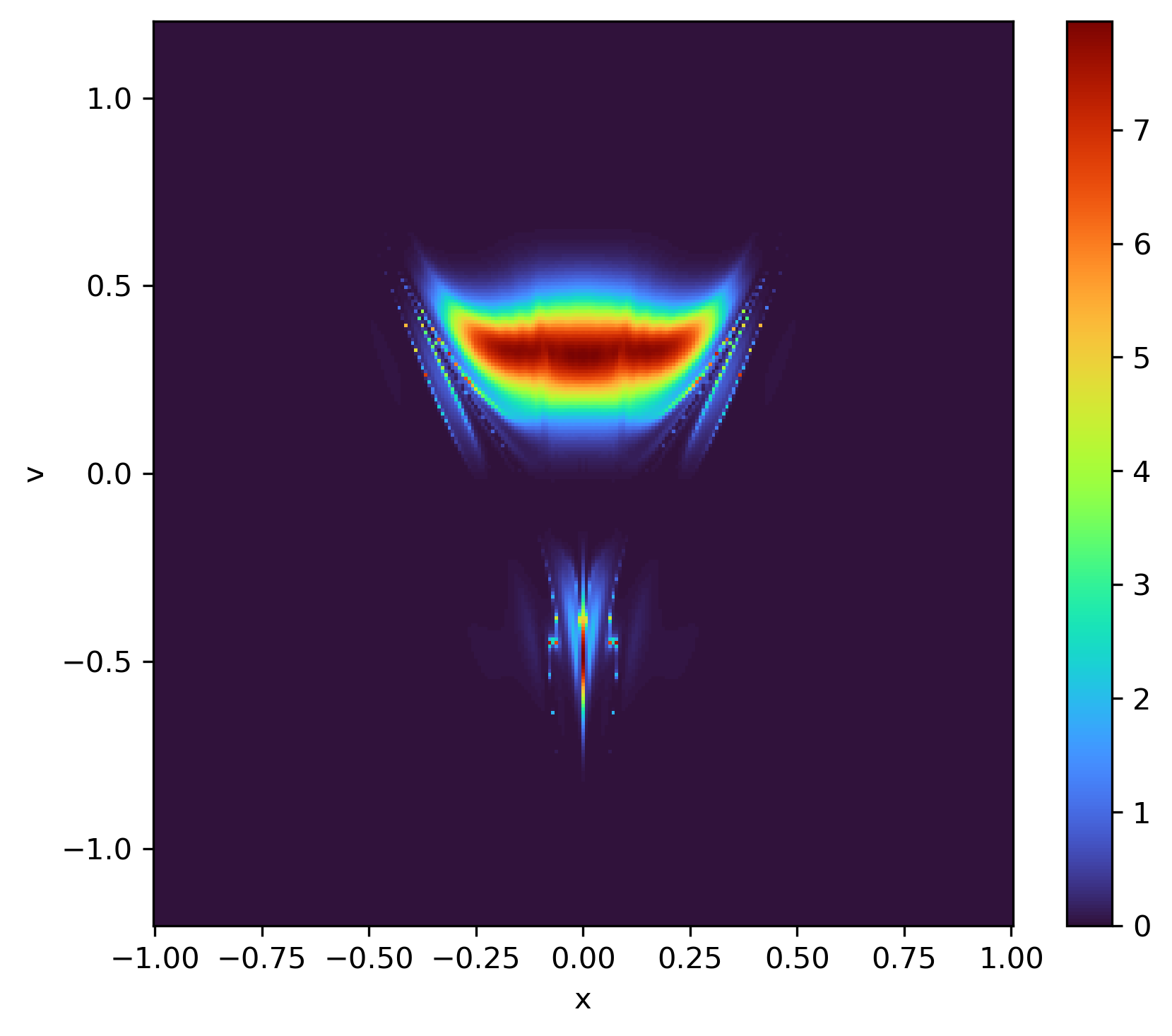}
\subcaption{\label{fig:filamentation_instability_f_vx_vy} 
NuFI-PC $f(50,\tfrac{L_x}{2},u,v)$
}
\end{subfigure}
\caption{\label{fig:filamentation_instability_f_nufi_ham_vs_pc}
Cross-section of distribution functions from the simulation of the 
\emph{filamentation instability} with NuFI-Ham (Lie-split) and NuFI-PC using 
a resolution of $32^3$ in phase-space and $\Delta t = \tfrac{1}{30}$. 
}
\end{figure}

In figure \ref{fig:energies_filamentation_instability_nufi_ham_vs_pc} we compare the energy evolution of NuFI-Ham and NuFI-PC. 
The observed growth rate is roughly $\gamma \approx 0.21$ for the magnetic 
field energy. The results for NuFI-Ham and NuFI-PC, i.\,e., the 
different Maxwell solvers overall agree well until $t \approx 40$ after which 
NuFI-Ham shows slightly more oscillations as does NuFI-PC with lower 
resolution. The additional oscillations for NuFI-Ham are likely due to 
overshoots in the spline representation as we did not use either Gauss cleaning 
nor filtering for this simulation, see section 
\ref{sec:spline_issues_and_gauss_filter}. That being said, the good agreement 
well into the non-linear phase suggests that the relatively low resolution of 
$32^3$ is sufficient for convergence in this case with both NuFI-Ham and 
NuFI-PC.
In figure \ref{fig:filamentation_instability_f_nufi_ham_vs_pc} we compare 
several cross sections of the distribution function at time $t=50$ between 
NuFI-Ham and NuFI-PC generated with a plotting resolution of $512 \times 512$. 
Also here we see a very good agreement between the two approaches.

\begin{figure}[h!]
\centering
\includegraphics[width=0.79\textwidth]{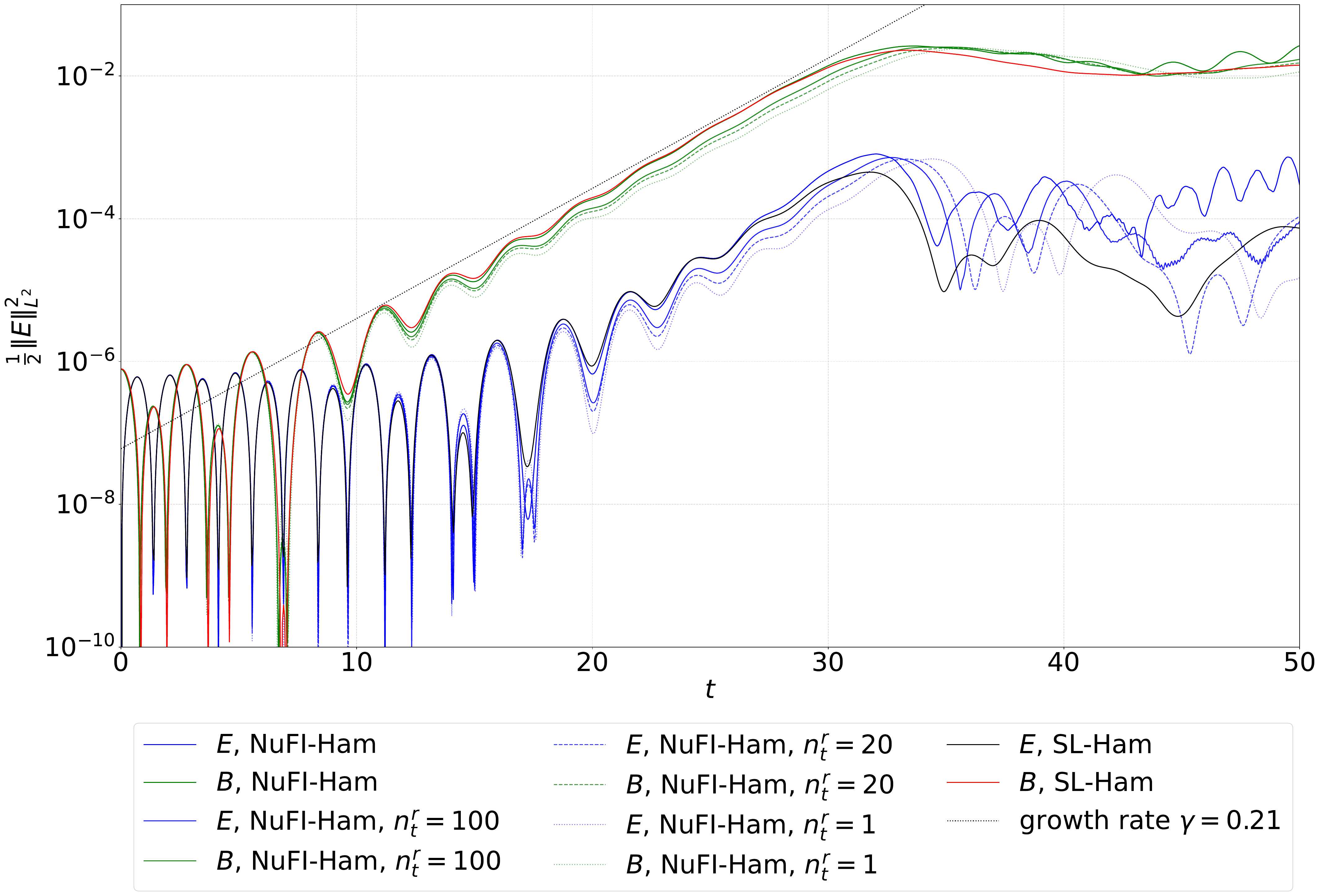} 
\caption{\label{fig:energies_filamentation_instability_nufi_vs_restarted_vs_sl}
Evolution of electric and magnetic energy over time for a simulation of the 
\emph{Filamentation instability} compared between NuFI-Ham (with/without 
restarts) and SL-Ham. All approaches used $32^3$ as phase-space resolution.
}
\end{figure}

In figure \ref{fig:energies_filamentation_instability_nufi_vs_restarted_vs_sl} we are comparing the results between NuFI-Ham, restarted NuFI-Ham and SL-Ham.
Any amount of restart leads to some degree of electro-magnetic energy 
underestimation. Notably the deviation for the electric energy is stronger.
While all results agree qualitatively for the linear growth stage, both the 
restarted NuFI-Ham and the SL-Ham simulations tend to slightly underestimate 
the electric and magnetic energy. 
Notably the restarted NuFI-Ham simulations all underestimate both the electric
and magnetic energy more than SL-Ham during the linear growth state:
the error for the electric energy is up to between $10^{-5}$ and $2\cdot 10^{-5}$ depending 
on the restart frequency $n_t^r$, where an increase in $n_t^r$ improves 
the prediction. After $t\approx 25$ SL-Ham starts deviating more than the 
restarted NuFI-Ham simulation and starts exhibiting an energy drift. 
The electric energy in the non-linear stage of the instability is predicted 
to be about $5\cdot 10^{-4}$ lower by SL-Ham than by NuFI-Ham as well as 
NuFI-Ham with $n_t^r = 20$ and $100$.

\begin{figure}[h!]
\centering
\begin{subfigure}{0.495\textwidth}
\includegraphics[width=0.99\textwidth]{Figures/1x2v/Filamentation_Instability/3d/NuFI-Ham/f_vx_vy_50}
\subcaption{\label{fig:filamentation_instability_f_vx_vy_nufi} 
NuFI-Ham $f(50,\tfrac{L_x}{2},u,v)$
}
\end{subfigure}
\begin{subfigure}{0.495\textwidth}
\includegraphics[width=0.99\textwidth]{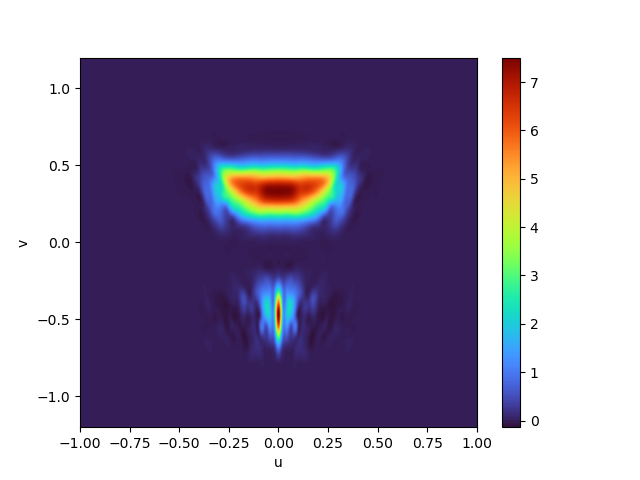}
\subcaption{\label{fig:filamentation_instability_f_vx_vy_nufi_ntr_100} 
NuFI-Ham $f(50,\tfrac{L_x}{2},u,v)$, $n_t^r=100$
}
\end{subfigure}
\begin{subfigure}{0.495\textwidth}
\includegraphics[width=0.99\textwidth]{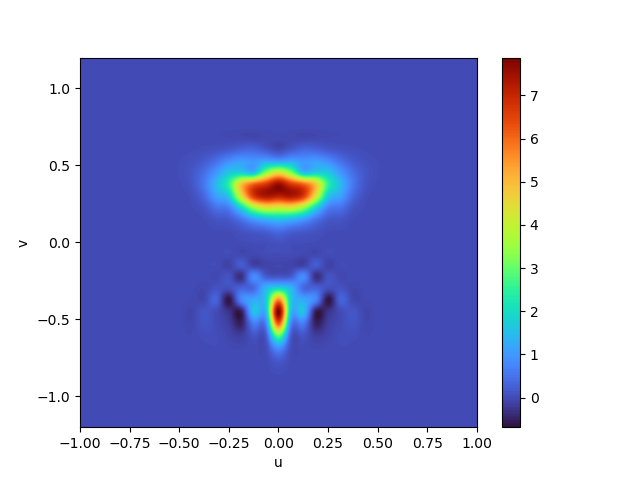}
\subcaption{\label{fig:filamentation_instability_f_vx_vy_nufi_ntr_1} 
NuFI-Ham $f(50,\tfrac{L_x}{2},u,v)$, $n_t^r=1$
}
\end{subfigure}
\begin{subfigure}{0.495\textwidth}
\includegraphics[width=0.99\textwidth]{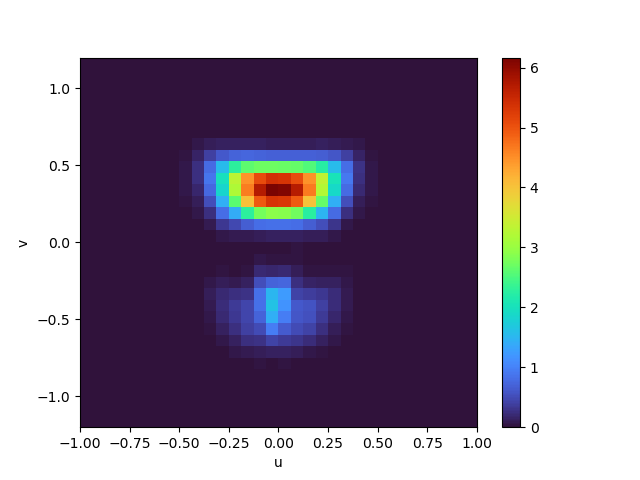}
\subcaption{\label{fig:filamentation_instability_f_vx_vy_sl_ham} 
SL-Ham $f(50,\tfrac{L_x}{2},u,v)$
}
\end{subfigure}
\caption{\label{fig:filamentation_instability_f_vx_vy_restarted_vs_sl}
$u-v$-cross-section of distribution functions from the simulation of the 
\emph{filamentation instability} with NuFI-Ham with different $n_t^r$ and 
SL-Ham using a resolution of $32^3$ in phase-space and 
$\Delta t = \tfrac{1}{30}$. 
}
\end{figure}

\begin{figure}[h!]
\centering
\begin{subfigure}{0.495\textwidth}
\includegraphics[width=0.99\textwidth]{Figures/1x2v/Filamentation_Instability/3d/NuFI-Ham/f_x_vy_50}
\subcaption{\label{fig:filamentation_instability_f_x_vy_nufi} 
NuFI-Ham $f(50,x,0,v)$
}
\end{subfigure}
\begin{subfigure}{0.495\textwidth}
\includegraphics[width=0.99\textwidth]{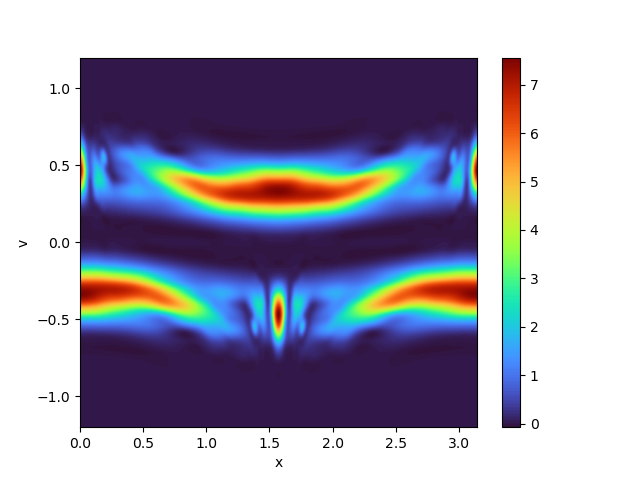}
\subcaption{\label{fig:filamentation_instability_f_x_vy_nufi_ntr_100} 
NuFI-Ham $f(50,x,0,v)$, $n_t^r=100$
}
\end{subfigure}
\begin{subfigure}{0.495\textwidth}
\includegraphics[width=0.99\textwidth]{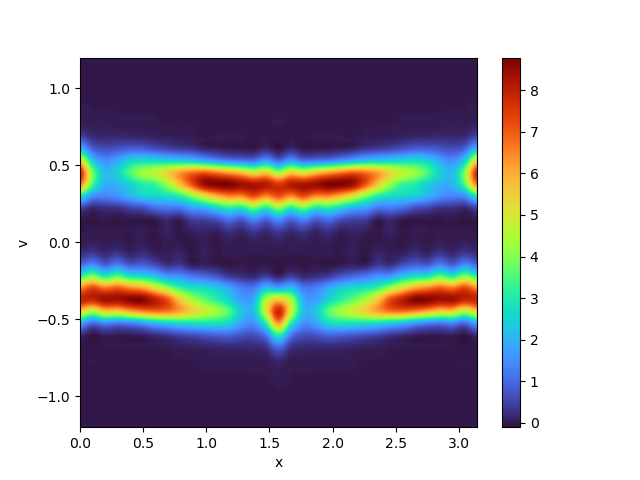}
\subcaption{\label{fig:filamentation_instability_f_x_vy_nufi_ntr_1} 
NuFI-Ham $f(50,x,0,v)$, $n_t^r=1$
}
\end{subfigure}
\begin{subfigure}{0.495\textwidth}
\includegraphics[width=0.99\textwidth]{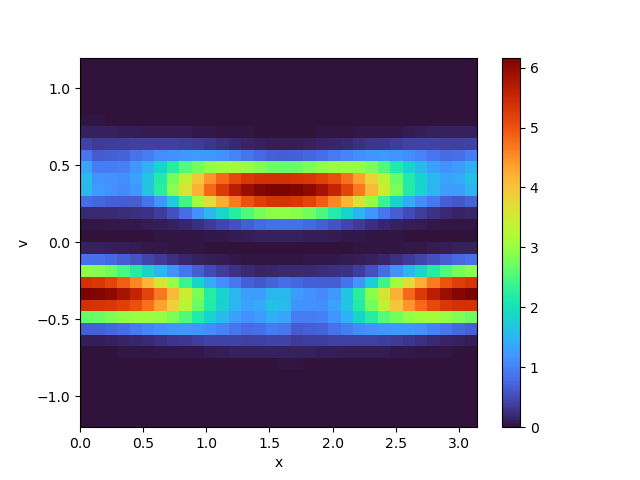}
\subcaption{\label{fig:filamentation_instability_f_x_vy_sl_ham} 
SL-Ham $f(50,x,0,v)$
}
\end{subfigure}
\caption{\label{fig:filamentation_instability_f_x_vy_restarted_vs_sl}
$x-v$-cross-section of distribution functions from the simulation of the 
\emph{filamentation instability} with NuFI-Ham with different $n_t^r$ and 
SL-Ham using a resolution of $32^3$ in phase-space and 
$\Delta t = \tfrac{1}{30}$. 
}
\end{figure}

In figure \ref{fig:filamentation_instability_f_vx_vy_restarted_vs_sl} we 
compare the $u-v$ cross-sections and in figure \ref{fig:filamentation_instability_f_x_vy_restarted_vs_sl} the 
$x-v$ cross-sections of the distribution function at $t=50$ between 
the different approaches. While clearly NuFI-Ham shows the most details, we can
also see that restarting with $n_t^r=100$ retains a significant amount of the
fine-scale features in the distribution function, i.\,e., NuFI-Ham shows a 
clear benefit even if only used as a sub-cycling routine. As SL-Ham is based 
upon a Finite-Volume scheme it only stores the average value of $f$ in a cell 
and therefore we cannot hope to capture structure below the grid level, which 
in turn is quite coarse here with a resolution of $32^3$. 
In fact in figure \ref{fig:filamentation_instability_f_x_vy_restarted_vs_sl}
we also see that the distribution function from SL-Ham looks (in parts) 
qualitatively different from the result of NuFI-Ham and also NuFI with 
$n_t^r=1$, which in turn is equivalent to a classical backwards Semi-Lagrangian
scheme with cubic interpolation.
%These deviations may explain why SL-Ham appears to be less accurate in 
%predicting the dynamics in the non-linear stage of the instability as these may 
%rely upon exactly these fine scale features. So we conclude that while the 
%numerical dissipation helps SL-Ham to capture the linear growth stage more 
%accurately, it also deteriorates the solution quality and predictive capability
%of SL for the dynamics in the non-linear phase, which are potentially more 
%interesting on a long time scale.

\begin{remark}
Currently we employ \enquote{simple} cubic interpolation for the restart of 
NuFI, see section \ref{sec:restarted_nufi_vm}, which does not explicitly 
enforce positivity of the distribution function nor preserves any of the 
moments. Therefore we observe some overshoots for NuFI with frequent restarts,
see e.g. figure \ref{fig:filamentation_instability_f_vx_vy_nufi_ntr_1}, which 
may contribute to loss of accuracy in the energy prediction, while SL-Ham 
uses a conservative Finite Volume scheme, which enforces positivity. We expect 
that introducing a positivity-preserving interpolation or approximation scheme 
for NuFI-restarts would further improve accuracy and stability, as well as make
SL-Ham and NuFI with $n_t^r=1$ more comparable.
\end{remark}

\subsection{Streaming Weibel Instability}
\label{sec:streaming_weibel}

The \textit{Streaming Weibel Instability} is a classic kinetic instability used to test 
electro-magnetic solvers for the Vlasov system. Here we consider the setup 
suggested by Cheng et al. and Kormann et al. \autocite{Cheng_2014, 
KORMANN2025118290}. 
The $y$ and $z$ spatial components as well as $v_z$ velocity directions are 
assumed to be uniform and therefore the system reduces to $1x 2v$. Additionally 
we can assume an uniform ion distribution and a periodic boundary in the 
$x$-direction. The initial electron distribution has a temperature anisotropy 
$T_y > T_x$:
\begin{equation}
\label{eqn:f0_streaming_weibel}
f_0(x,u,v) = \tfrac{1}{2\pi \sigma} \exp\left( -\tfrac{u^2}{2\sigma^2} \right) \left( \delta \exp\left( \tfrac{( v - v_{1})^2}{2\sigma^2} \right)
+ (1- \delta) \exp\left( \tfrac{( v - v_{2})^2}{2\sigma^2} \right) \right),
\end{equation}
where $x \in [0, \tfrac{2\pi}{\theta}]$ and $\theta = 0.2$. The initial 
electric field is 0 and the initial magnetic field is set to
\begin{equation}
\label{eqn:B0_streaming_weibel}
\BB_0(x) = \begin{pmatrix}
0 \\ 0 \\ \beta \sin(\theta x)
\end{pmatrix}.
\end{equation}

\begin{figure}[h!]
\centering
\includegraphics[width=0.69\textwidth]{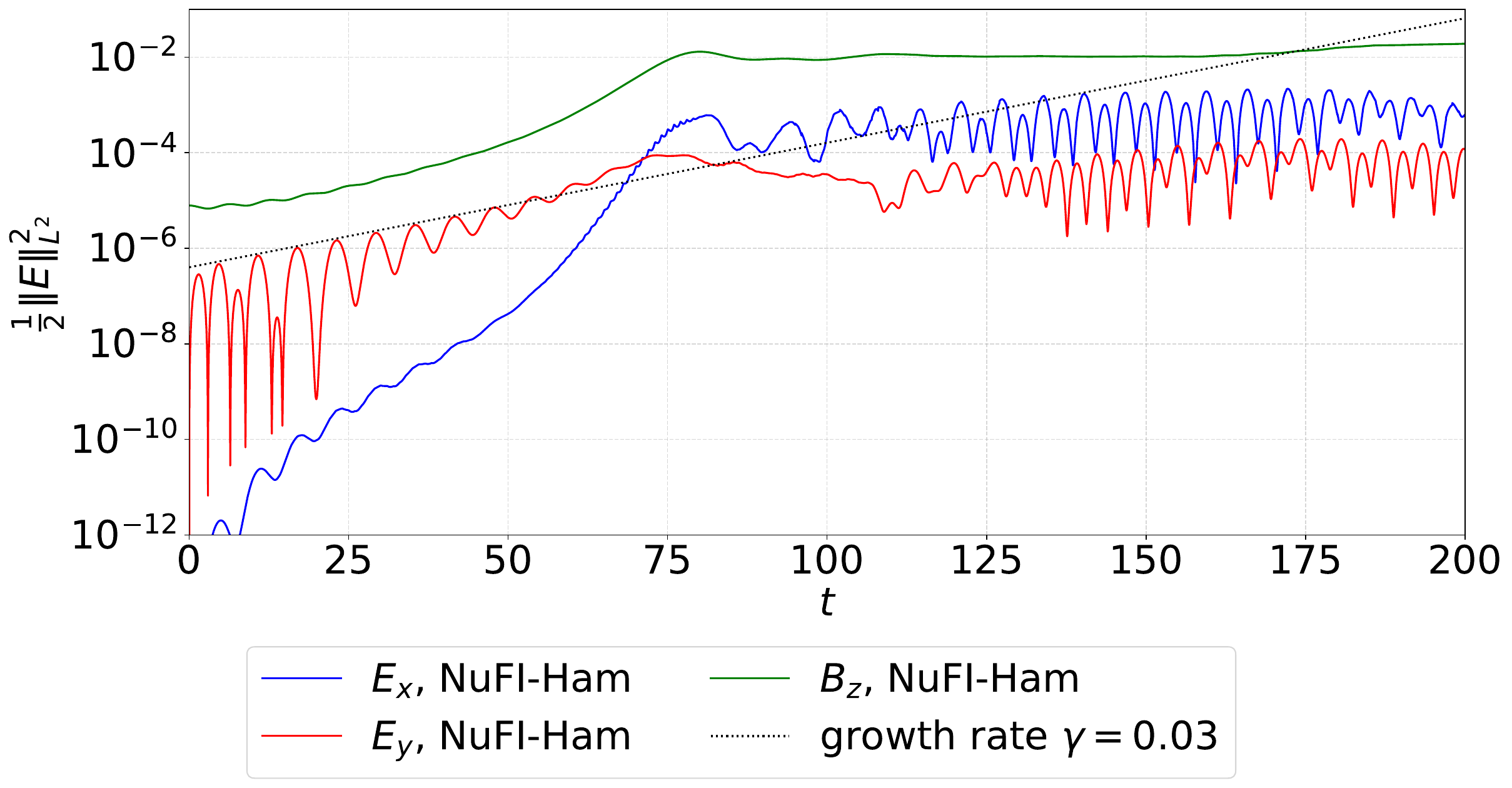}
\caption{\label{fig:streaming_weibel_energies}
Evolution of the energies associated to the components of the electric and 
magnetic fields over time for the \emph{streaming Weibel instability} compared 
between NuFI-Ham and NuFI-PC. The analytically predicted growth rate 
$\gamma = 0.03$ of the electric energy matches with the simulation results.
}
\end{figure}

We truncate the velocity domain to $[-0.5, 0.5] \times [-1.2, 1.2]$. 
In figure \ref{fig:streaming_weibel_energies} we consider the evolution of 
electro-magnetic energies for a simulation with a phase-space resolution of
$32 \times 64 \times 64$, a time-step of $\Delta t = \tfrac{1}{100}$ and a
restart frequency of $n_t^r = 100$. The linear growth with
growth rate $\gamma = 0.03$ of the $E_y$ component is well reproduced by the 
NuFI simulation. The dynamics produced by both approaches are overall in 
good agreement with the reference simulation from Kormann et 
al.~\autocite{KORMANN2025118290}. 

\begin{figure}[h!]
\centering
\begin{subfigure}{0.325\textwidth}
\includegraphics[width=0.99\textwidth]{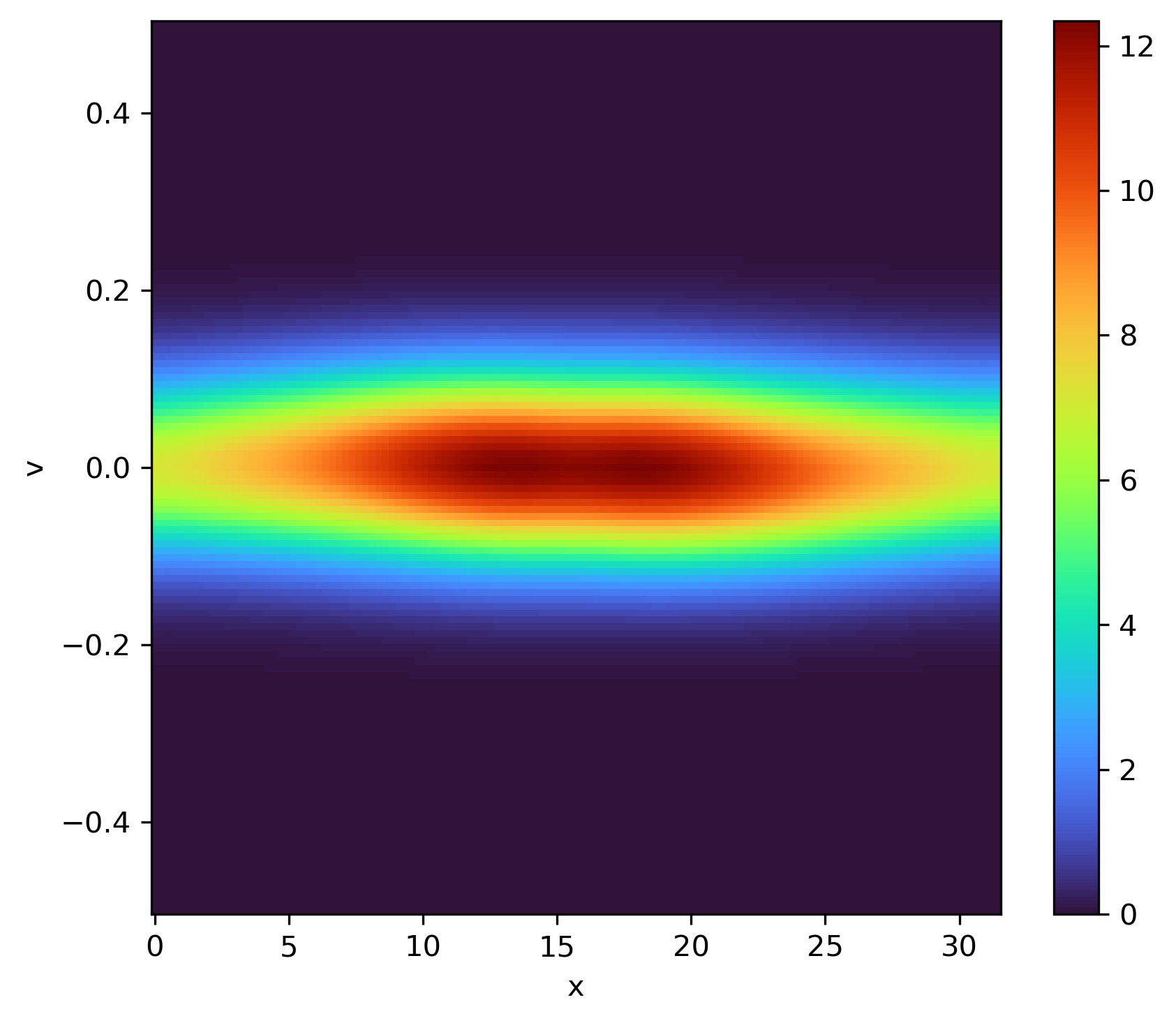}
\subcaption{\label{fig:streaming_weibel_f_x_vx_t_50} 
$f(50,\cdot, \cdot,0)$
}
\end{subfigure}
\begin{subfigure}{0.325\textwidth}
\includegraphics[width=0.99\textwidth]{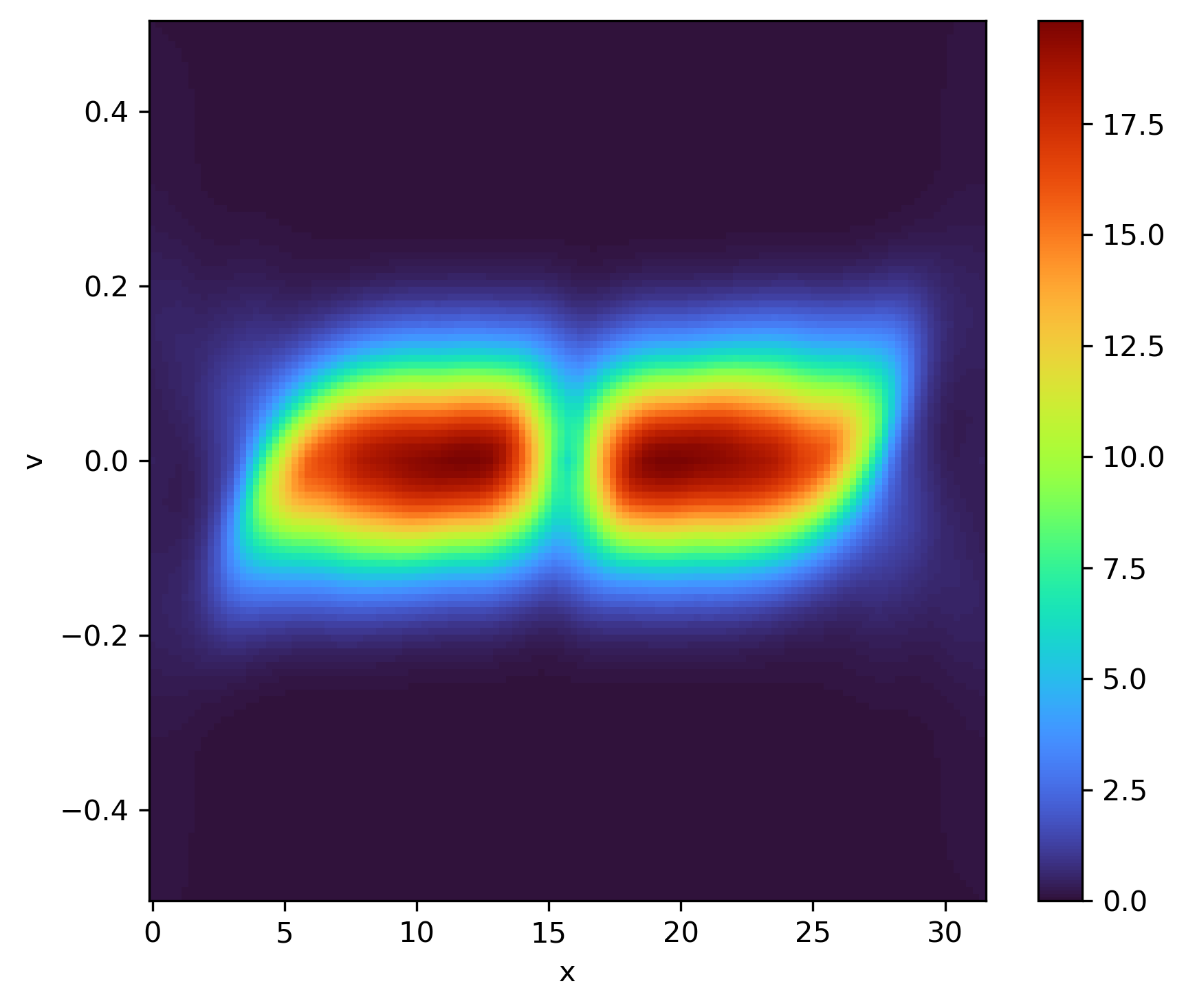}
\subcaption{\label{fig:streaming_weibel_f_x_vx_t_100} 
$f(100,\cdot, \cdot,0)$
}
\end{subfigure}
\begin{subfigure}{0.325\textwidth}
\includegraphics[width=0.99\textwidth]{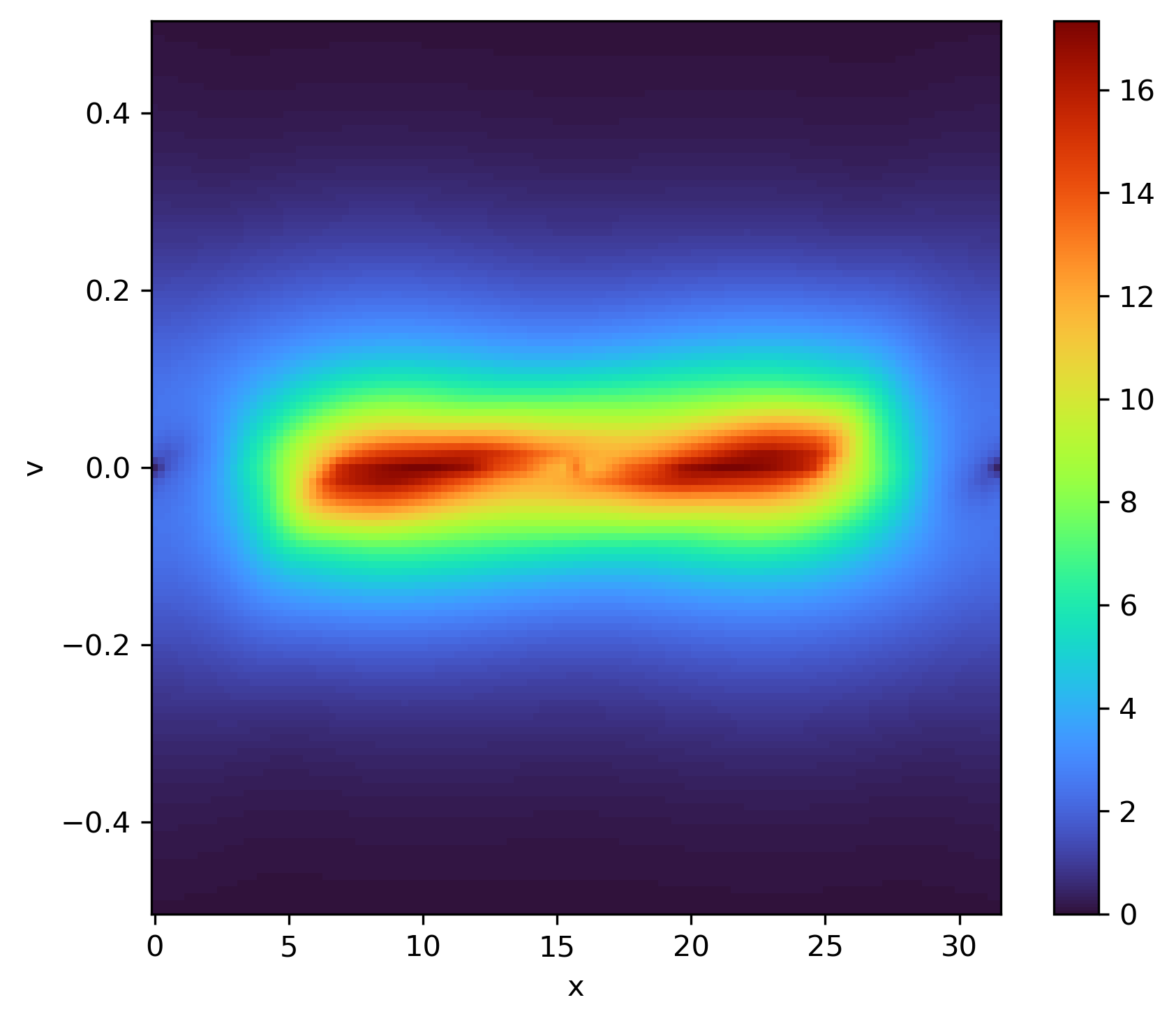}
\subcaption{\label{fig:streaming_weibel_f_x_vx_t_300} 
$f(300,\cdot, \cdot,0)$
}
\end{subfigure}
\centering
\begin{subfigure}{0.325\textwidth}
\includegraphics[width=0.99\textwidth]{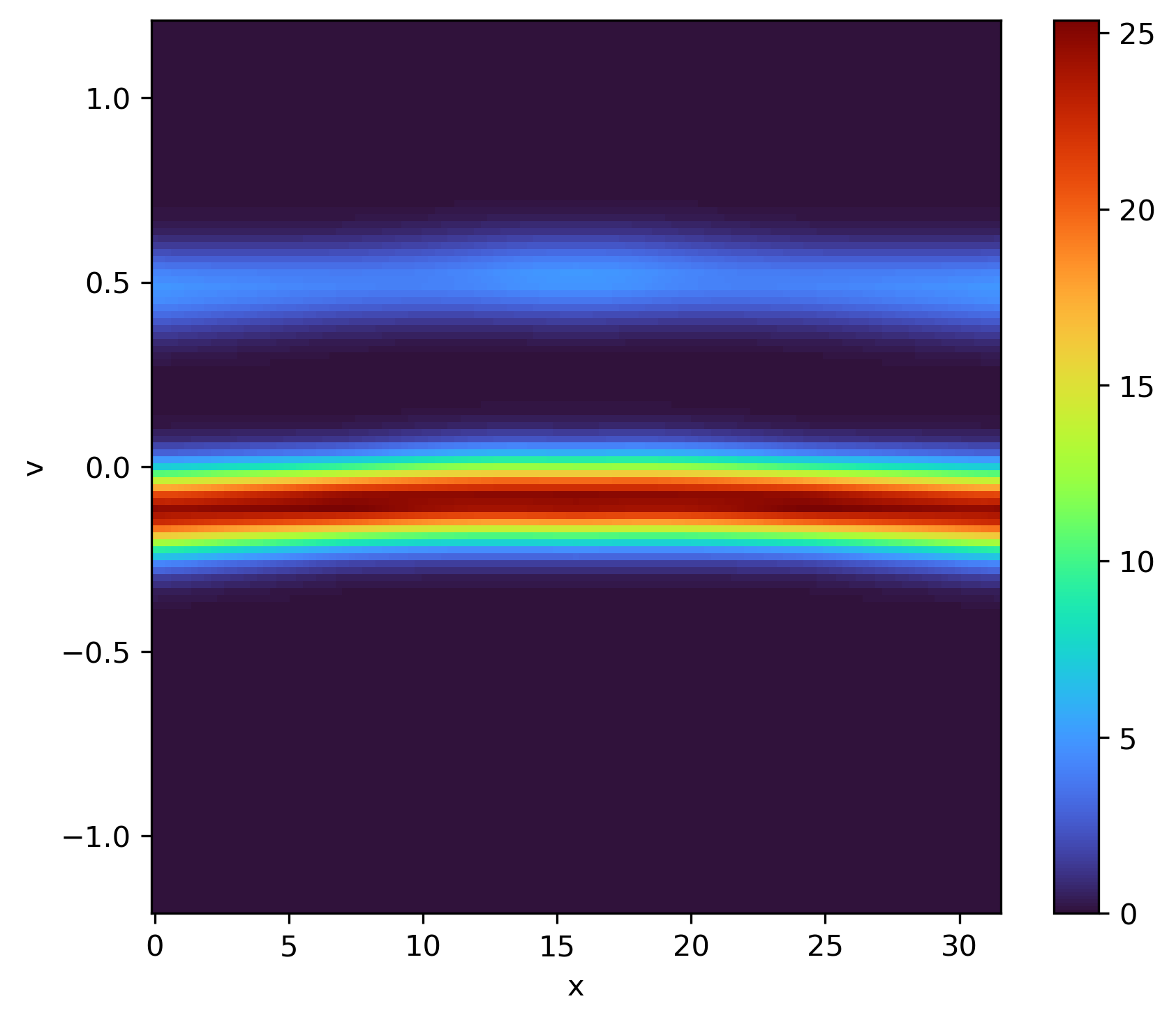}
\subcaption{\label{fig:streaming_weibel_f_x_vy_t_50} 
$f(50,\cdot,0,\cdot)$
}
\end{subfigure}
\begin{subfigure}{0.325\textwidth}
\includegraphics[width=0.99\textwidth]{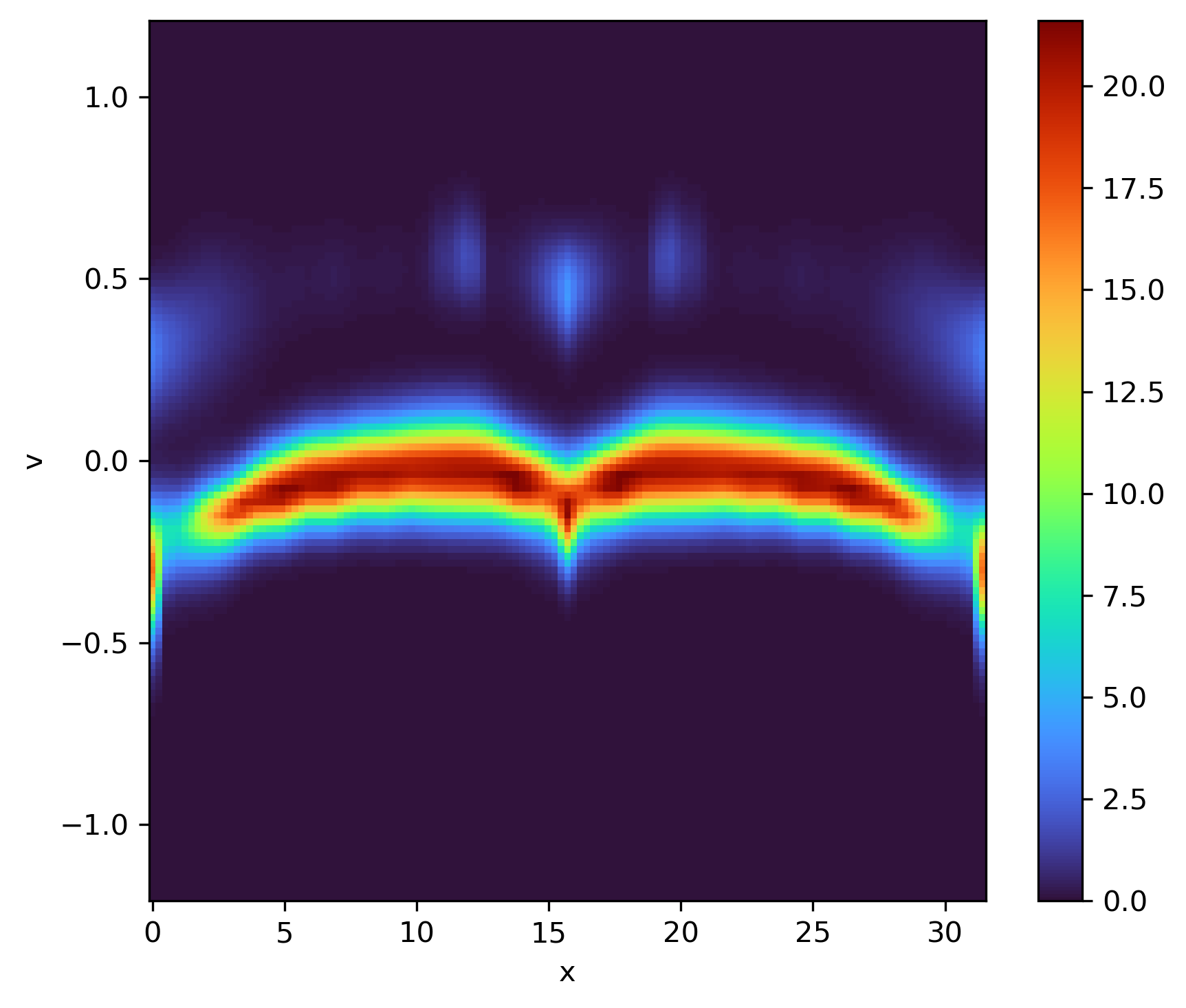}
\subcaption{\label{fig:streaming_weibel_f_x_vy_t_100} 
$f(100,\cdot,0,\cdot)$
}
\end{subfigure}
\begin{subfigure}{0.325\textwidth}
\includegraphics[width=0.99\textwidth]{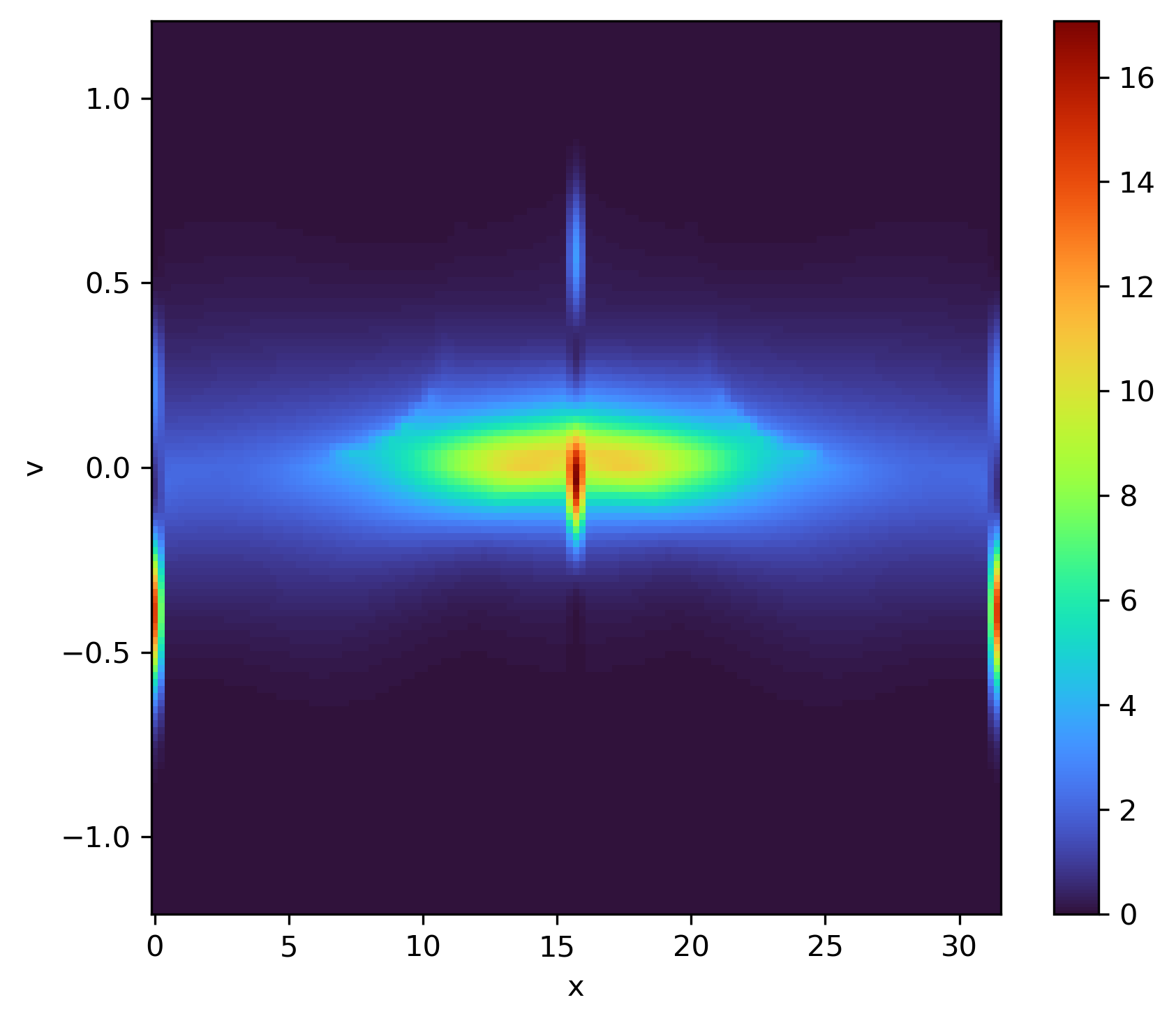}
\subcaption{\label{fig:streaming_weibel_f_x_vy_t_300} 
$f(1000,\cdot,0,\cdot)$
}
\end{subfigure}
\caption{\label{fig:streaming_weibel_f_x_vx_vy_crosssections}
Cross-sections of the velocity distribution function $f$ for the 
\emph{streaming Weibel instability} at $u = 0$ (top) and 
at $v=0$ (bottom) at times $t=50,100$ and $300$ (from left to right).  
}
\end{figure}

\begin{figure}[h!]
\centering
\begin{subfigure}{0.49\textwidth}
\includegraphics[width=0.99\textwidth]{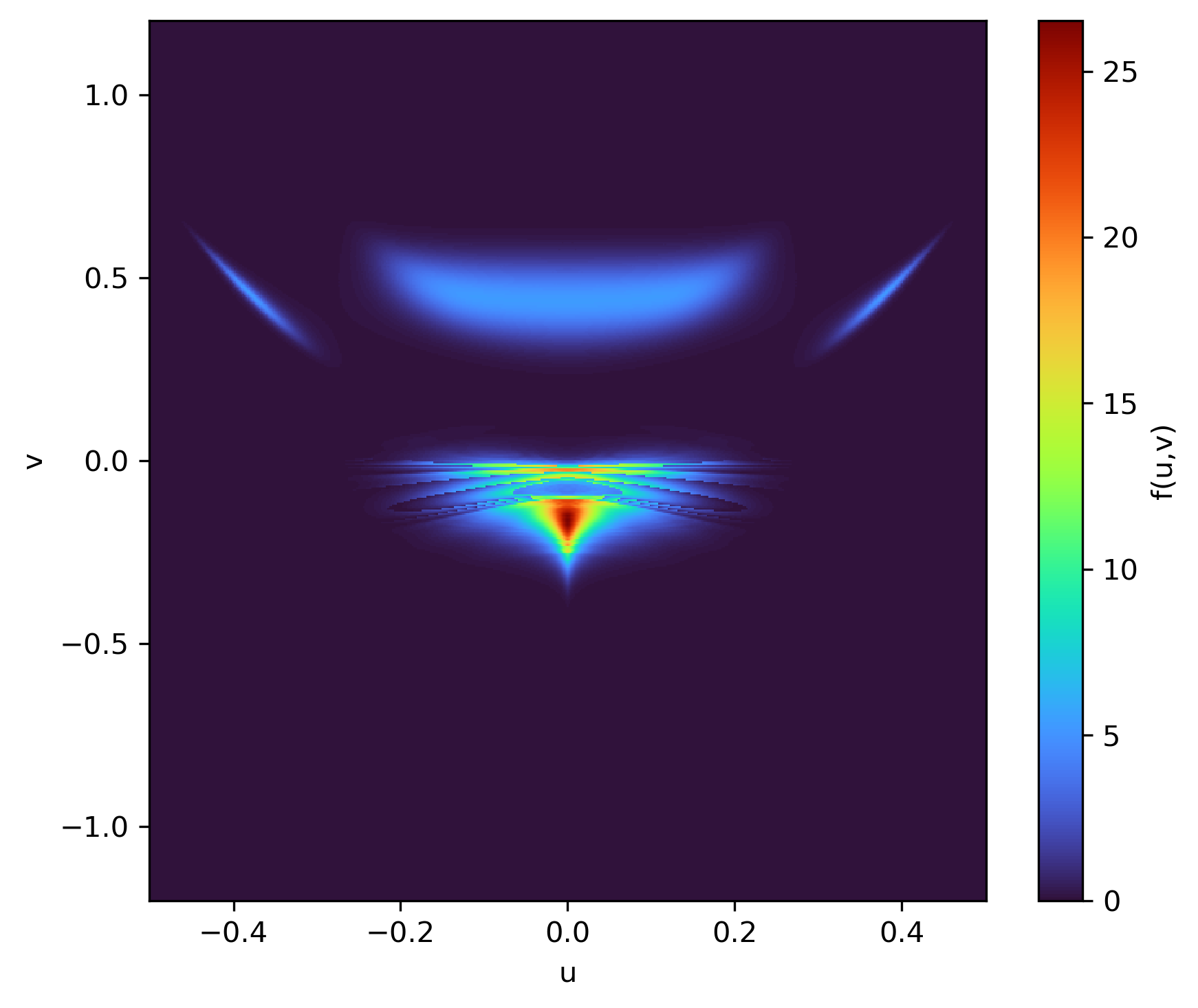}
\subcaption{\label{fig:streaming_weibel_f_u_v_x_Lx_half}
$f(100,\tfrac{\pi}{\theta}, \cdot,\cdot)$
}
\end{subfigure}
\begin{subfigure}{0.49\textwidth}
\includegraphics[width=0.99\textwidth]{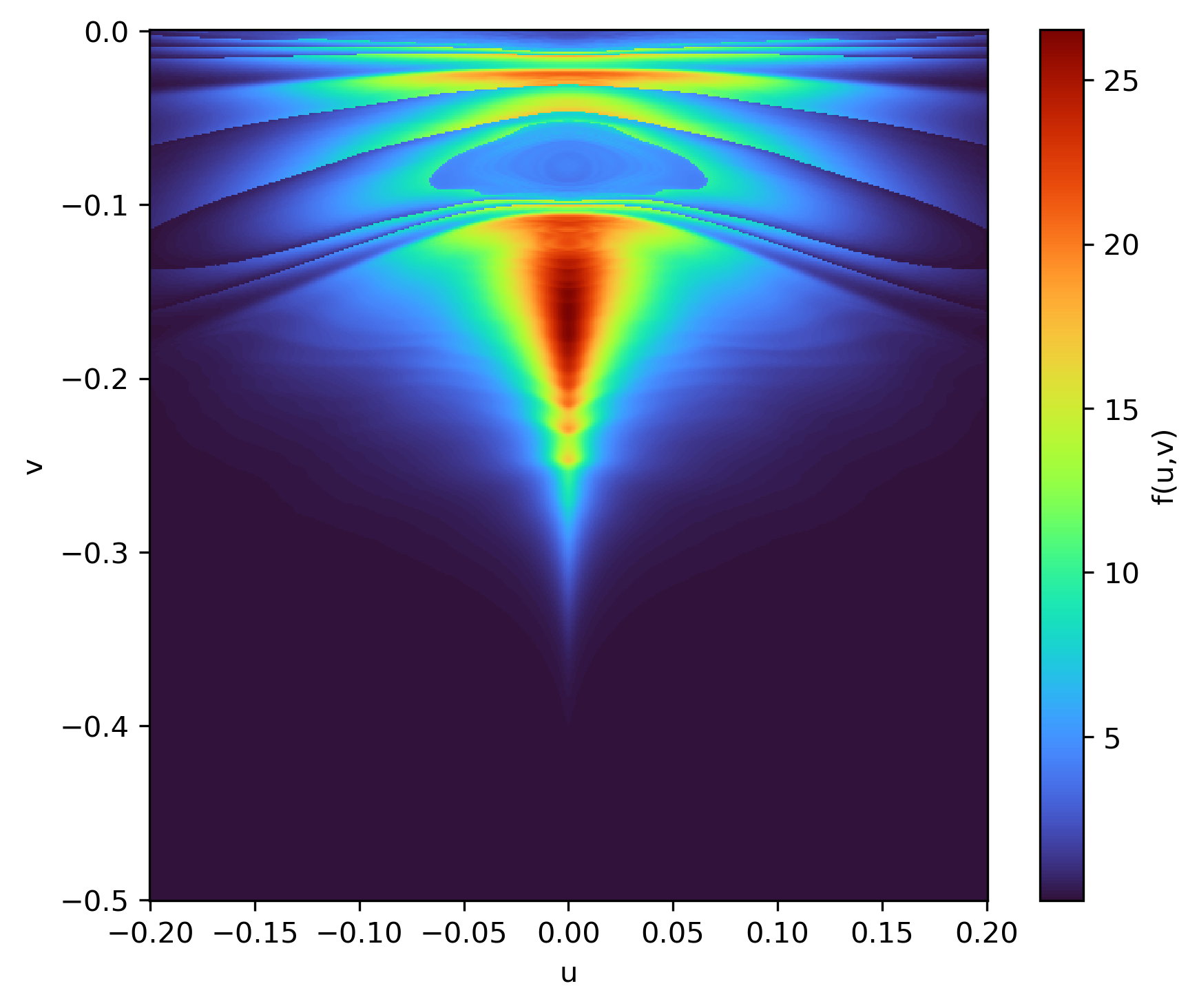}
\subcaption{\label{fig:streaming_weibel_f_u_v_x_Lx_half_zoomed}
Zoomed into $f(100,\tfrac{\pi}{\theta}, \cdot,\cdot)$
}
\end{subfigure}
\caption{\label{fig:streaming_weibel_f_u_v_middle}
Velocity distribution function $f$ for the \emph{streaming Weibel instability}
in the middle of the spatial domain ($x=\tfrac{\pi}{\theta}$) at the time $t=100$. Left we see the full velocity space and on the right we zoom into the velocity-space structure around $u=0$ and $v \le 0$.
}
\end{figure}

\begin{figure}[h!]
\centering
\begin{subfigure}{0.49\textwidth}
\includegraphics[width=0.99\textwidth]{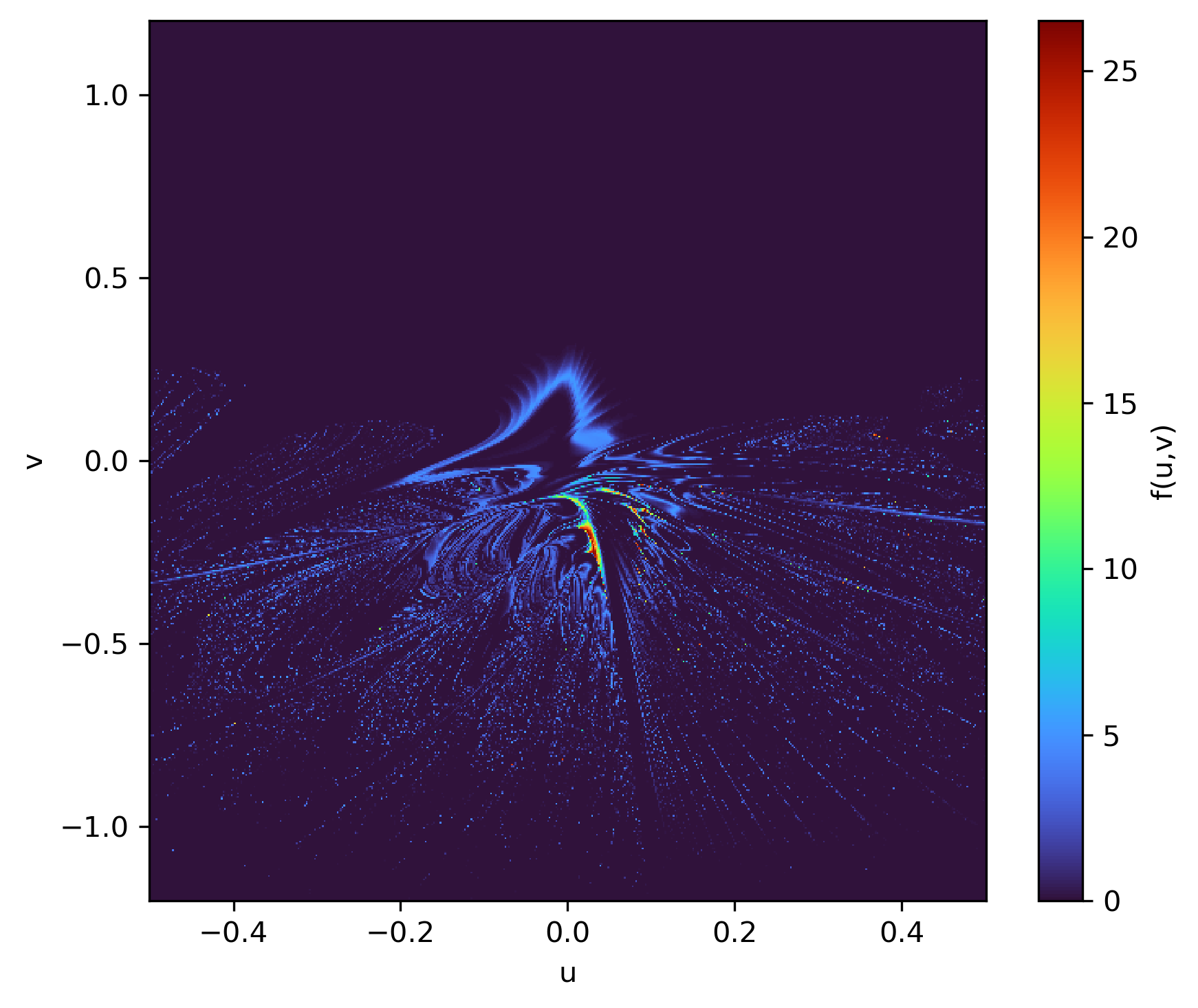}
\subcaption{\label{fig:streaming_weibel_f_u_v_x_0_05}
Full view ($32 \times 32$ cells) of $f(100,0.05, \cdot,\cdot)$
}
\end{subfigure}
\begin{subfigure}{0.49\textwidth}
\includegraphics[width=0.99\textwidth]{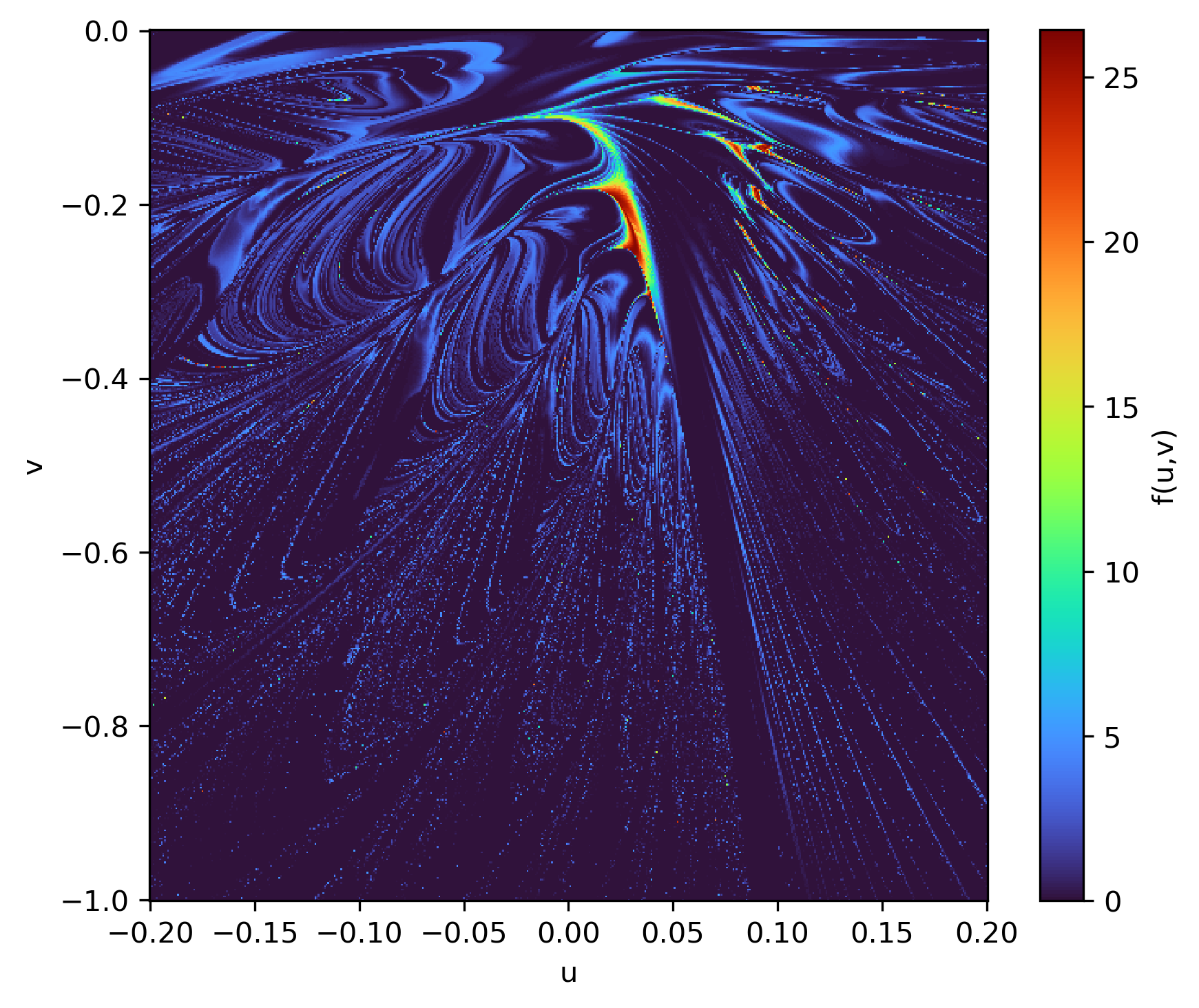}
\subcaption{\label{fig:streaming_weibel_f_u_v_x_0_05_zoomed}
Zoom to $13 \times 13$ cell block of $f(100,0.05, \cdot,\cdot)$
}
\end{subfigure}
\begin{subfigure}{0.49\textwidth}
\includegraphics[width=0.99\textwidth]{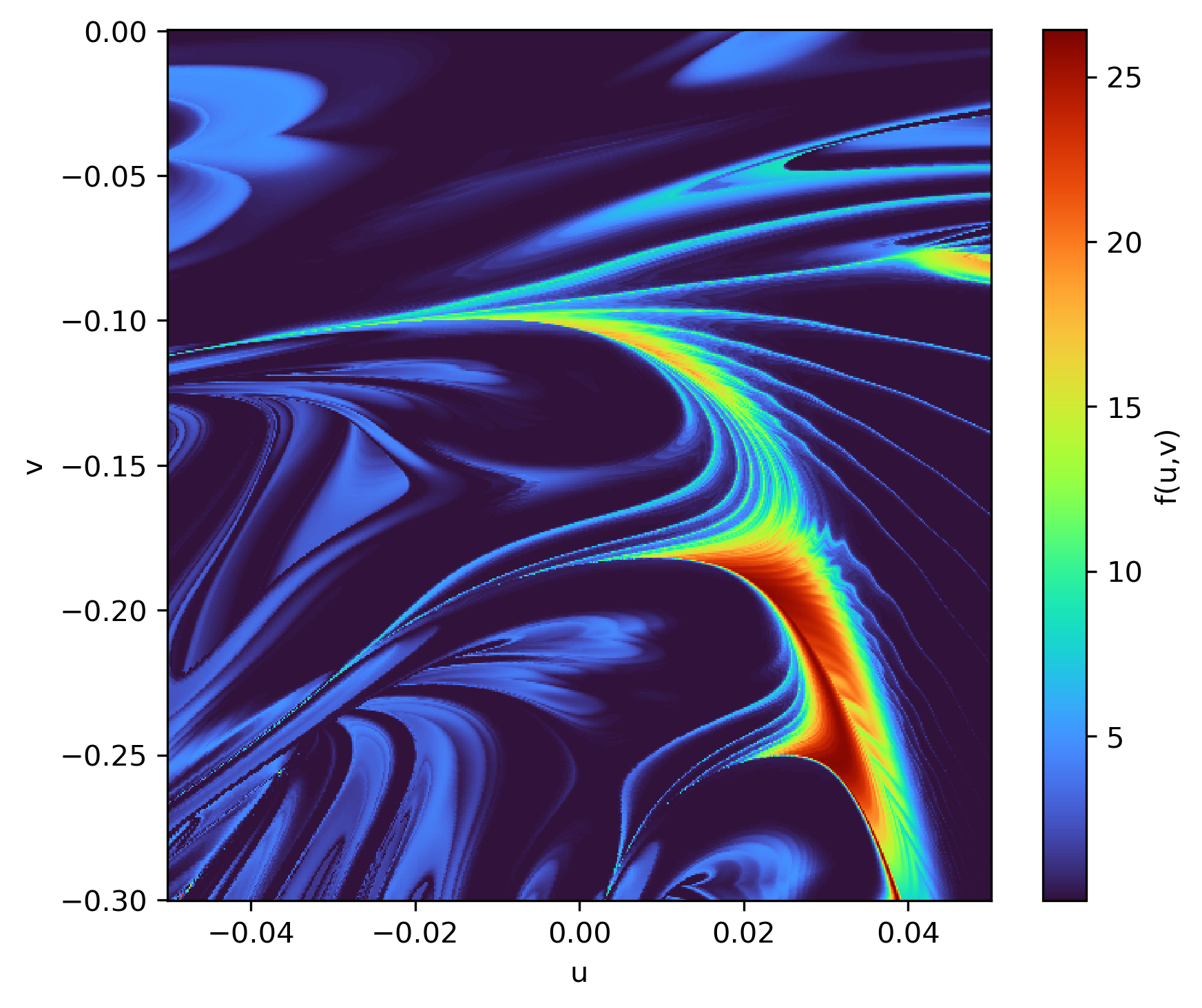}
\subcaption{\label{fig:streaming_weibel_f_u_v_x_0_05_more_zoomed}
Zoom to $3 \times 4$ cell block of $f(100,0.05, \cdot,\cdot)$
}
\end{subfigure}
\begin{subfigure}{0.49\textwidth}
\includegraphics[width=0.99\textwidth]{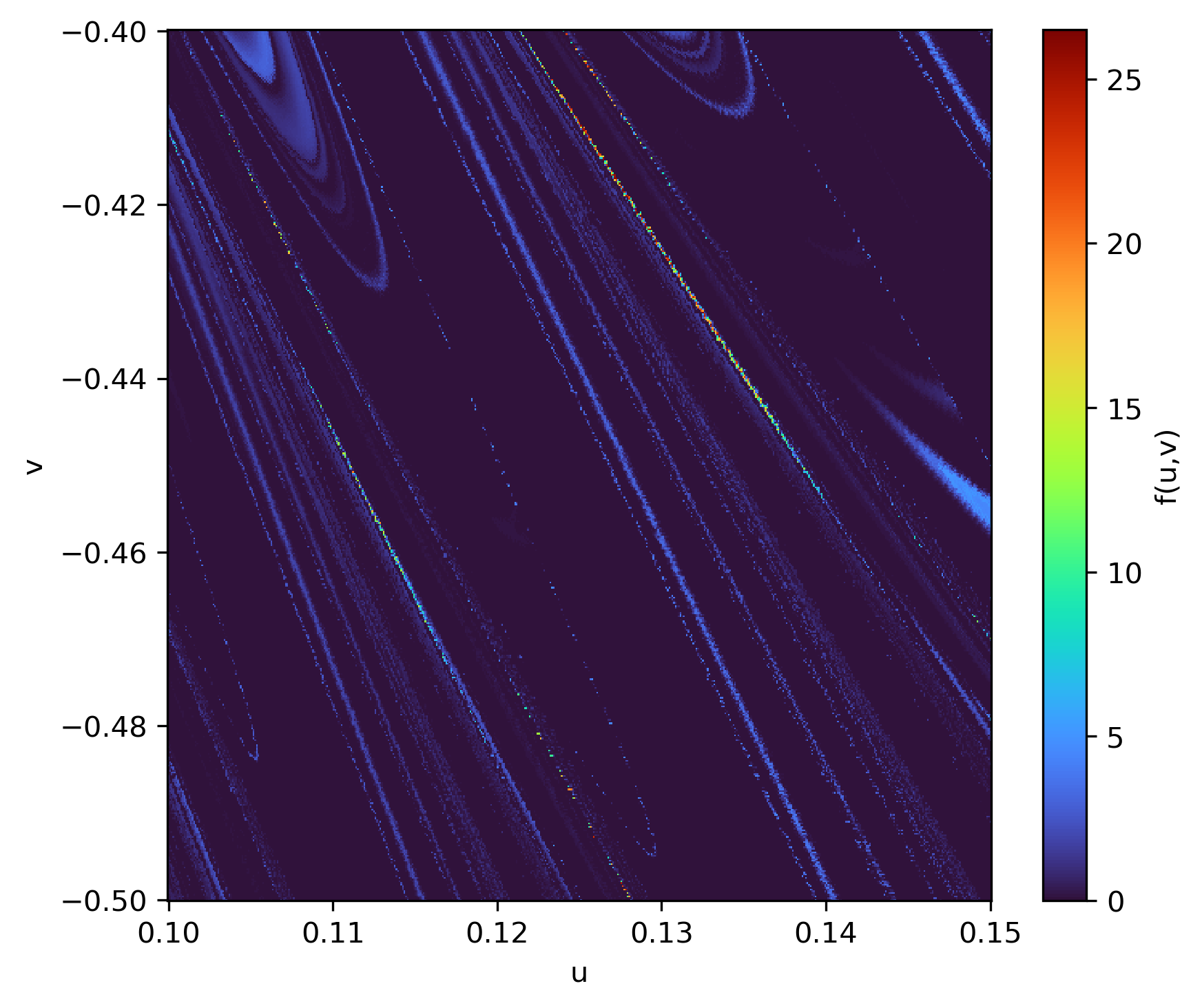}
\subcaption{\label{fig:streaming_weibel_f_u_v_x_0_05_even_more_zoomed}
Zoom to $2 \times 2$ cell block of $f(100,0.05, \cdot,\cdot)$
}
\end{subfigure}
\caption{\label{fig:streaming_weibel_f_u_v_left_boundary_zoomed}
Velocity distribution function $f$ for the \emph{streaming Weibel instability}
at the left boundary $x=0.05$ and time $t=100$ while zooming increasingly in.
The method is able to clearly resolve sub-grid structures far
beyond the capability of any state-of-the-art Semi-Lagrangian solver.
}
\end{figure}

This kinetic instability is particularly interesting as it is prone to develop
very fine scale features, which we can now look at in detail with NuFI. In
figure \ref{fig:streaming_weibel_f_x_vx_vy_crosssections} we show
cross-sections of the distribution function in at $u=0$ and $v=0$ respectively
for time $t= 50, 100$ and $200$. Furthermore, in figure
\ref{fig:streaming_weibel_f_u_v_middle}, we also show the distribution function
in the middle of the spatial domain.
The displayed distribution functions are visually in good agreement reference
solution in Cheng et al. \autocite{Cheng_2014}
Even with the low phase-space resolution used for this simulation we still
retain good sub-grid resolution, also called the \emph{zoom property}, which is
characteristic to kinetic solvers directly discretizing the flow map like NuFI
or CMM.~\autocite{Krah2023905}
To showcase this we also zoomed deeper into the low density region of the
velocity space around the left spatial domain boundary, see figure
\ref{fig:streaming_weibel_f_u_v_left_boundary_zoomed}. Even when zooming below
the grid resolution, we still see a noise-free and finely resolved distribution
function, which is beyond the capability of any other state-of-the-art method.

Finally, let us note that we chose a relatively fine time-step as we noticed 
that this instability is particularly prone to induce oscillations in our 
spline representation. This is potentially caused by the fact that NuFI -- in 
contrast to other methods -- is able to capture the particularly strong 
filamentation in the distribution function for this instability. 
In addition to the remedies which we discussed in section 
\ref{sec:spline_issues_and_gauss_filter}, we noticed that reducing the time 
step size also further increases the numerical stability of our scheme.

\subsection{Rotated Filamentation Instability in 2x3v}

As a final test case we consider a higher-dimensional setup including 
mobile ions to test the feasibility of running simulations beyond 1x2v as 
well as to verify that the ion-electron coupling works as intended. To this end
we opted to build upon the previously discussed \textit{filamentation 
instability}, see section \ref{sec:filamentation_instability}. We simply rotate 
both the beams and the perturbation through the magnetic field by 45$^\circ$
in the $x$-$y$ plane. This way the growth rate of the instability remains 
unchanged so we have a simple method to verify the correctness of our results,
however, because the fields are now also depending on $y$ we genuinely need
the 2x3v phase-space and therefore can test the full set of Maxwell equations
as well as multi-dimensional Vlasov dynamics. The electron initial data is now
\begin{align}
f_0^e(u,v,w) = \frac{1}{2(2\pi \vth^e)^{3/2} }
\exp\left(-\frac{v_s^2+w^2}{2\vth^e}\right)  \left( \exp\left( \frac{(v_r - v_b)^2}{2\vth^e} \right) 
+ \exp\left( \frac{(v_s - v_b)^2}{2\vth^e} \right) \right),
\label{eqn:f0_e_rotated_filament_instab}
\end{align}

where we set $v_r = \tfrac{u - v}{\sqrt{2}}$, $v_s = \tfrac{u + v}{\sqrt{2}}$,
the beam speed $v_b = 0.4$ and electron thermal velocity to $\vth^e = 0.1$. 
The magnetic field is set to
\begin{equation}
\label{eqn:B0_rotated_filamentation_instab}
\BB_0(x,y) = \begin{pmatrix}
0 \\ 0 \\ \beta \sin(k \tfrac{x + y}{\sqrt{2}})
\end{pmatrix},
\end{equation}

where $\beta = 10^{-3}$ and $k=2$ as before in section 
\ref{sec:filamentation_instability}. The initial electric field is 0. 

As we also want to include mobile ions, but for comparability remain on the 
electron-scales, we set $m_e = 1$ and $m_i = 25$. 
Note that we artificially use a much smaller mass ratio of only 25 (as opposed 
to the real mass ratio of about 1836), which drastically increases the effect
of ions on the dynamics and decreases the scale separation, which in turn 
allows us to test the code with fewer degrees of freedom and less time steps.
For the ions we chose the same temperature, i.\,e., $T_i = T_e$, which leads us
to $\vth^i = 0.1/\sqrt{25} = 0.02$. The ions start from an equilibrium
\begin{equation}
\label{eqn:f0_i_rotated_filament_instab}
f_0(u,v,w) = \exp\left(\frac{u^2+v^2+w^2}{2\vth^i}\right).
\end{equation}

We ran the simulation with a resolution of $64\times 64\times 64\times 64 
\times 32$, $\Delta t = \tfrac{1}{50}$ (with 1st order Lie splitting) and a 
restart frequency of $n_t^r = 20$.

\begin{figure}[h!]
\centering
\begin{subfigure}{0.49\textwidth}
\includegraphics[width=0.99\textwidth]{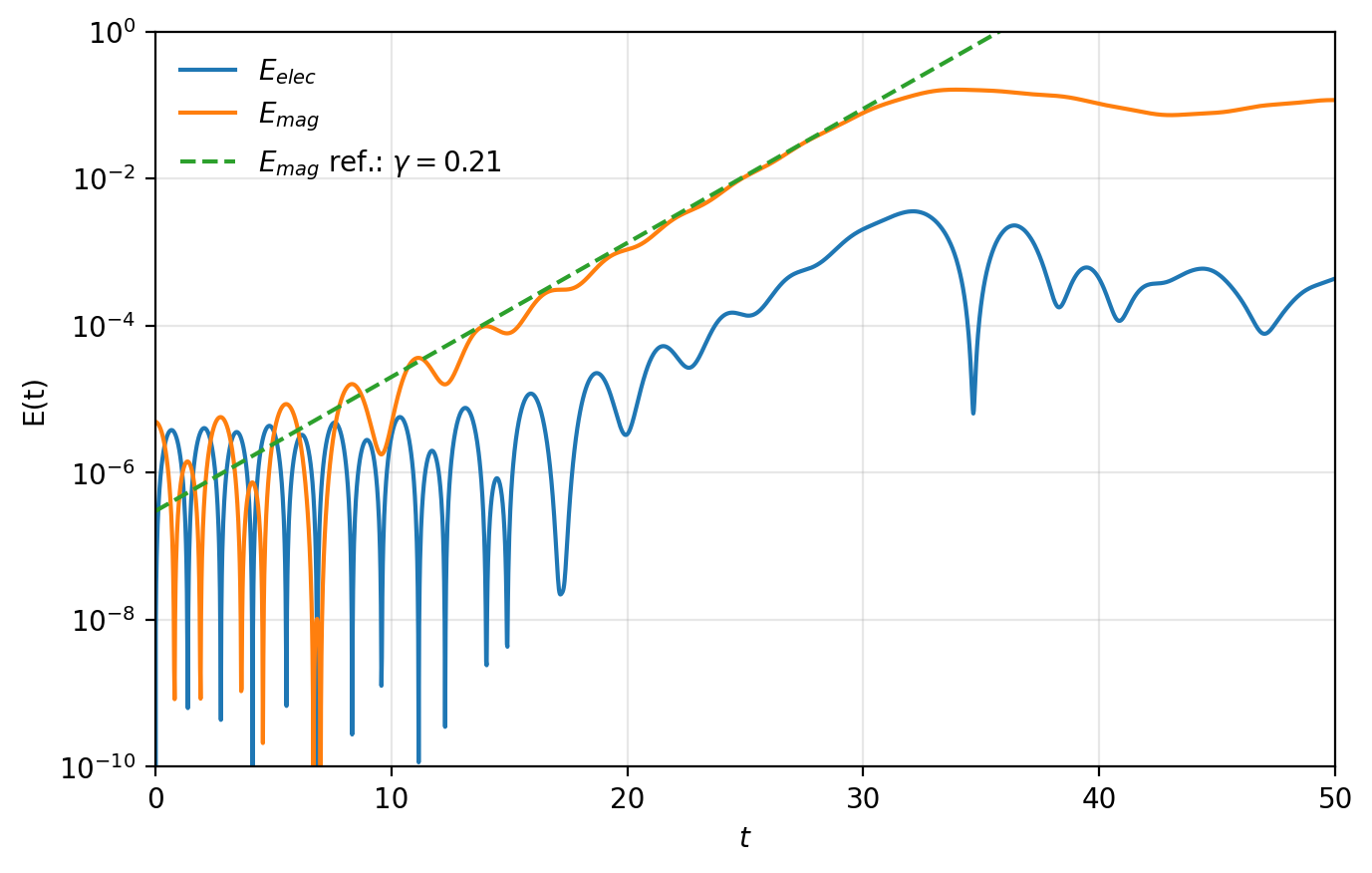}
\subcaption{\label{fig:energy_dynamics_2x3v_rotated_filament}
Evolution of electric and magnetic energy
}
\end{subfigure}
\begin{subfigure}{0.49\textwidth}
\includegraphics[width=0.99\textwidth]{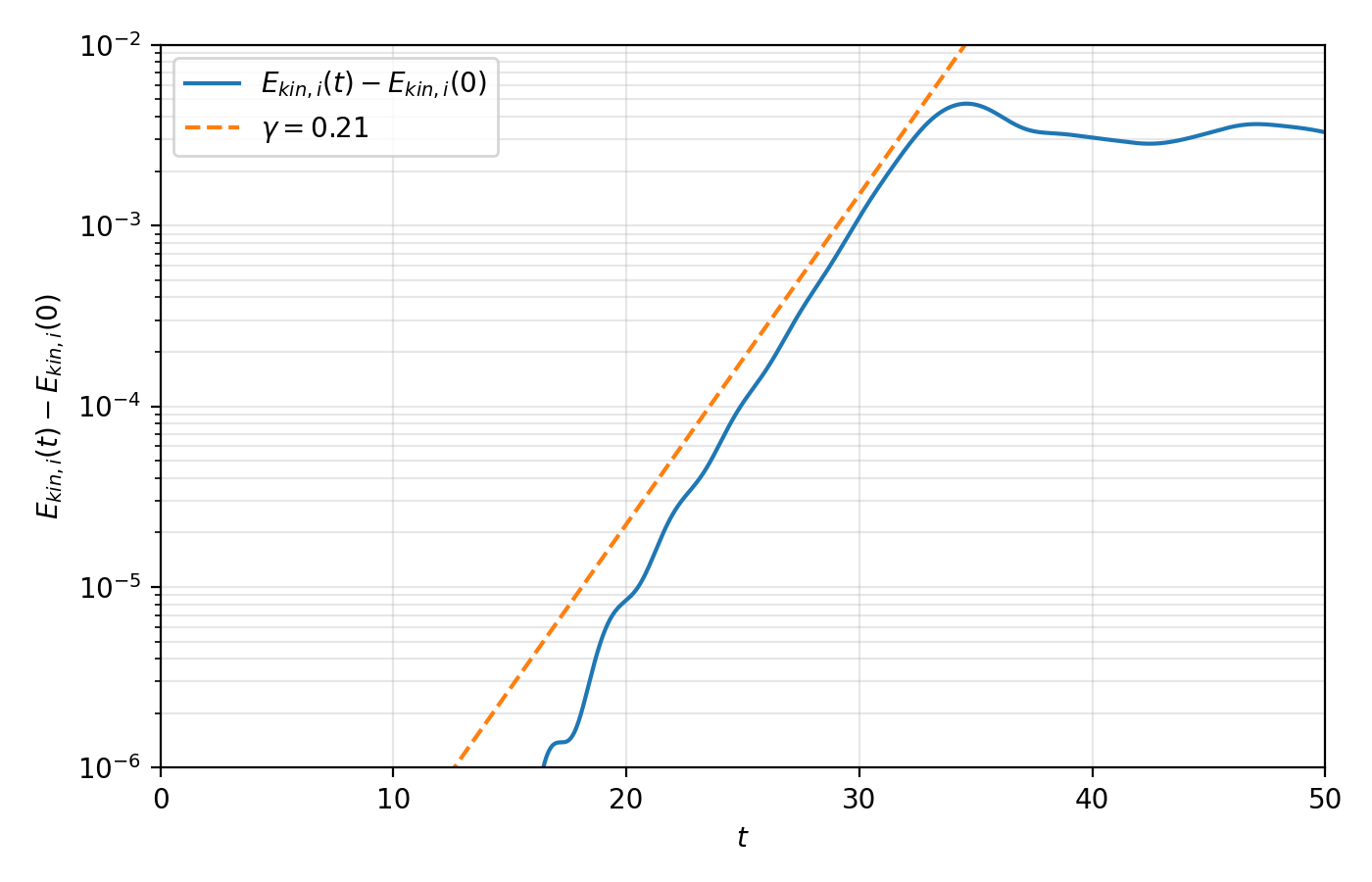}\subcaption{\label{fig:ion_kinetic_energy_growth_2x3v_rotated_filament}
Ion kinetic energy evolution 
}
\end{subfigure}
\begin{subfigure}{0.49\textwidth}
\includegraphics[width=0.99\textwidth]{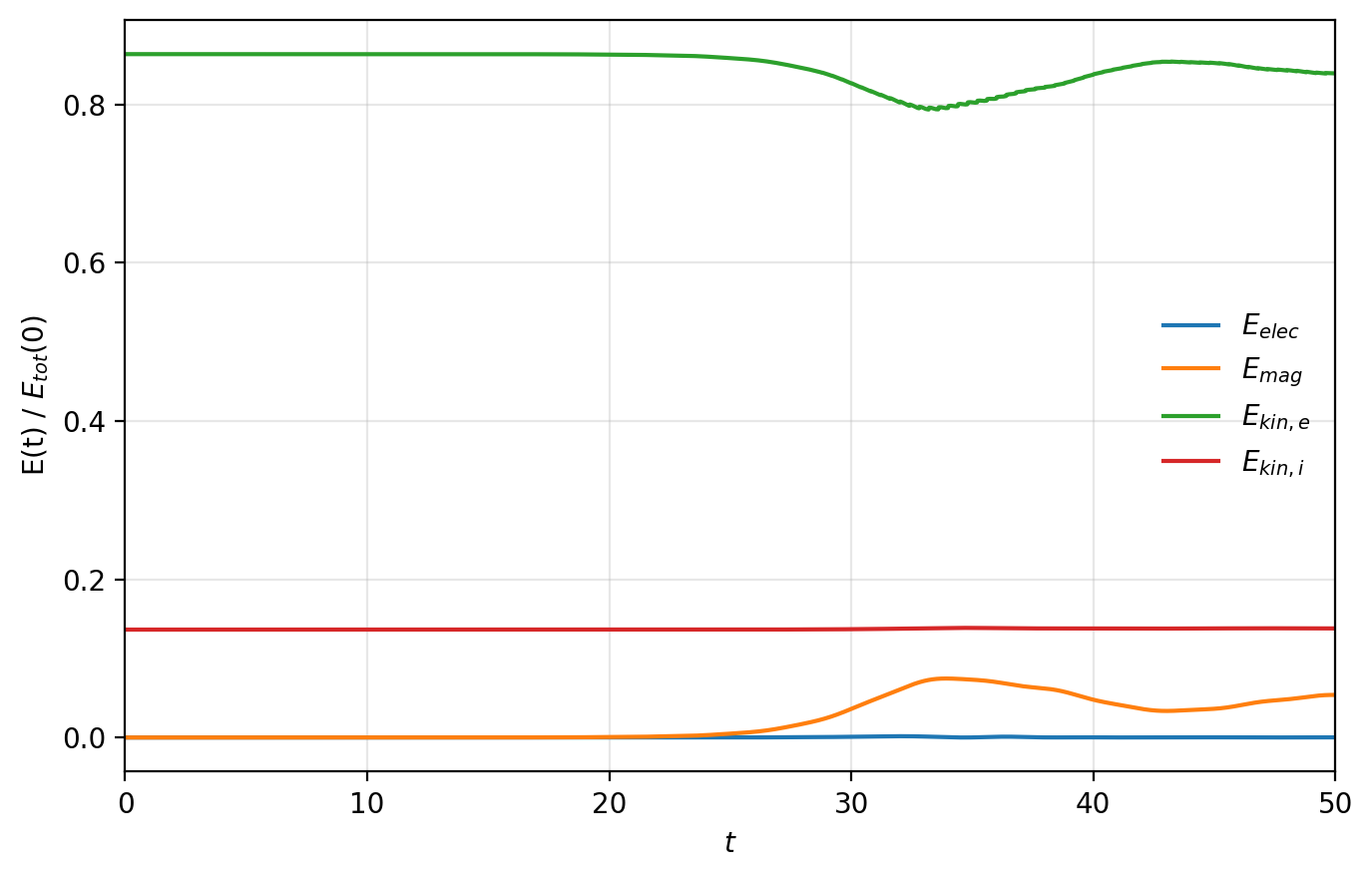}\subcaption{\label{fig:energy_normalized_energies_2x3v_rotated_filament}
Relative share of each energy component
}
\end{subfigure}
\begin{subfigure}{0.49\textwidth}
\includegraphics[width=0.99\textwidth]{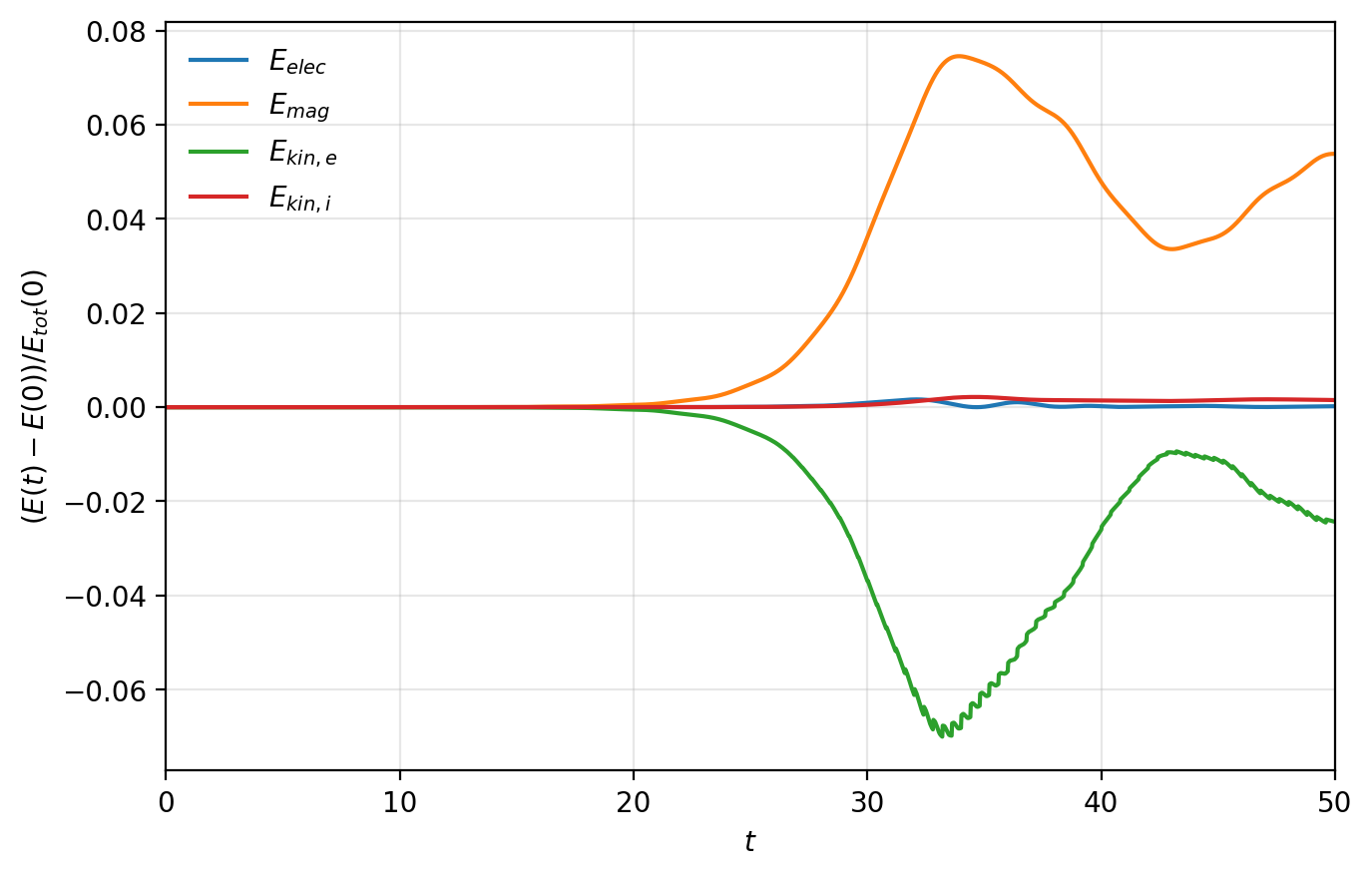}\subcaption{\label{fig:energy_energy_changes_2x3v_rotated_filament}
Relative deviation of each energy component
}
\end{subfigure}
\caption{\label{fig:energies_2x3v_rotated_filament_instab}
Energy evolution for the 2x3v \textit{rotated filamentation instability}. 
The top left panel shows the evolution of electric and magnetic energies, the 
top right panel shows how the difference of the initial and current ion kinetic 
energy evolves over time, the bottom panels show time-evolution of the 
deviation and just the energies scaled by the total energy respectively.
}
\end{figure}

\begin{figure}[h!]
\centering
\begin{subfigure}{0.49\textwidth}
\includegraphics[width=\textwidth]{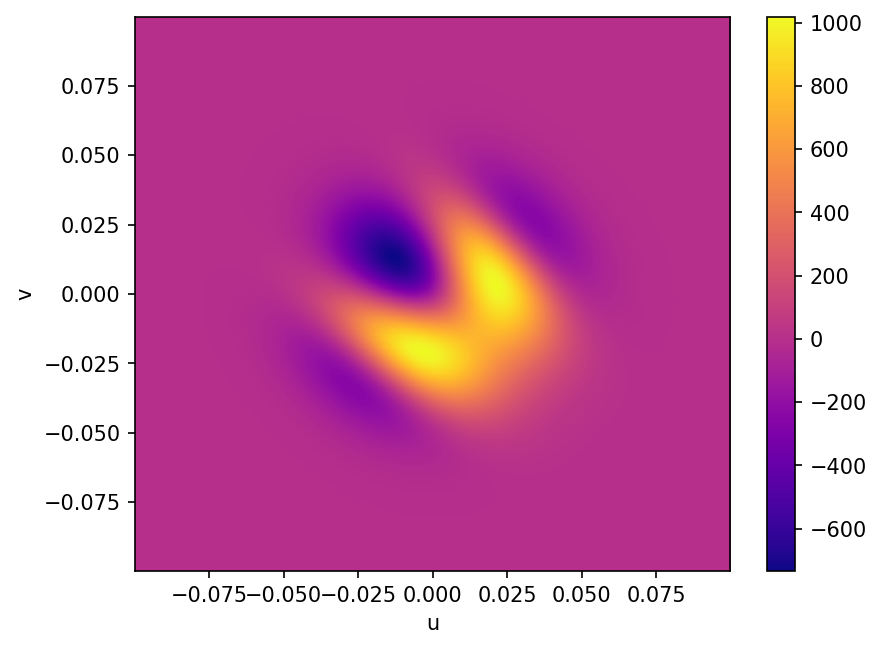}
\subcaption{\label{fig:f_e_50_uv_2x3v_rotated_filament}
$(f^i - f_0^i)(50,L/2,L/2,\cdot,\cdot,0)$
}
\end{subfigure}
\begin{subfigure}{0.49\textwidth}
\includegraphics[width=0.95\textwidth]{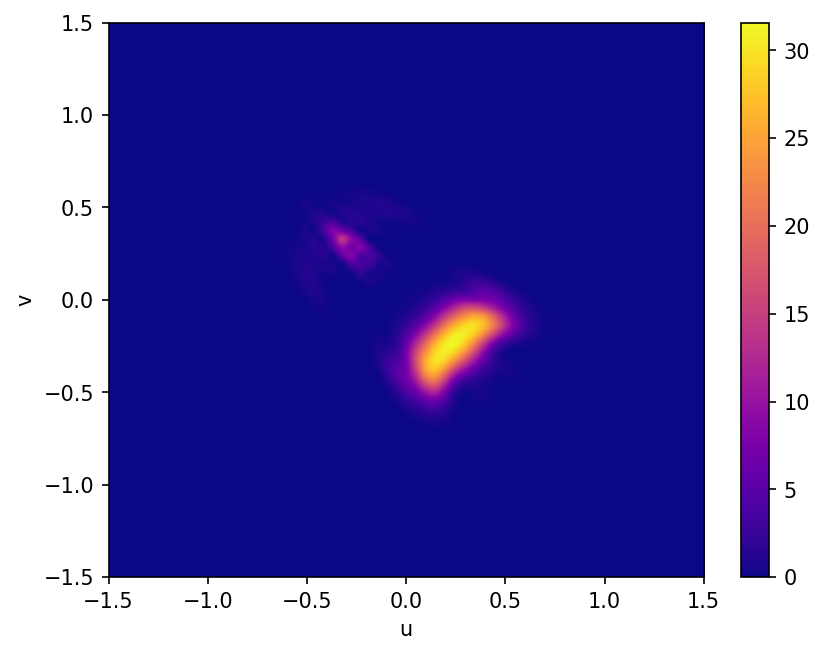}\subcaption{\label{fig:f_i_diff_50_uv_2x3v_rotated_filament}
$f^e(50,L/2,L/2,\cdot,\cdot,0)$
}
\end{subfigure}
\begin{subfigure}{0.49\textwidth}
\includegraphics[width=0.95\textwidth]{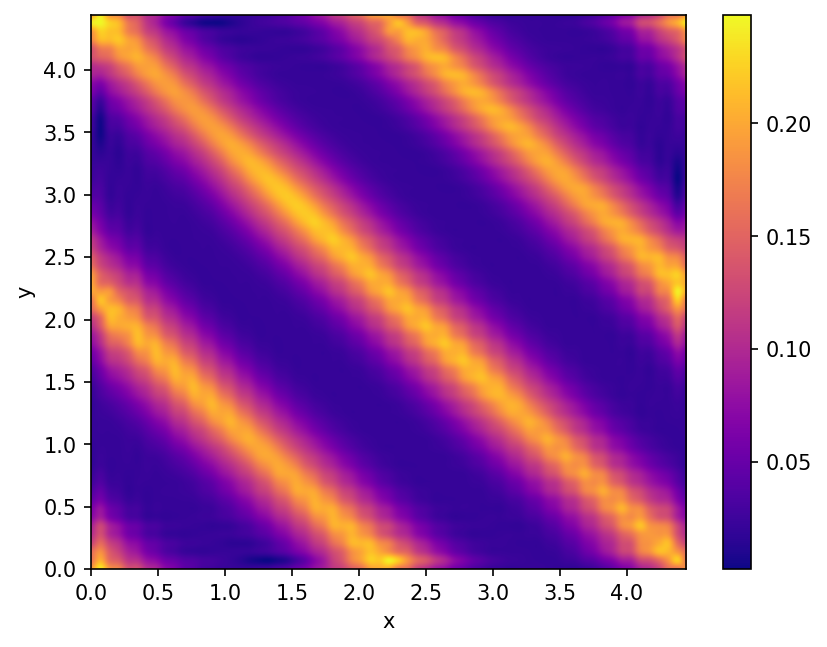}\subcaption{\label{fig:f_e_50_xy_2x3v_rotated_filament}
$f^i(50,\cdot,\cdot,0,0,0)$
}
\end{subfigure}
\begin{subfigure}{0.49\textwidth}
\includegraphics[width=0.95\textwidth]{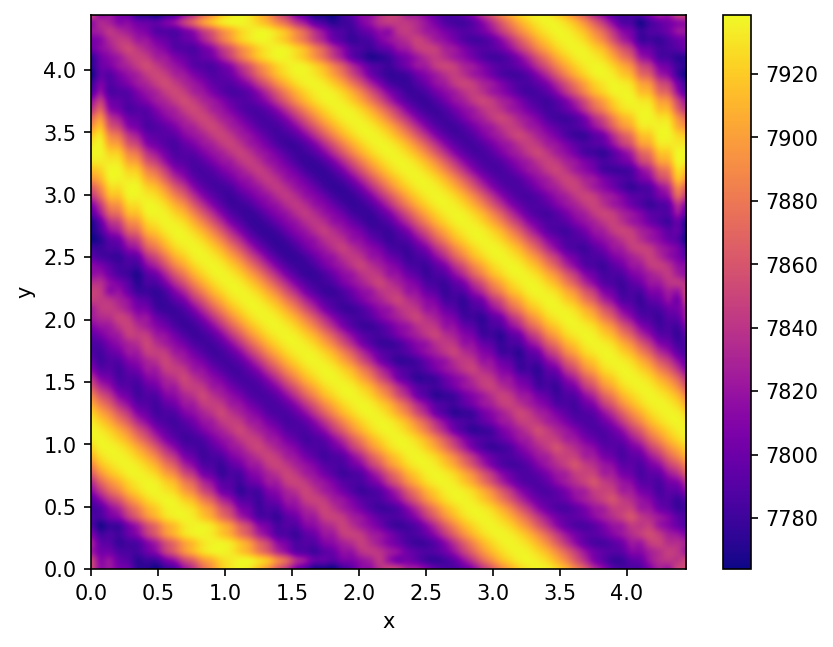}\subcaption{\label{fig:f_i_50_xy_2x3v_rotated_filament}
$f^e(50,\cdot,\cdot,0,0,0)$
}
\end{subfigure}
\caption{\label{fig:f_2x3v_rotated_filament_instab}
Slices of the ion (left) and electron (right) distribution functions at times
$t = 50 \omega_{p,e}^{-1}$. Top we show a $u$-$v$ slice at $x=y=L/2$ as well 
as $w=0$ and on the bottom a $x$-$y$ slice at $u=v=w=0$. For the ion $u$-$v$
slice we actually show the difference to the initial Maxwellian for better
visibility of the changes.
}
\end{figure}

The growth rate from the 1x2v \textit{filamentation instability}, section 
\ref{sec:filamentation_instability}, is in fact well reproduced as can be seen 
in \ref{fig:energy_dynamics_2x3v_rotated_filament}. Because the ions still have 
a substantially higher mass than the electrons, it is expected that their 
influence on the dynamics is still relatively small and in particular the 
linear growth rate of the instability remains almost unchanged. In the figures
\ref{fig:energy_energy_changes_2x3v_rotated_filament} and 
\ref{fig:energy_normalized_energies_2x3v_rotated_filament} we see that the 
dynamics is still strongly dominated by the electrons, however, in the non-
linear phase the ions also absorb some energy. In fact in figure 
\ref{fig:ion_kinetic_energy_growth_2x3v_rotated_filament} we see that the ion
kinetic energy also grows with the expected growth rate confirming that while
the role of the ions for the overall dynamics is indeed minor, they do partake 
in it. 
In figure \ref{fig:f_2x3v_rotated_filament_instab} we also show case slices of
the ion and electron distribution functions. The intially prescribed 45$^\circ$
rotation is clearly visible when comparing to the result from the 1x2v 
counterpart. Additionally we see that the ions indeed follow the dynamics of the electrons, albeit with smaller variation of the distribution function as 
is expected due to the large mass difference.

Note that over the simulation period we have a total energy error of roughly 
3 \%, which is due to the relatively frequent restarts, which currently are 
neither mass nor energy-preserving. 

\subsection{Conservation properties}
\label{sec:num_test_conservation_properties}

\subsubsection{NuFI time-stepping}

Following the literature~\autocite{CROUSEILLES2015224, KORMANN2025118290}, 
we first consider \emph{Landau Damping}, see section 
\ref{sec:weak_landau_damping}, to test the conservation properties 
of our approach. While a physically highly relevant phenomena, the (isolated) 
dynamics of \emph{Landau Damping} are less complicated and therefore 
well-suited as a first stress-test close to linear theory. For this relatively 
simple case we only consider the 1st order in time Lie-splitting for both NuFI 
and SL.

\begin{figure}[h!]
\centering
\begin{subfigure}{0.49\textwidth}
\includegraphics[width=0.99\textwidth]{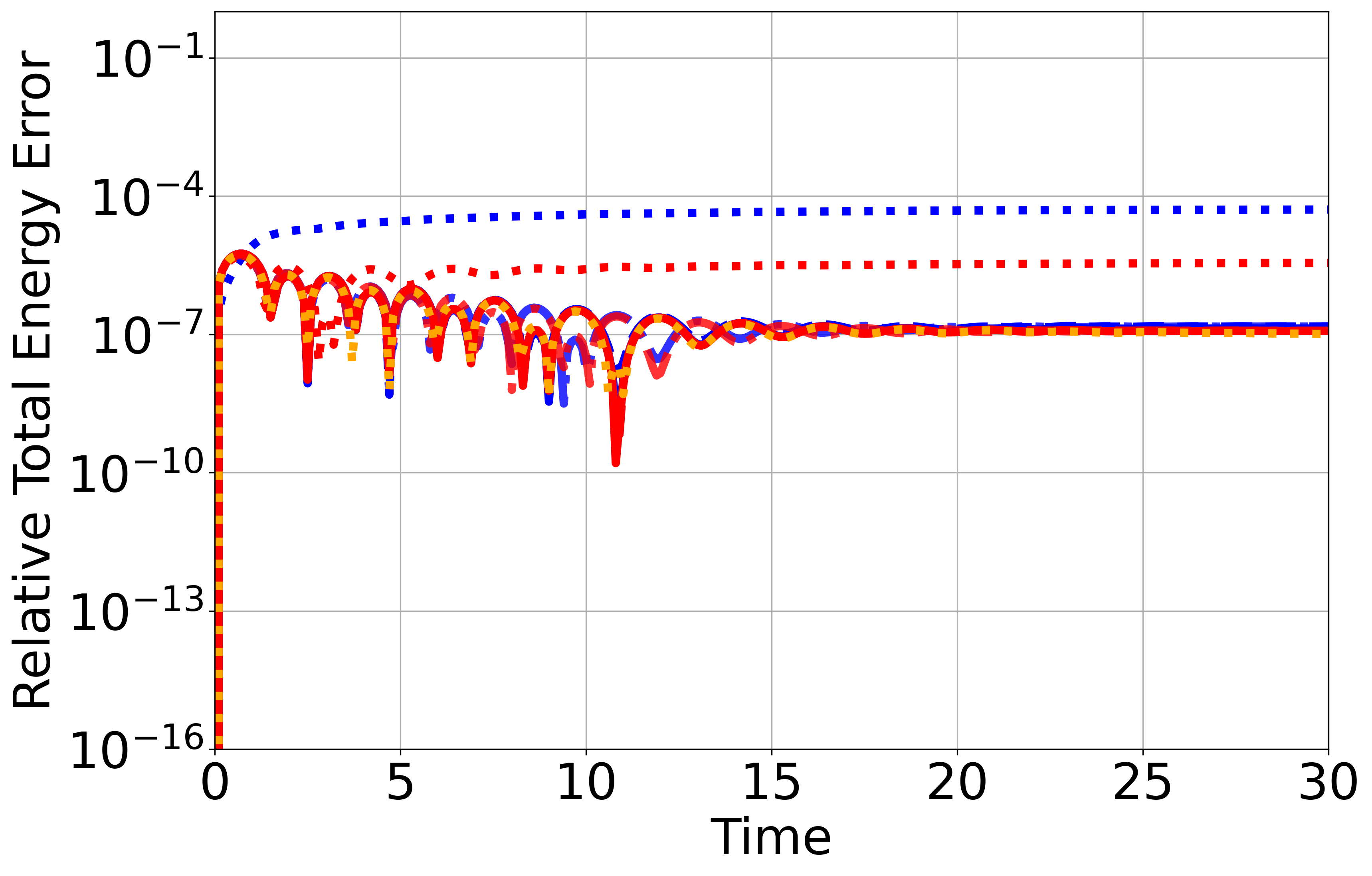}
\subcaption{\label{fig:tot_energy_conserve_landau_damp}
Total Energy
}
\end{subfigure}
\begin{subfigure}{0.49\textwidth}
\includegraphics[width=0.99\textwidth]{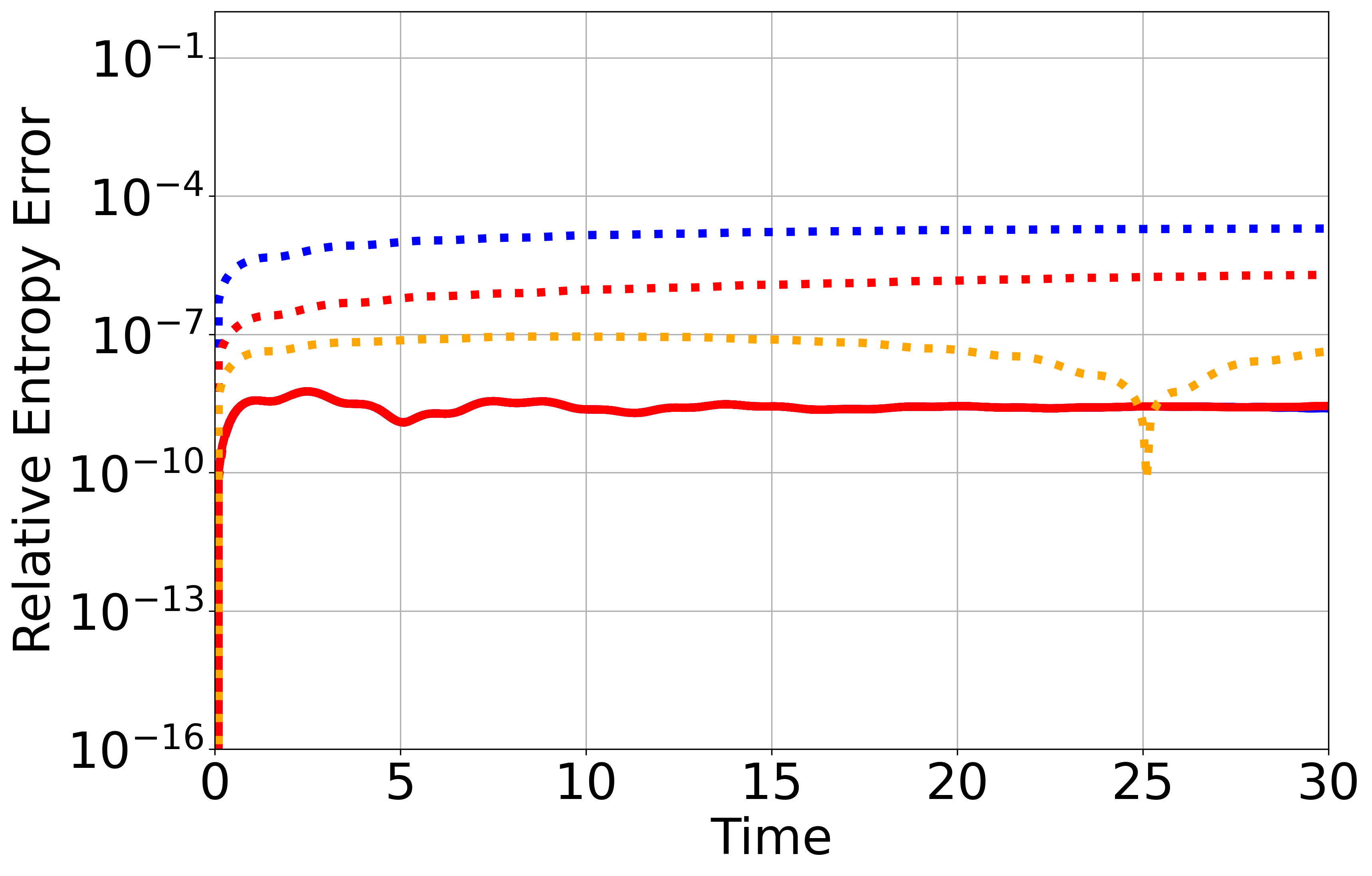}
\subcaption{\label{fig:entropy_conserve_landau_damp}
Entropy
}
\end{subfigure}
\begin{subfigure}{0.49\textwidth}
\includegraphics[width=0.99\textwidth]{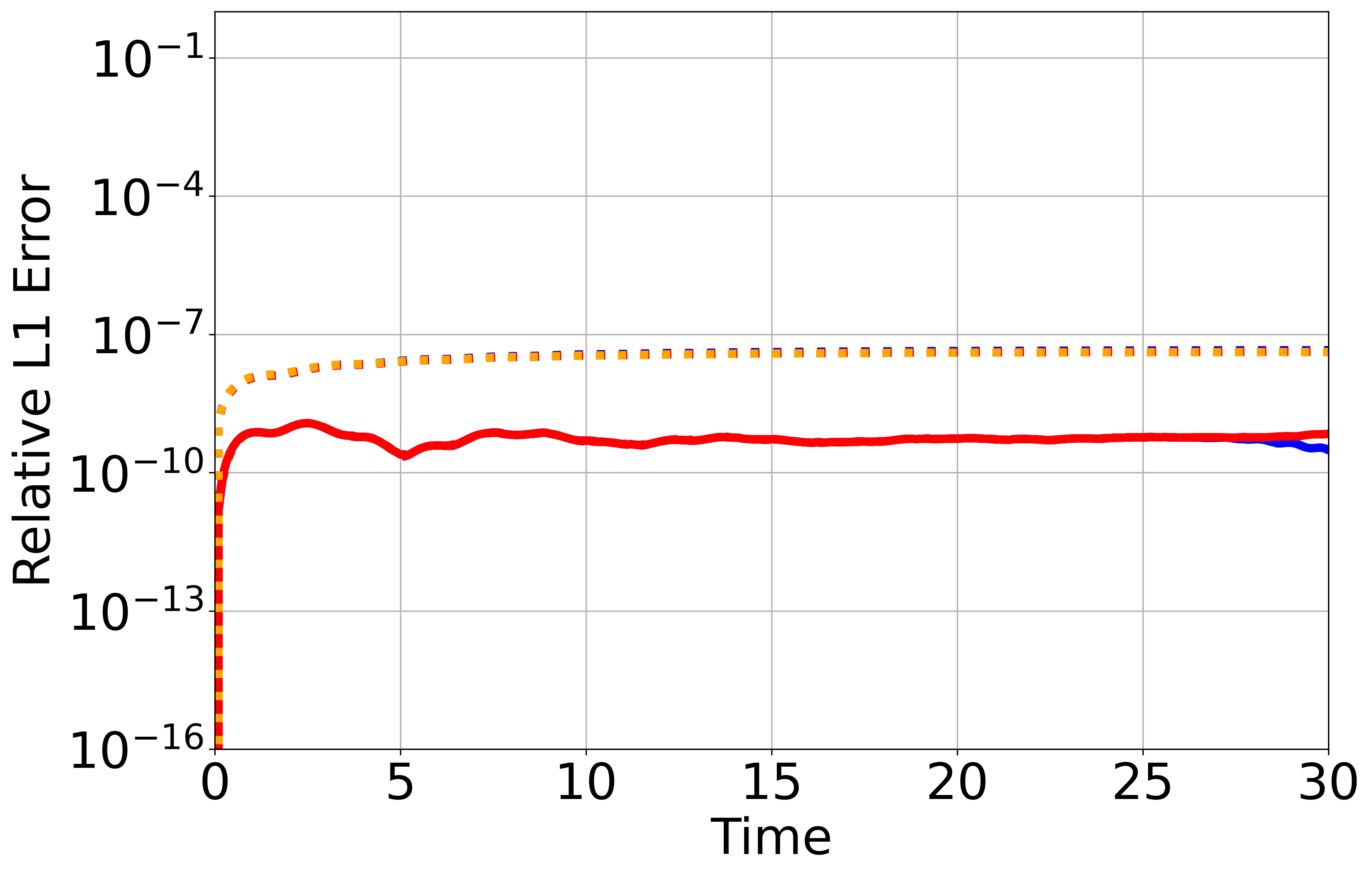}
\subcaption{\label{fig:l1_norm_conserve_landau_damp}
$L^1$-norm
}
\end{subfigure}
\begin{subfigure}{0.49\textwidth}
\includegraphics[width=0.99\textwidth]{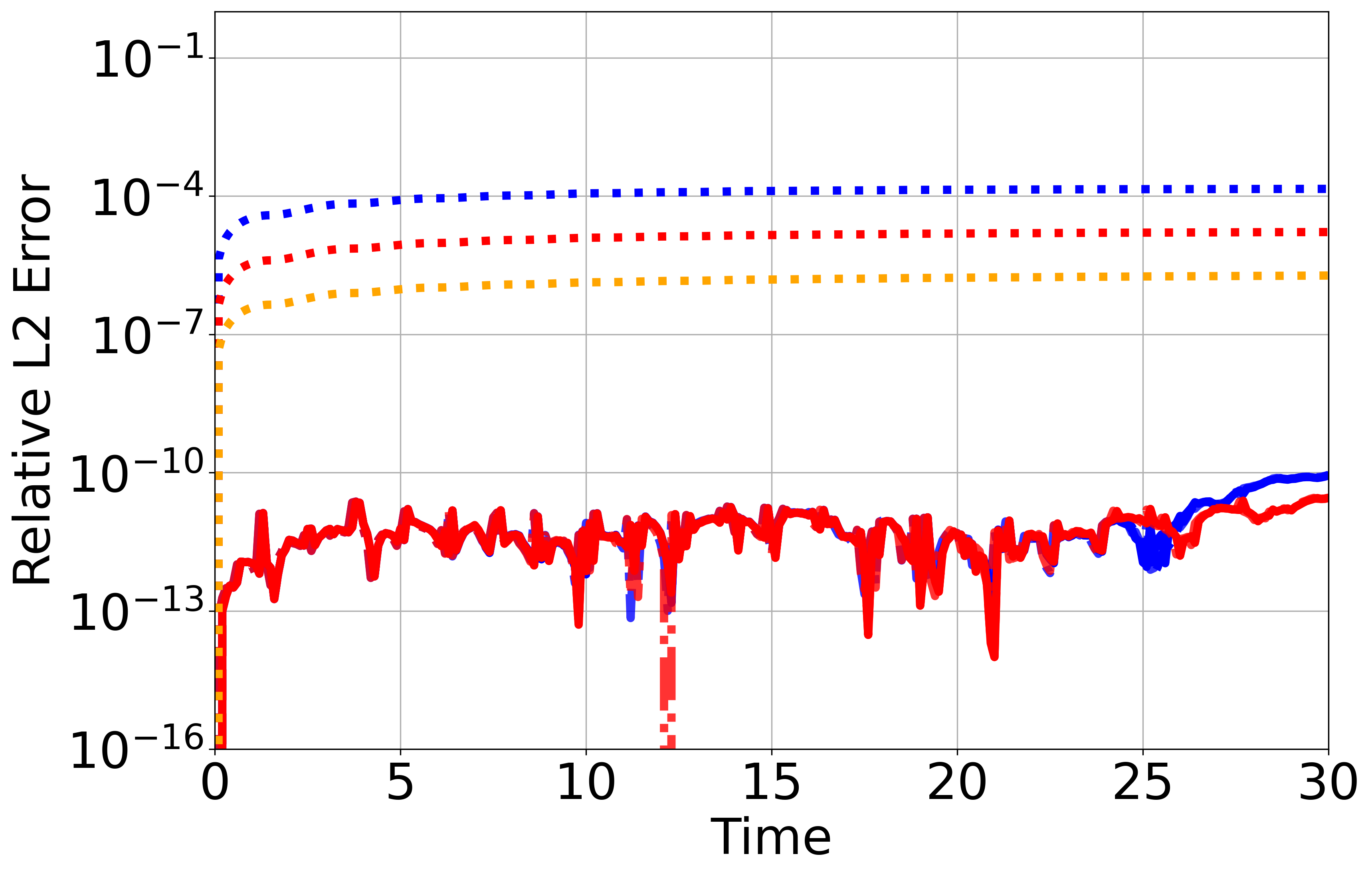}
\subcaption{\label{fig:l2_norm_conserve_landau_damp}
$L^2$-norm
}
\end{subfigure}
\begin{subfigure}{0.99\textwidth}
\includegraphics[width=0.99\textwidth]{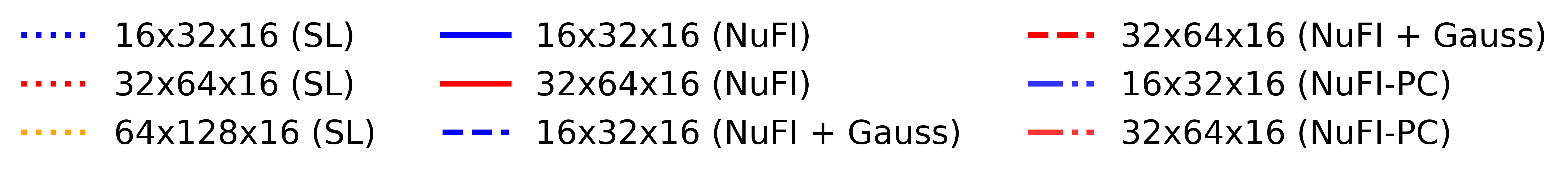}
\end{subfigure}
\caption{\label{fig:conservation_properties_landau_damping}
Conservation of total energy, entropy as well as $L^1$- and $L^2$-norm for 
the \emph{weak Landau Damping} test case compared between NuFI-Ham, NuFI-PC and 
Semi-Lagrangian with Lie splitting.
}
\end{figure}

In figure \ref{fig:conservation_properties_landau_damping} we look at the 
relative deviation from the start value of total energy, entropy as well as 
$L^1$ and $L^2$-norm throughout the simulation. The graphs of NuFI-Ham and 
NuFI-PC show almost identical behavior during the simulation time. The relative 
error remains at a constant level or, in the case of total energy, even 
declines slightly. Notably Gauss cleaning does not introduce notable errors in
any of the observed quantities, however, as \emph{Landau Damping} is an 
electro-static effect a natural good agreement for the Gauss law is somewhat 
expected and therefore the potential corrections to $\EE$ (and their respectively 
introduced errors) likely remain small.
Also note that the agreement of the errors in entropy as well as $L^p$-norms
is expected for NuFI, especially in a case close to equilibrium, as technically
those quantities are preserved analytically, i.e., the observed error is the
quadrature error made when numerically evaluating the involved integrals, which 
in turn uses the same number of quadrature points for all simulations.

While the Semi-Lagrangian approach also preserves all observed quantities well,
the observed errors for SL-Ham are substantially higher than for all NuFI 
approaches. Towards the end of the simulation NuFI has a 3 orders of magnitude 
smaller total energy error and for the $L^2$-norm the error is even between 4 
and 6 orders of magnitude smaller for NuFI compared to SL-Ham depending on the
resolution.

\begin{figure}[h!]
\centering
\includegraphics[width=0.99\textwidth]{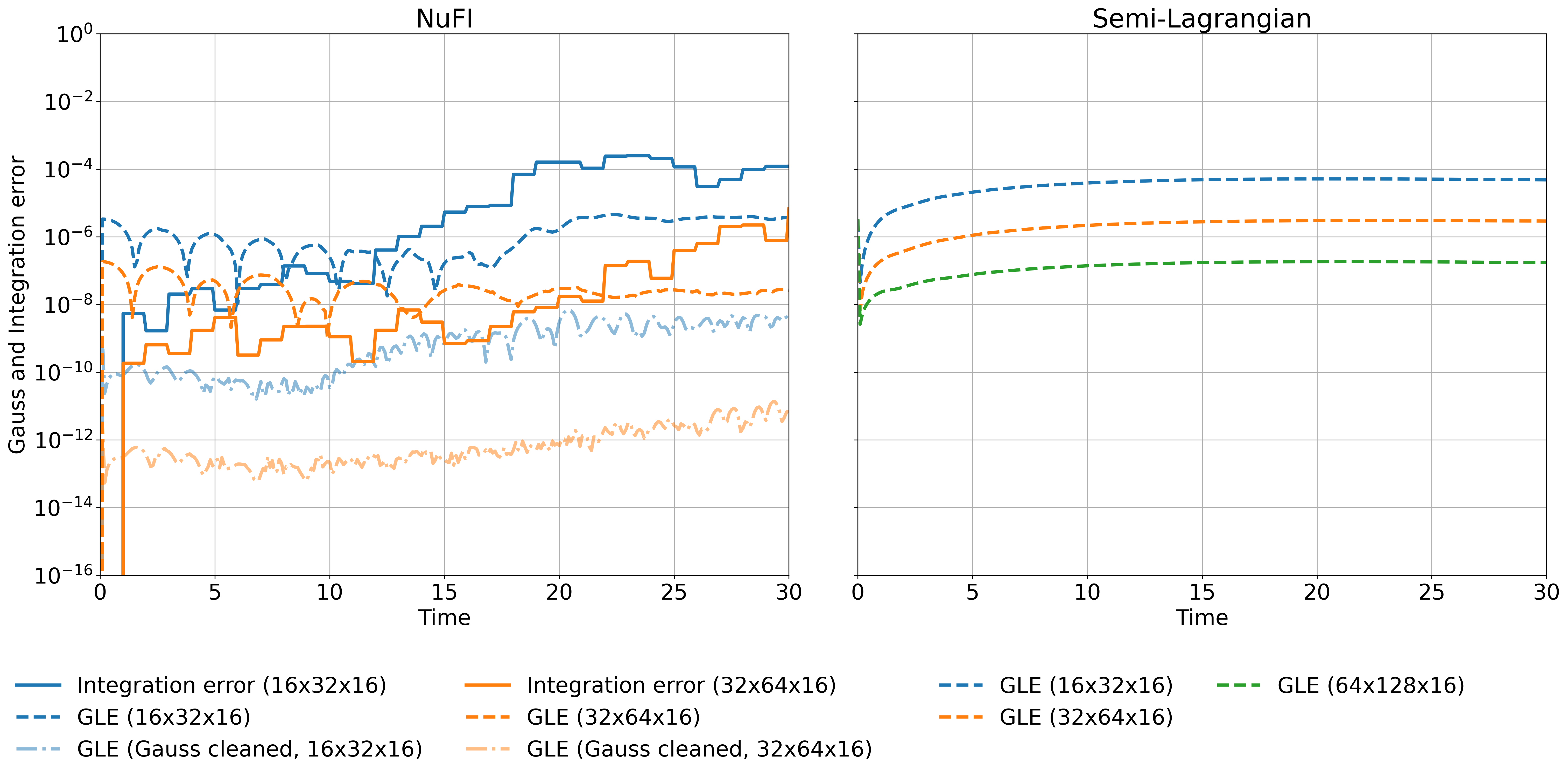}
\caption{\label{fig:gle_landau_damping}
Gauss law error (GLE) and integration error for the \emph{weak Landau Damping} 
test case compared between NuFI and Semi-Lagrangian with Lie splitting.
}
\end{figure}

In figure \ref{fig:gle_landau_damping} we look at the Gauss law error (GLE)
and integration error (IE). The GLE is about 1 order of magnitude
better for NuFI-Ham when comparing with SL-Ham with the same resolution. 
When additionally also introducing Gauss cleaning in NuFI the GLE
can be further reduced by 3-4 orders of magnitude. 

In this case, the GLE is initially slightly higher than the 
integration error but with advanced dynamics the distribution function also 
becomes more complicated and hence the integration error overtakes the GLE. 
Note that initially the GLE is dominated by the space discretization and in the 
later stage is bounded from above through the integration error as expected 
from theory.

\begin{figure}[h!]
\centering
\begin{subfigure}{0.49\textwidth}
\includegraphics[width=0.99\textwidth]{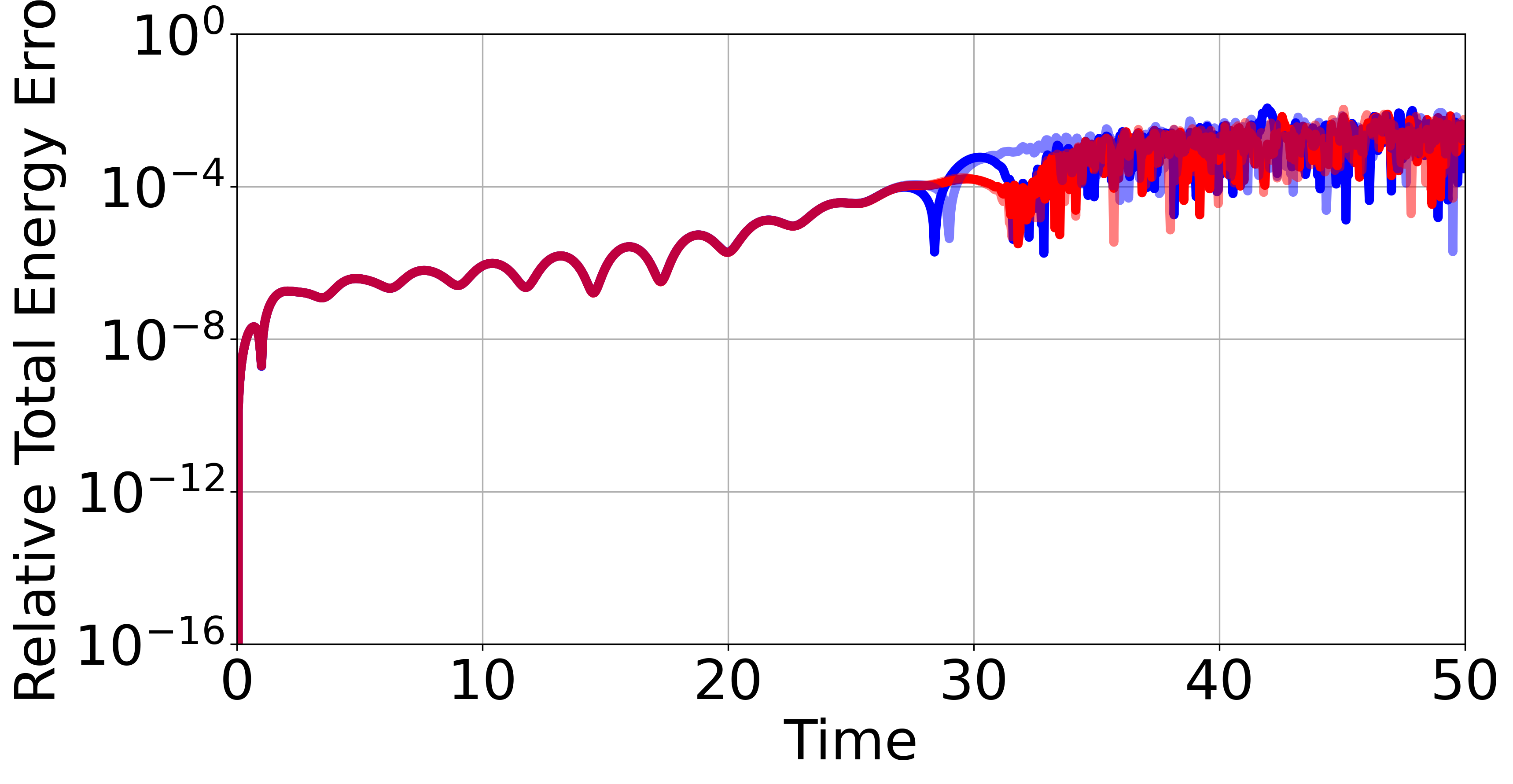}
\subcaption{\label{fig:nufi_tot_energy_conserve_filamentation_instab}
Total Energy, NuFI
}
\end{subfigure}
\begin{subfigure}{0.49\textwidth}
\includegraphics[width=0.99\textwidth]{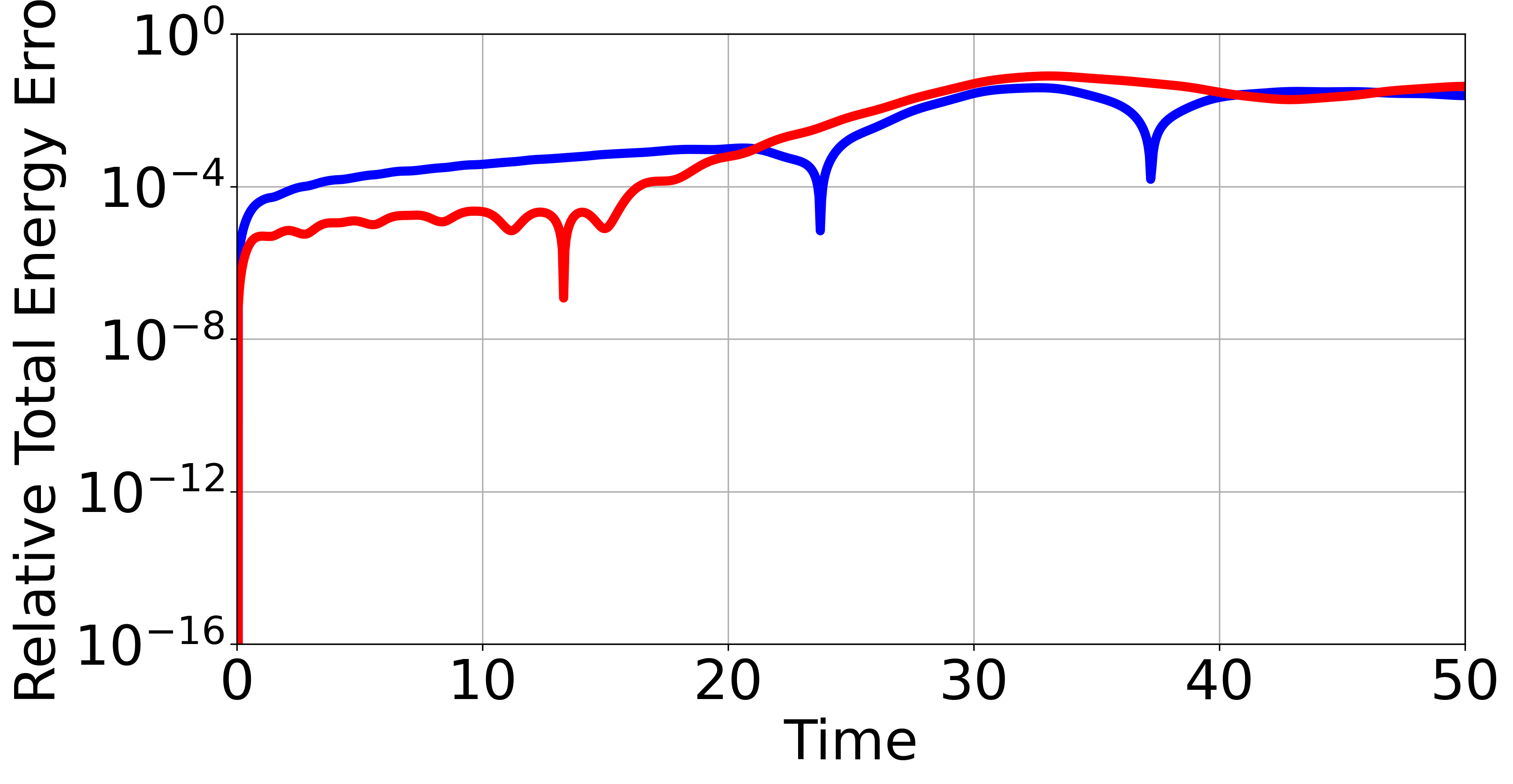}
\subcaption{\label{fig:sl_tot_energy_conserve_filamentation_instab}
Total Energy, SL
}
\end{subfigure}
\begin{subfigure}{0.49\textwidth}
\includegraphics[width=0.99\textwidth]{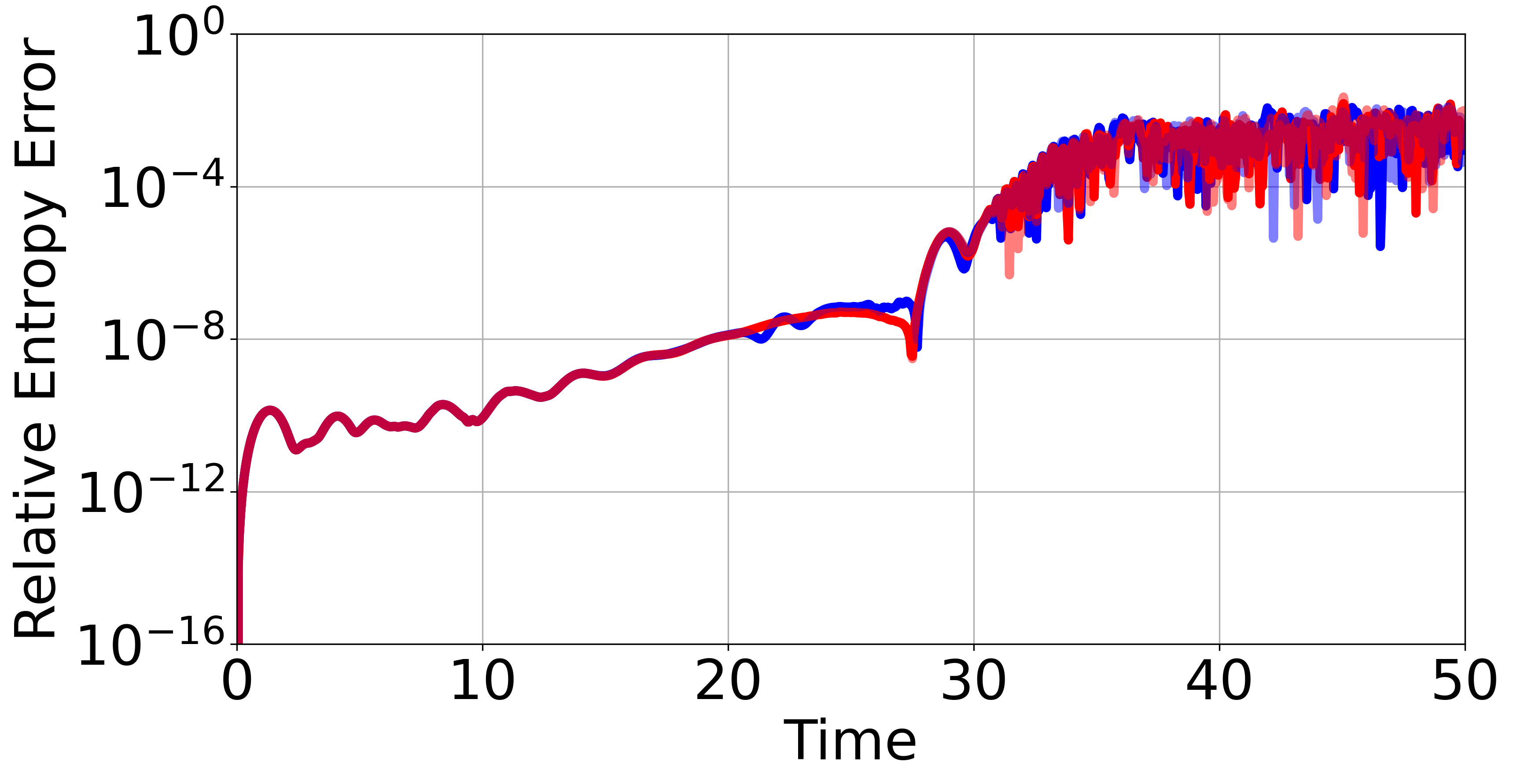}
\subcaption{\label{fig:nufi_entropy_conserve_filamentation_instab}
Entropy, NuFI
}
\end{subfigure}
\begin{subfigure}{0.49\textwidth}
\includegraphics[width=0.99\textwidth]{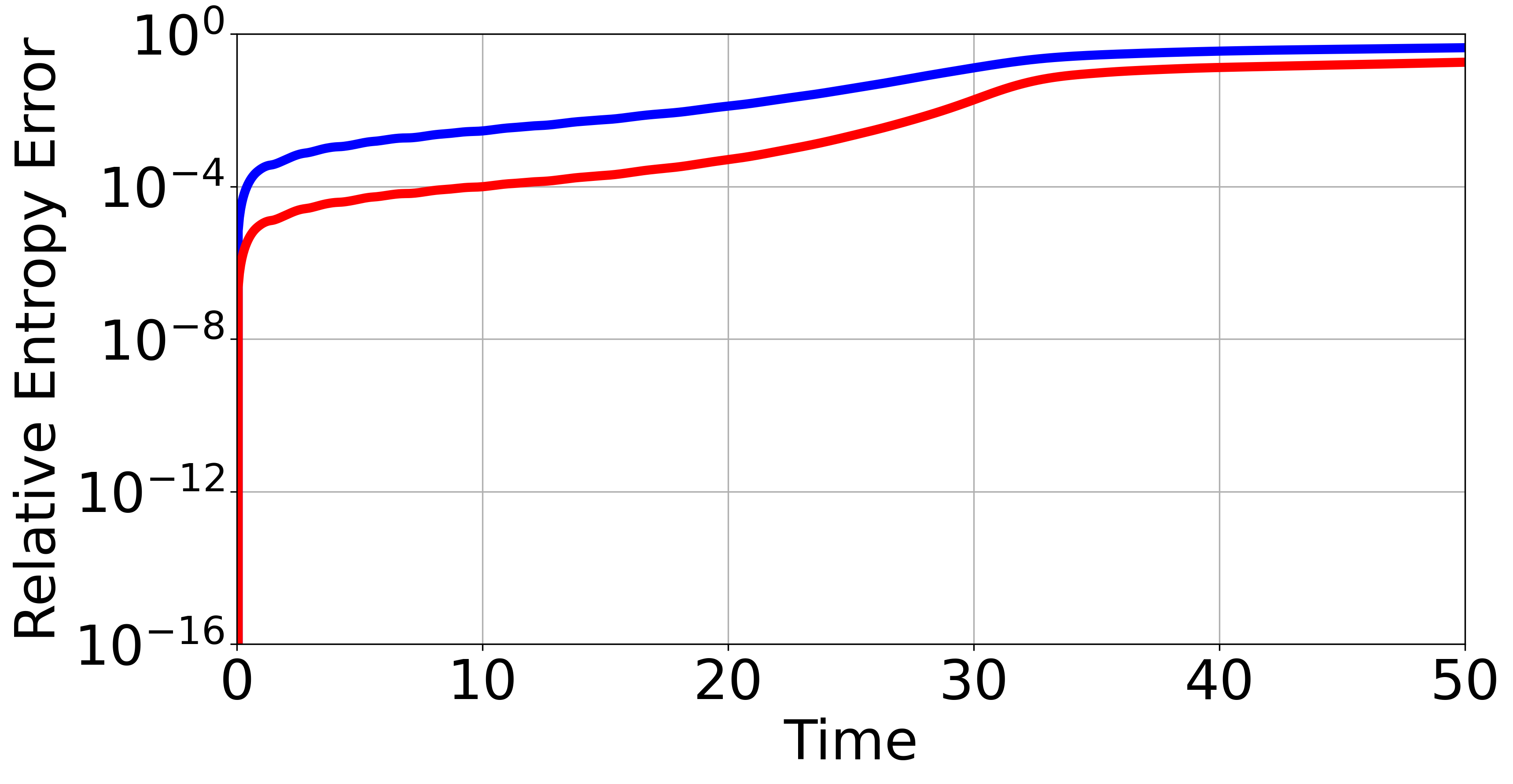}
\subcaption{\label{fig:sl_entropy_conserve_filamentation_instab}
Entropy, SL
}
\end{subfigure}
\begin{subfigure}{0.49\textwidth}
\includegraphics[width=0.99\textwidth]{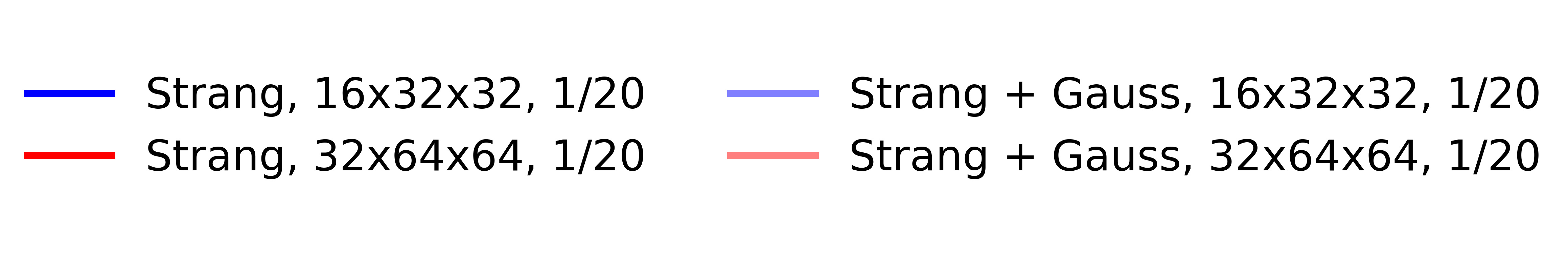}
\end{subfigure}
\begin{subfigure}{0.49\textwidth}
\includegraphics[width=0.99\textwidth]{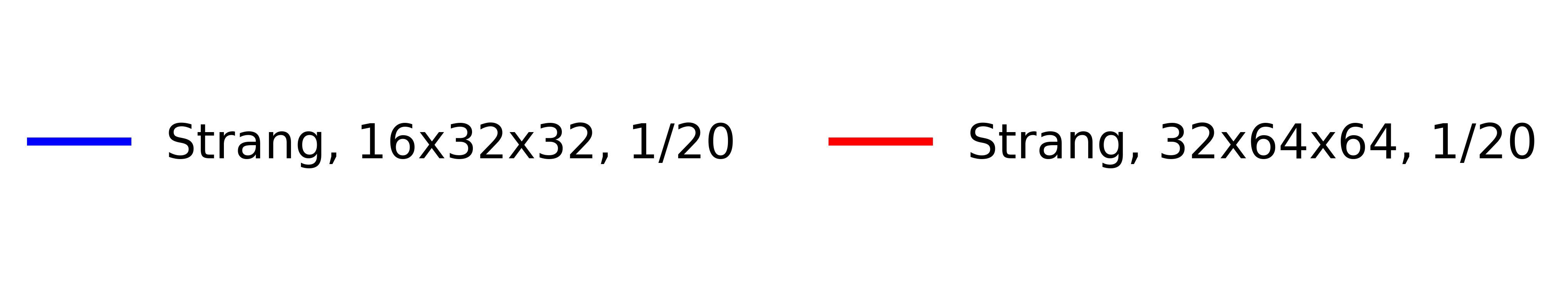}
\end{subfigure}
\caption{\label{fig:conservation_energy_entropy_properties_filamentation_instab}
Conservation of total energy and entropy for the \emph{filamentation 
instability} test case  compared between NuFI-Ham and Semi-Lagrangian with 2nd 
order Strang splitting.
}
\end{figure}

Next we consider the \emph{filamentation instability}, see section 
\ref{sec:filamentation_instability}, which is a genuine electro-magnetic test 
case. Here we test the 2nd order in time Strang splitting. In figure 
\ref{fig:conservation_energy_entropy_properties_filamentation_instab} we look
at the conservation of total energy and entropy.

As already observed for the previous example, in the linear phase of the 
dynamics until $t\approx 20$ the relative deviation for NuFI remains below 
$10^{-5}$ for the total energy, while SL-Ham here already shows errors of up to
$10^{-3}$. In the non-linear phase the error for NuFI remains below 
$1\%$ for all times while for SL-Ham the error grows up to $5\%$. For the 
higher resolution run the relative energy error at $T=50$ for SL-Ham reaches 
$4\%$ and for NuFI it is $0.2 \%$, i.e., roughly 20 times smaller.

For the conservation of entropy the picture is even clearer as the relative 
error of NuFI for both resolutions remains below $1\%$ while for SL-Ham the 
error exceeds $43 \%$ for the lower and $18 \%$ for the higher resolution.

\begin{figure}[h!]
\centering
\begin{subfigure}{0.49\textwidth}
\includegraphics[width=0.99\textwidth]{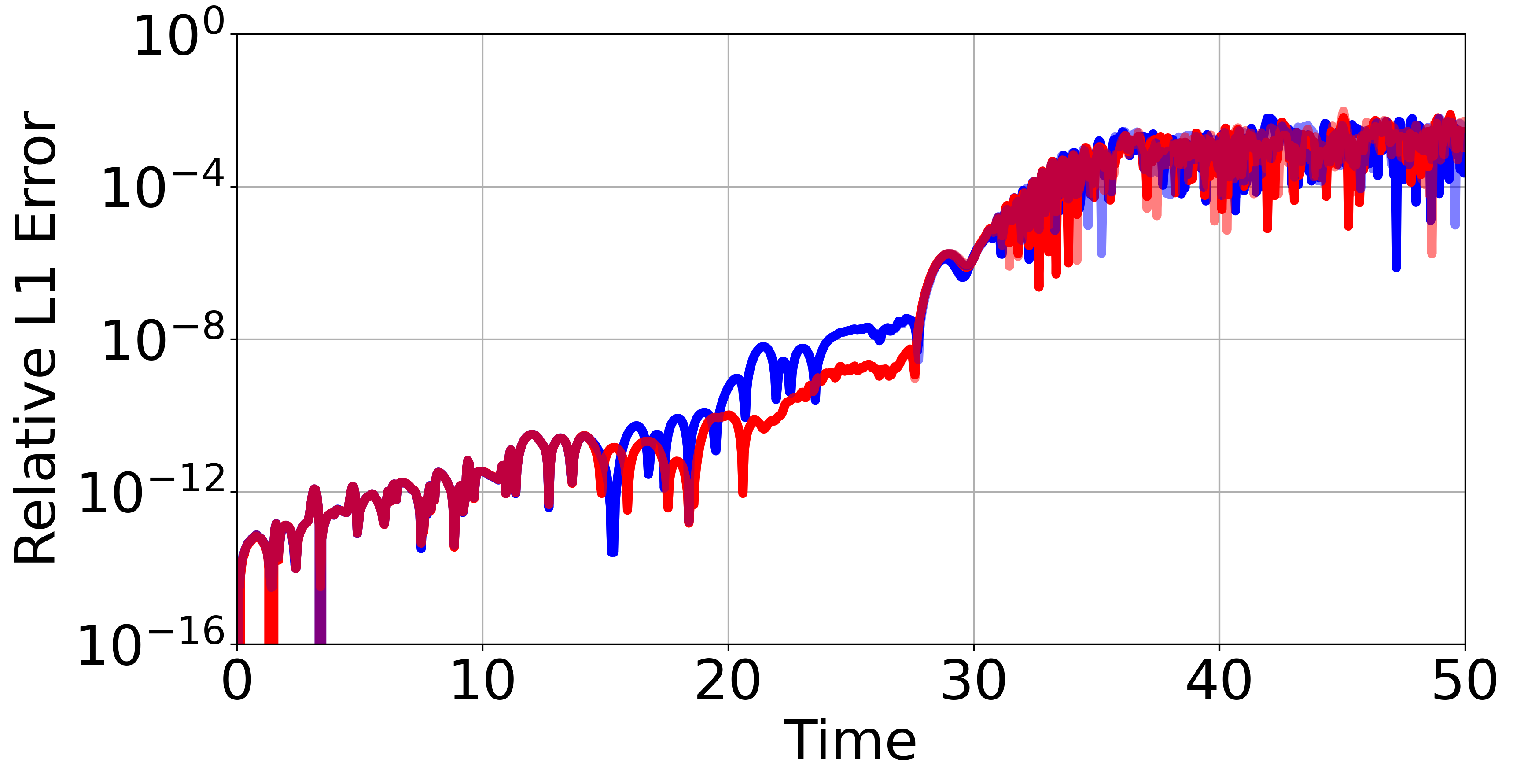}
\subcaption{\label{fig:nufi_l1_conserve_filamentation_instab}
$L^1$-norm, NuFI
}
\end{subfigure}
\begin{subfigure}{0.49\textwidth}
\includegraphics[width=0.99\textwidth]{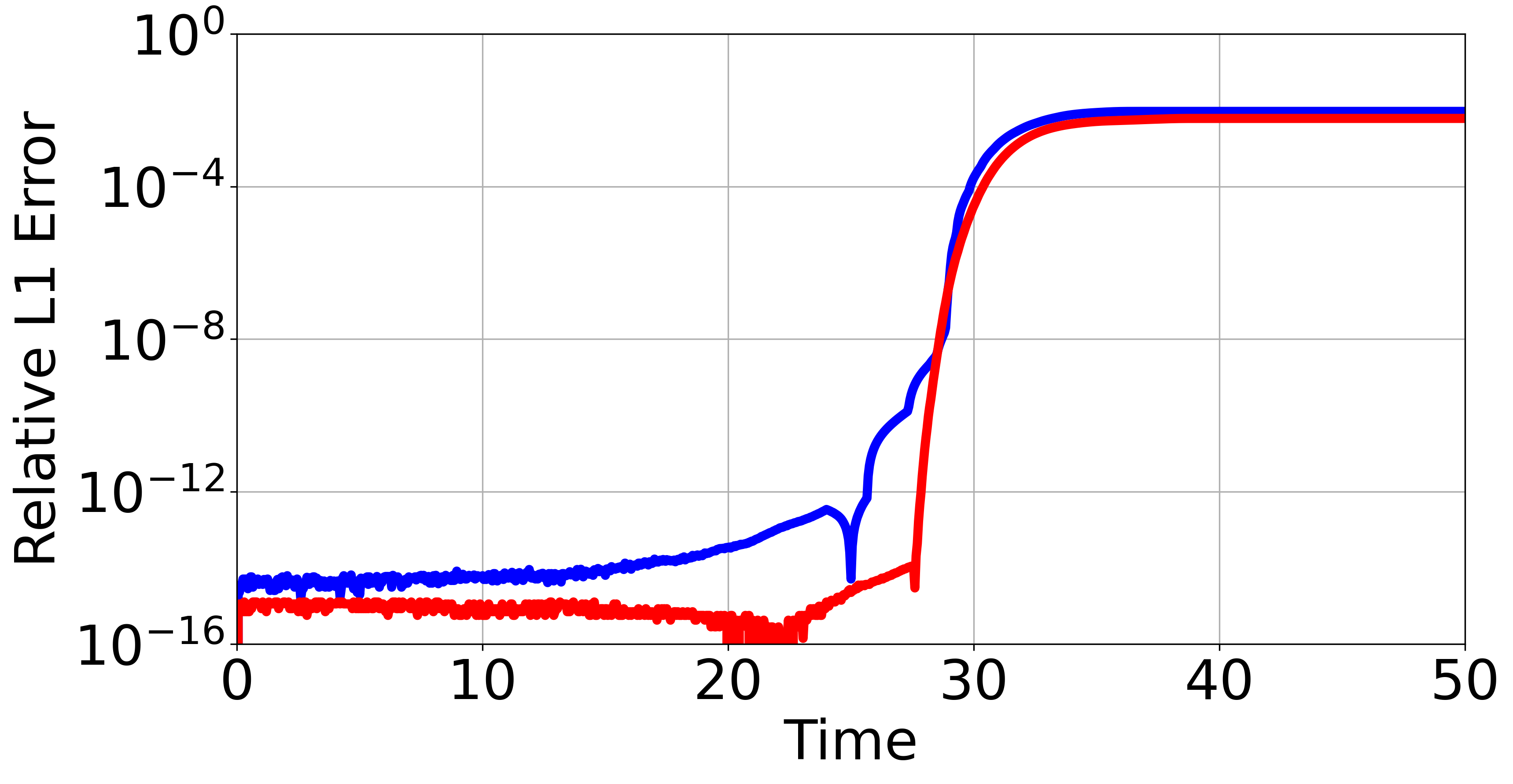}
\subcaption{\label{fig:sl_l1_conserve_filamentation_instab}
$L^1$-norm, SL
}
\end{subfigure}
\begin{subfigure}{0.49\textwidth}
\includegraphics[width=0.99\textwidth]{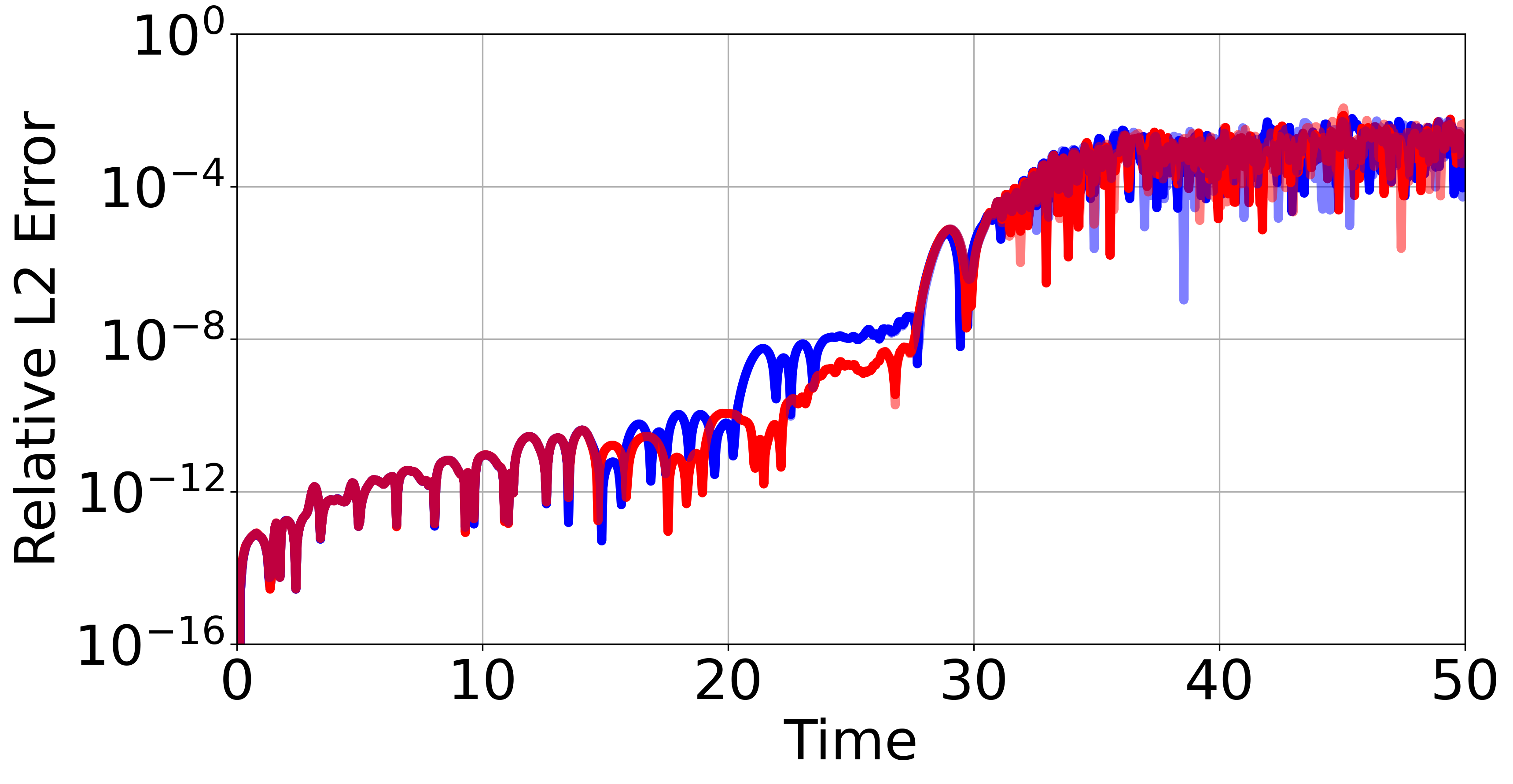}
\subcaption{\label{fig:nufi_l2_conserve_filamentation_instab}
$L^2$-norm, NuFI
}
\end{subfigure}
\begin{subfigure}{0.49\textwidth}
\includegraphics[width=0.99\textwidth]{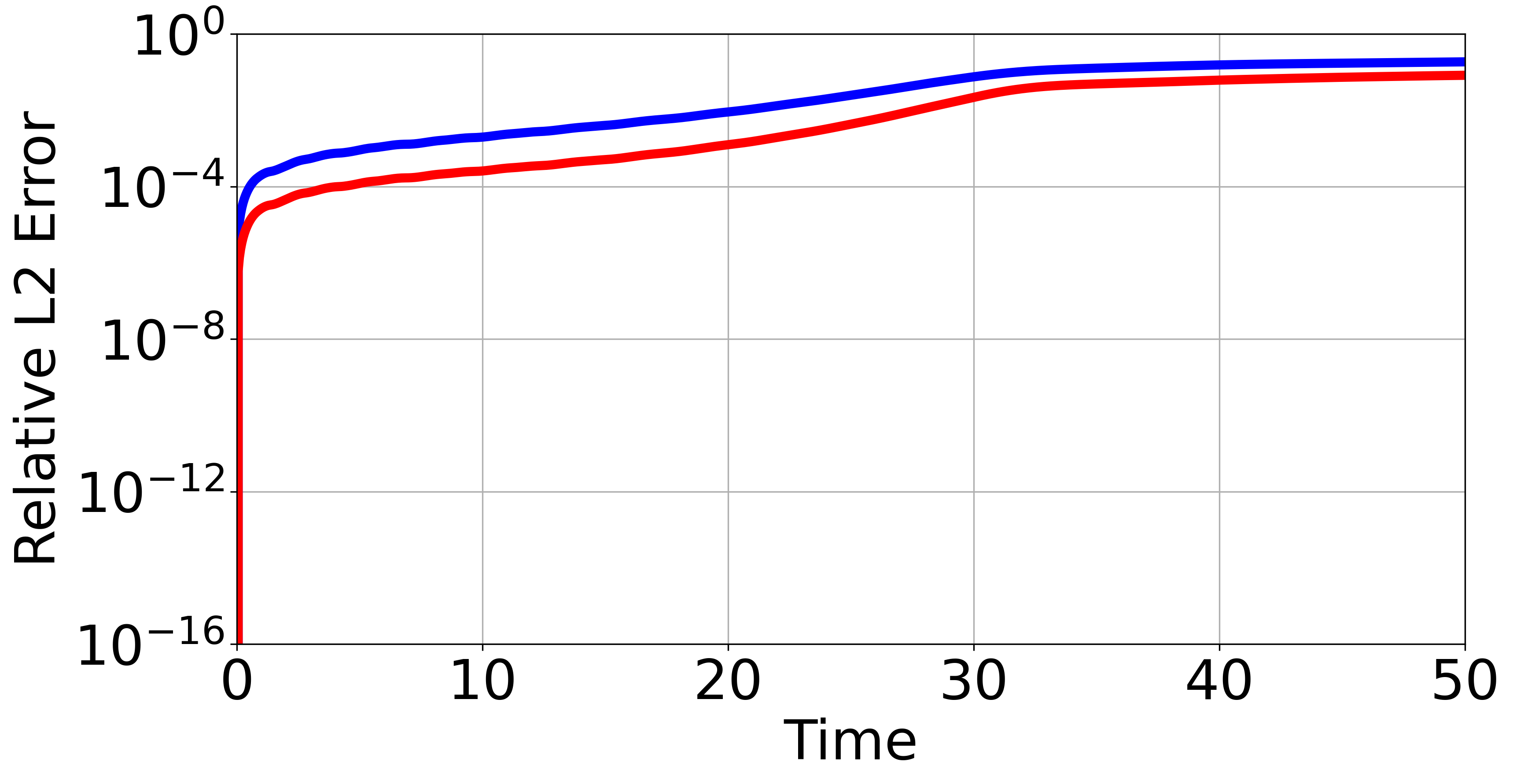}
\subcaption{\label{fig:sl_l2_conserve_filamentation_instab}
$L^2$-norm, SL
}
\end{subfigure}
\begin{subfigure}{0.49\textwidth}
\includegraphics[width=0.99\textwidth]{Figures/Conservation_Properties/Filamentation_Instability/Figures/legend_nufi}
\end{subfigure}
\begin{subfigure}{0.49\textwidth}
\includegraphics[width=0.99\textwidth]{Figures/Conservation_Properties/Filamentation_Instability/Figures/legend_sl}
\end{subfigure}
\caption{\label{fig:conservation_lp_properties_filamentation_instab}
Conservation of $L^1$- and $L^2$-norm for the \emph{filamentation 
instability} test case compared between NuFI-Ham and Semi-Lagrangian with 2nd 
order Strang splitting.
}
\end{figure}

In figure \ref{fig:conservation_lp_properties_filamentation_instab} we look
at the conservation of the $L^1$- and $L^2$-norm. Here the already observed
trends for the total energy and entropy are confirmed. The $L^1$-norm, 
i.e., mass, is well conserved by both approaches with a maximum error 
of $1\%$, but the same is only true for the $L^2$-norm conservation by NuFI 
whereas SL-Ham exhibits errors of up to $10\%$ for late times. 

Additionally, whether Gauss law cleaning is applied or not does not seem to 
noticeably effect the error of conservation laws for NuFI. In particular, for 
the conservation of total energy this is a good sign as it numerically confirms
that Gauss cleaning does not break the conservation properties of NuFI.

\begin{figure}[h!]
\centering
\includegraphics[width=0.99\textwidth]{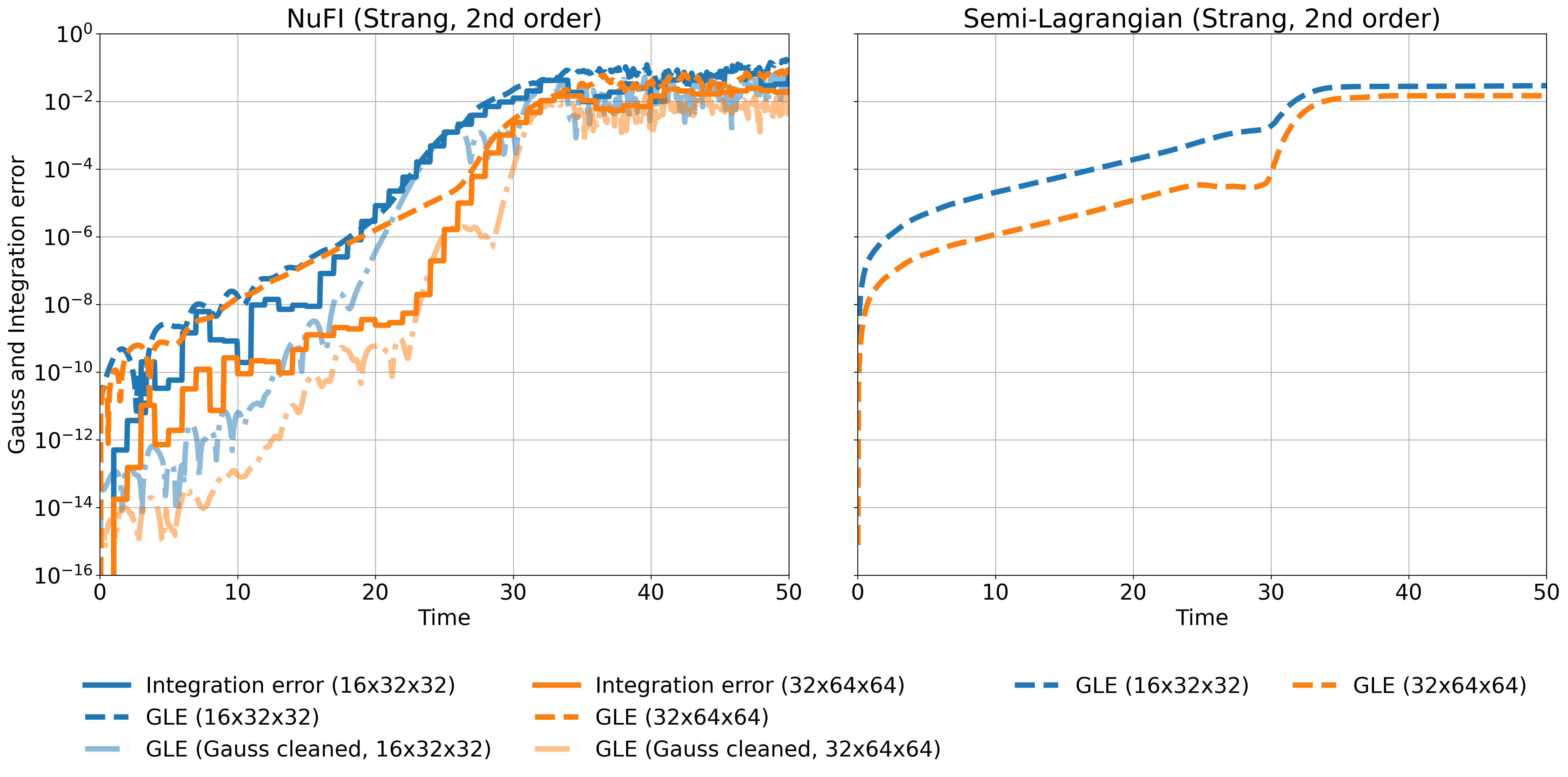}
\caption{\label{fig:gle_filamentation_instab}
Gauss law error (GLE) and integration error for the \emph{filamentation 
instability} test case compared between NuFI-Ham and Semi-Lagrangian with 2nd 
order Strang splitting.
}
\end{figure}

In figure \ref{fig:gle_filamentation_instab} we now also look at the Gauss law 
and integration error for the \emph{filamentation instability}. In the linear
phase the GLEs are substantially smaller for NuFI-Ham compared to
SL-Ham. In the non-linear stage, however, the Gauss law preservation is 
slightly better for SL-Ham than for NuFI with Gauss-cleaning. The low 
resolution simulation exhibits a GLE of up to $8\%$ for NuFI-Ham 
and $3\%$ for SL-Ham. For the high resolution simulations NuFI-Ham has a Gauss 
law error of up to $2\%$ and $1.5\%$ for SL-Ham. Using Gauss cleaning reduces 
the GLE back below $1\%$ for the high resolution simulation, 
however, the cleaning is significantly less effective than in the linear phase. 
The figure also clearly shows the relation between the GLE and IE. Note also 
that due to the lack of numerical dissipation in NuFI the distribution 
functions preserve more fine scale features and therefore integration is also 
less accurate, which is a potential explanation for the slightly higher GLE for 
NuFI in the late non-linear stage.

\subsubsection{Restarted NuFI}
\label{sec:conservation_properties_restarted_nufi}

When we restart NuFI, see section \ref{sec:restarted_nufi_vm}, we break the
conservation properties of the pure NuFI time-stepping. In this section we want
numerically quantify this effect for the Vlasov--Maxwell extension of NuFI. 
Note that in previous work we have shown for the electro-static case that while 
restarts indeed break the exact conservation properties, the overall algorithm 
is still physically more accurate when comparing to a Semi-Lagrangian approach 
with comparable discretization~\autocite{wilhelm2025restartingnumericalflowiteration, multi_species_kin_instability_nufi}.
In the following we focus on the restarted version of NuFI-Ham. NuFI-PC behaves
analogously under restart.

\begin{remark}
When taking the limit $n_t^r \rightarrow 1$ then NuFI collapses back to a 
Semi-Lagrangian scheme. If we consider NuFI-Ham with exact-in-time Fourier 
integration and cubic interpolation for the restarts then the resulting scheme 
is similar to what Crousseilles et al. suggested in their work. The main 
difference is that they chose a 3rd order conservative Finite-Volume update 
formula for the $\HamE$ and $\HamB$ substeps.~\autocite{CROUSEILLES2015224}  
In fact we observe comparable results for NuFI-Ham with $n_t^r=1$ and SL-Ham,
however, mass and $L^2$-norm conservation are slightly better for their 
conservative SL scheme, which is to be expected as their chosen Finite Volume 
scheme is built to improve mass-preservation.

Note that we do not take up any measures in our current restart procedure to 
increase the quality of the interpolant. Similar to SL-Ham we could also 
use an approximation which takes more information such as mass or entropy of 
$f$ into account and thereby also improve the conservation properties 
of the restarted NuFI approach. 
However, investigating such approximations goes beyond the scope of the present 
manuscript and will be left for future work.
\end{remark}

\begin{figure}[h!]
\centering
\begin{subfigure}{0.49\textwidth}
\includegraphics[width=0.99\textwidth]{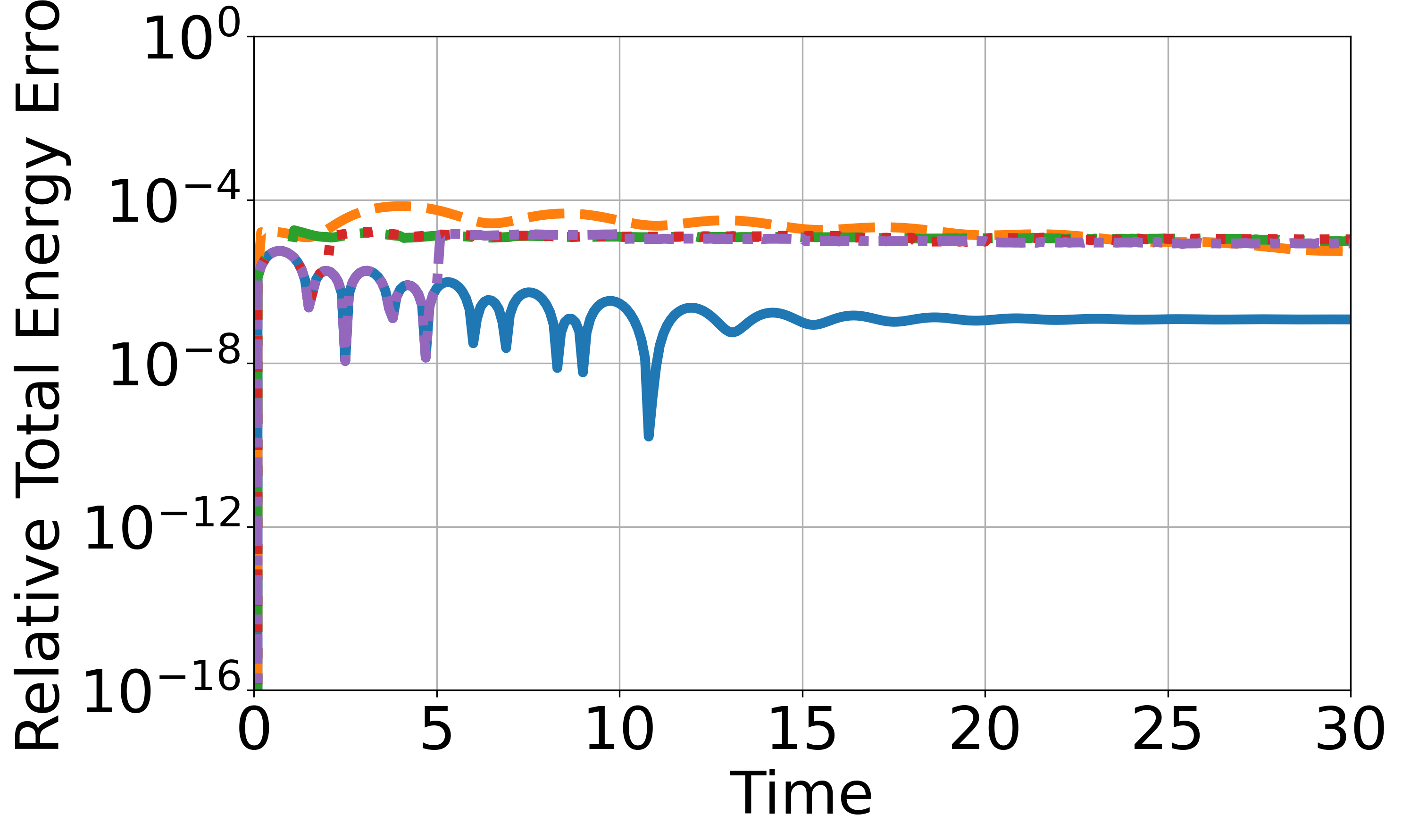}
\subcaption{\label{fig:restarted_tot_energy_conserve_landau_damp}
Total Energy
}
\end{subfigure}
\begin{subfigure}{0.49\textwidth}
\includegraphics[width=0.99\textwidth]{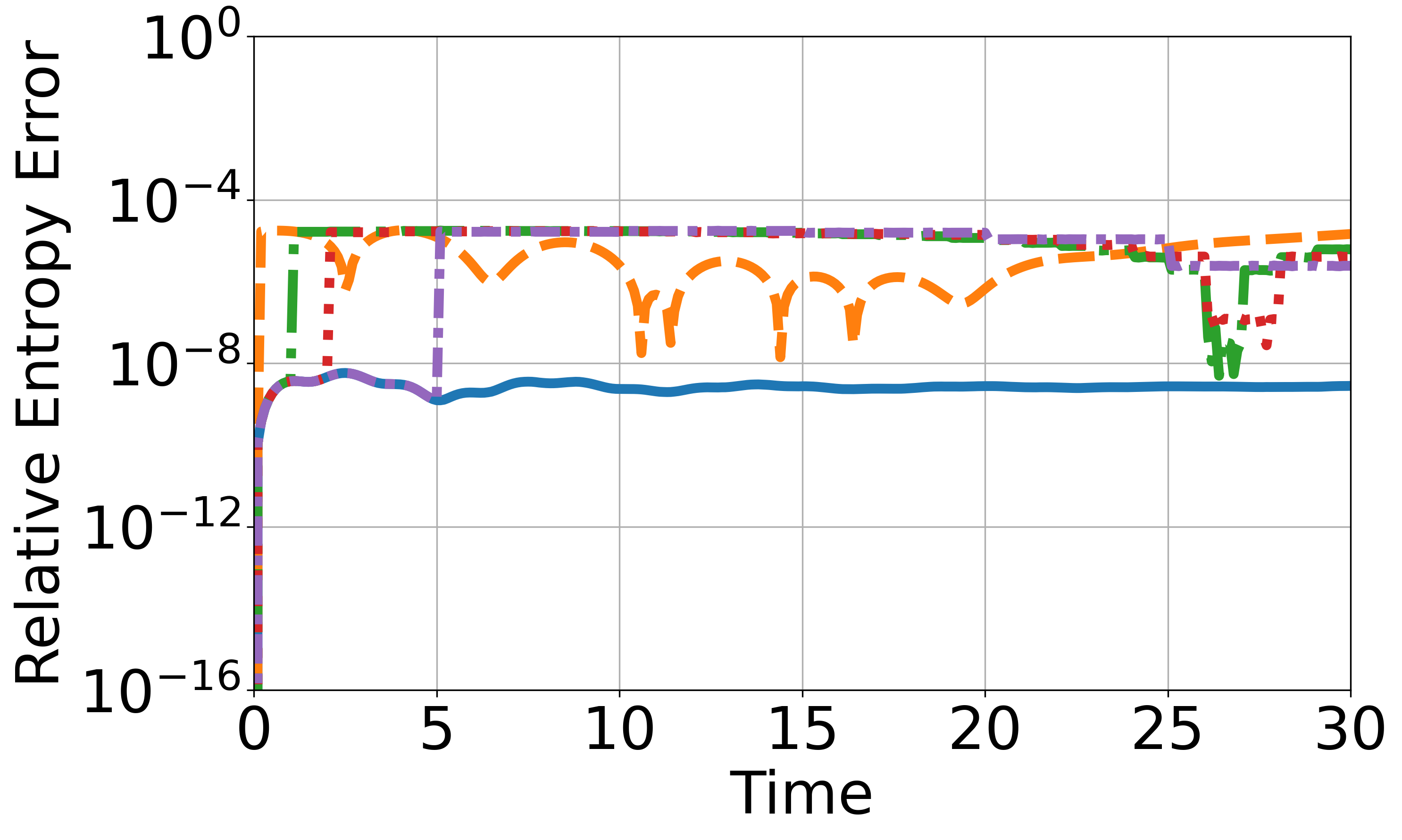}
\subcaption{\label{fig:restarted_entropy_conserve_landau_damp}
Entropy
}
\end{subfigure}
\begin{subfigure}{0.49\textwidth}
\includegraphics[width=0.99\textwidth]{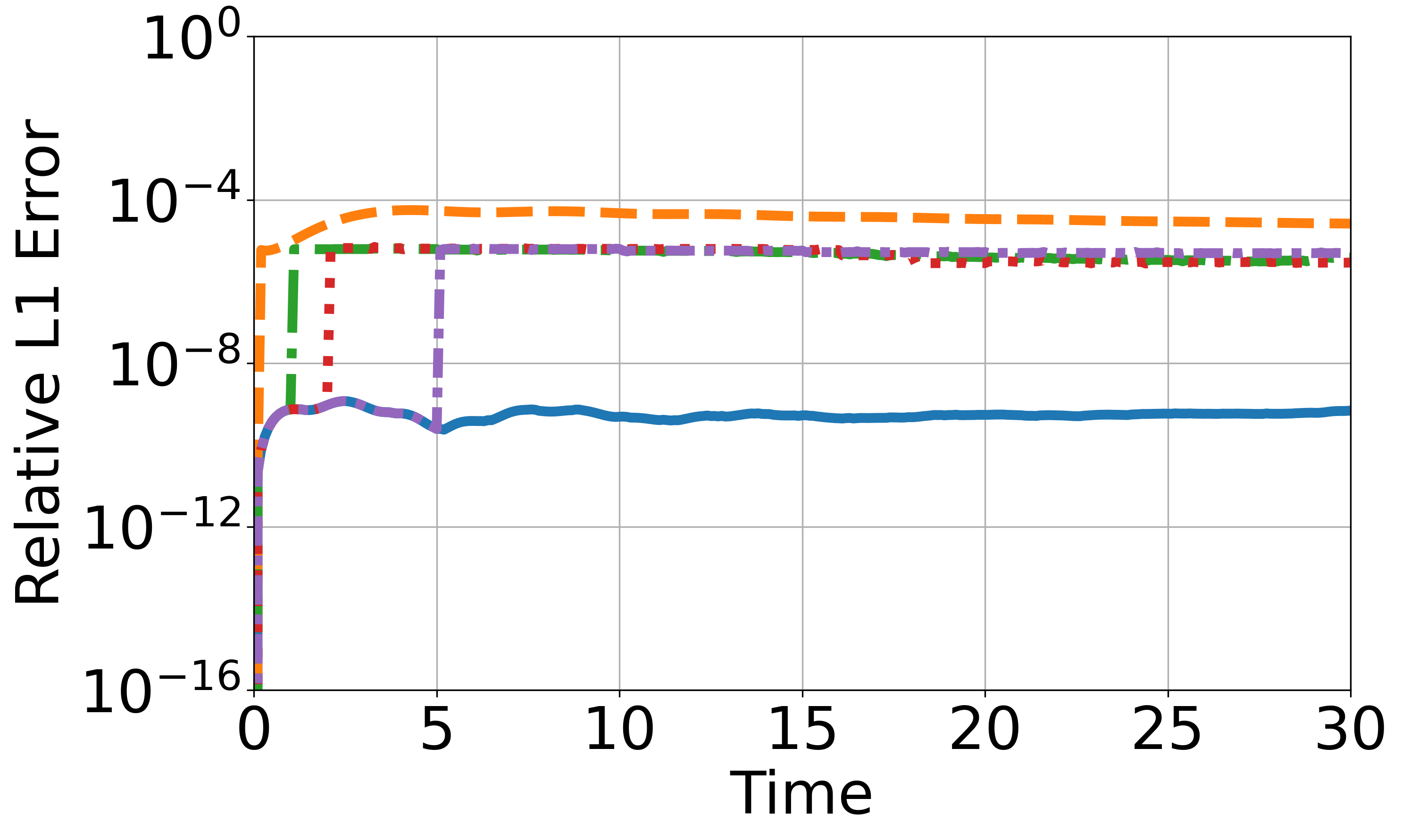}
\subcaption{\label{fig:restarted_l1_norm_conserve_landau_damp}
$L^1$-norm
}
\end{subfigure}
\begin{subfigure}{0.49\textwidth}
\includegraphics[width=0.99\textwidth]{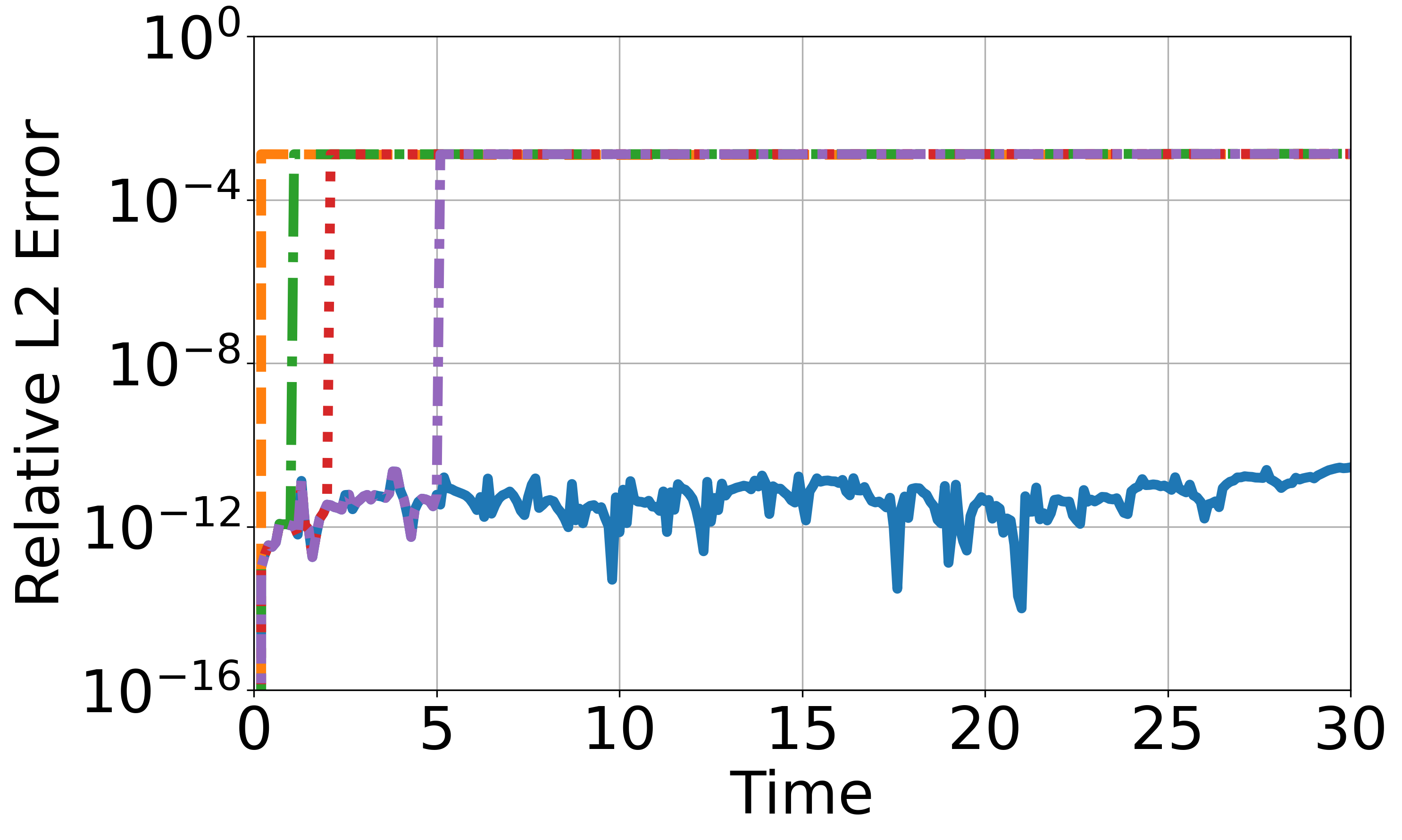}
\subcaption{\label{fig:restarted_l2_norm_conserve_landau_damp}
$L^2$-norm
}
\end{subfigure}
\begin{subfigure}{0.99\textwidth}
\centering
\includegraphics[width=0.59\textwidth]{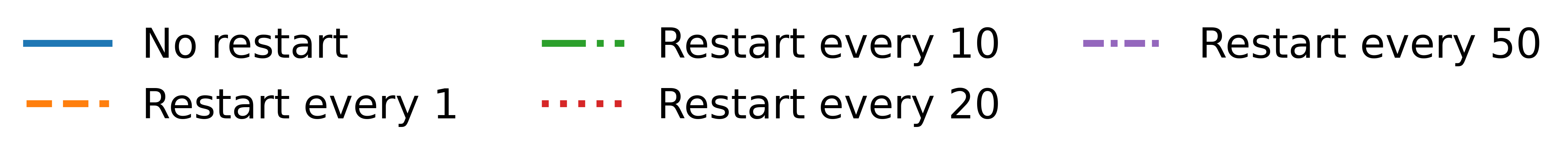}
\end{subfigure}
\caption{\label{fig:restarted_conservation_properties_landau_damping}
Conservation of total energy, entropy as well as $L^1$- and $L^2$-norm for 
the \emph{weak Landau Damping} test case compared between different restart 
times for NuFI-Ham.
}
\end{figure}

In figure \ref{fig:restarted_conservation_properties_landau_damping} we look
at how the conservation properties of the numerical solution are affected by
different restart frequencies for the \emph{Landau Damping} test case. To this
end we fix the phase-space resolution to $32 \times 64 \times 16$ and a 
time-step of $\Delta t = \tfrac{1}{10}$, while varying the restart frequency 
$n_t^r \in \{ 1,10,20,50\}$.
As is to be expected when using restarts the conservation errors are 
substantially larger than for the non-restarted NuFI. The relative errors for 
total energy increase by 2 orders of magnitude and by up to 3 orders of 
magnitude for entropy, arriving at a similar order of magnitude as SL-Ham. 
For the \emph{Landau Damping} case we do not see large differences in the 
quality of conservation between the different restart frequencies, however, 
the simulations with $n_t^r > 1$ are better in preserving the total energy 
by up to 1 order of magnitude until $t \approx 20$ and mass for all times when 
comparing to $n_t^r=1$. Notably for the conservation of entropy the 
preservation with $n_t^r=1$ is best until $t \approx 25$ after which less 
frequent restarts become advantageous again.
That being said, the observed errors for \emph{Landau Damping} were quite 
small for all approaches, including the Semi-Lagrangian ones and likely 
dominated by the underlying phase-space discretization.

\begin{figure}[h!]
\centering
\begin{subfigure}{0.49\textwidth}
\includegraphics[width=0.99\textwidth]{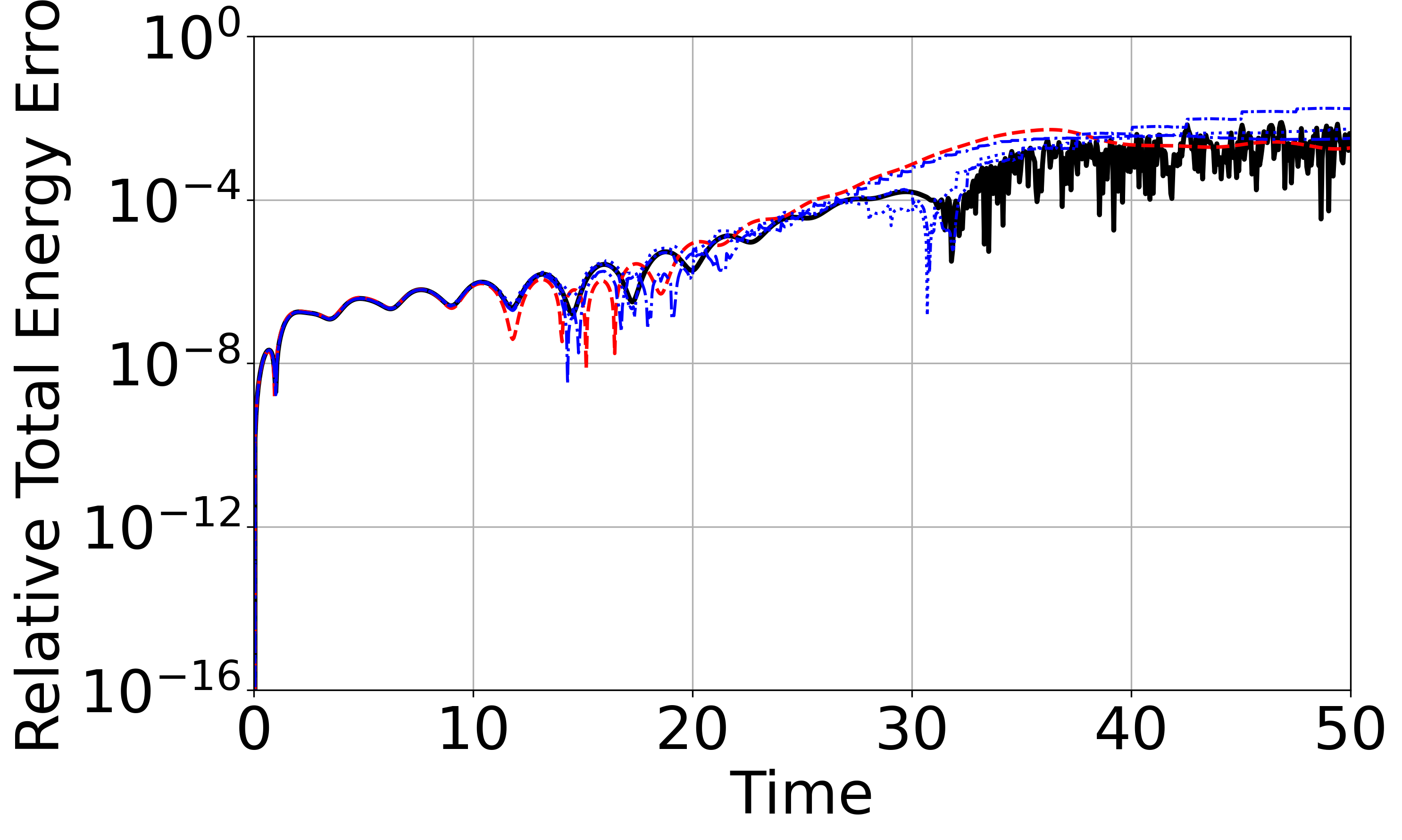}
\subcaption{\label{fig:restarted_tot_energy_conserve_filamentation_instab}
Total Energy
}
\end{subfigure}
\begin{subfigure}{0.49\textwidth}
\includegraphics[width=0.99\textwidth]{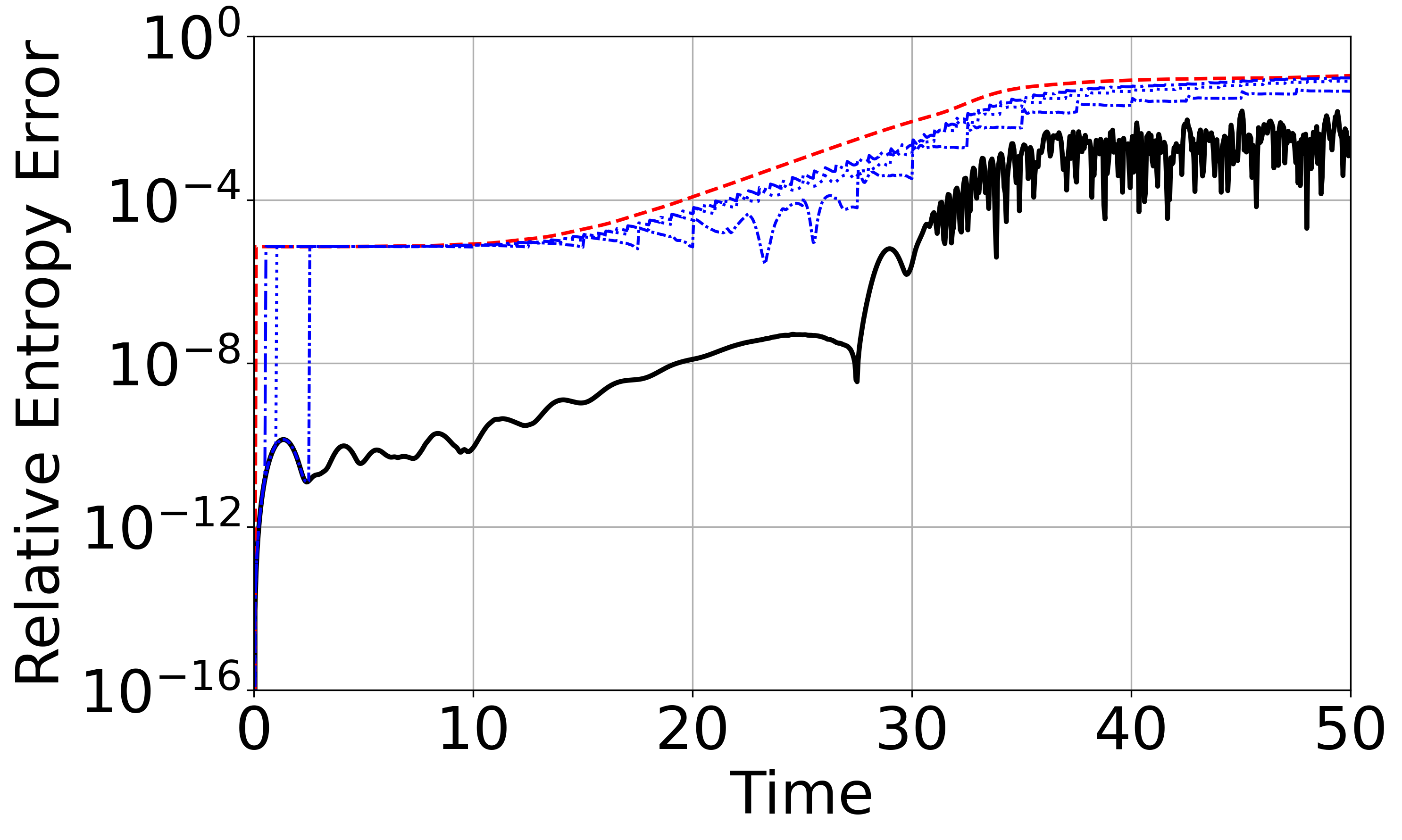}
\subcaption{\label{fig:restarted_entropy_conserve_filamentation_instab}
Entropy
}
\end{subfigure}
\begin{subfigure}{0.49\textwidth}
\includegraphics[width=0.99\textwidth]{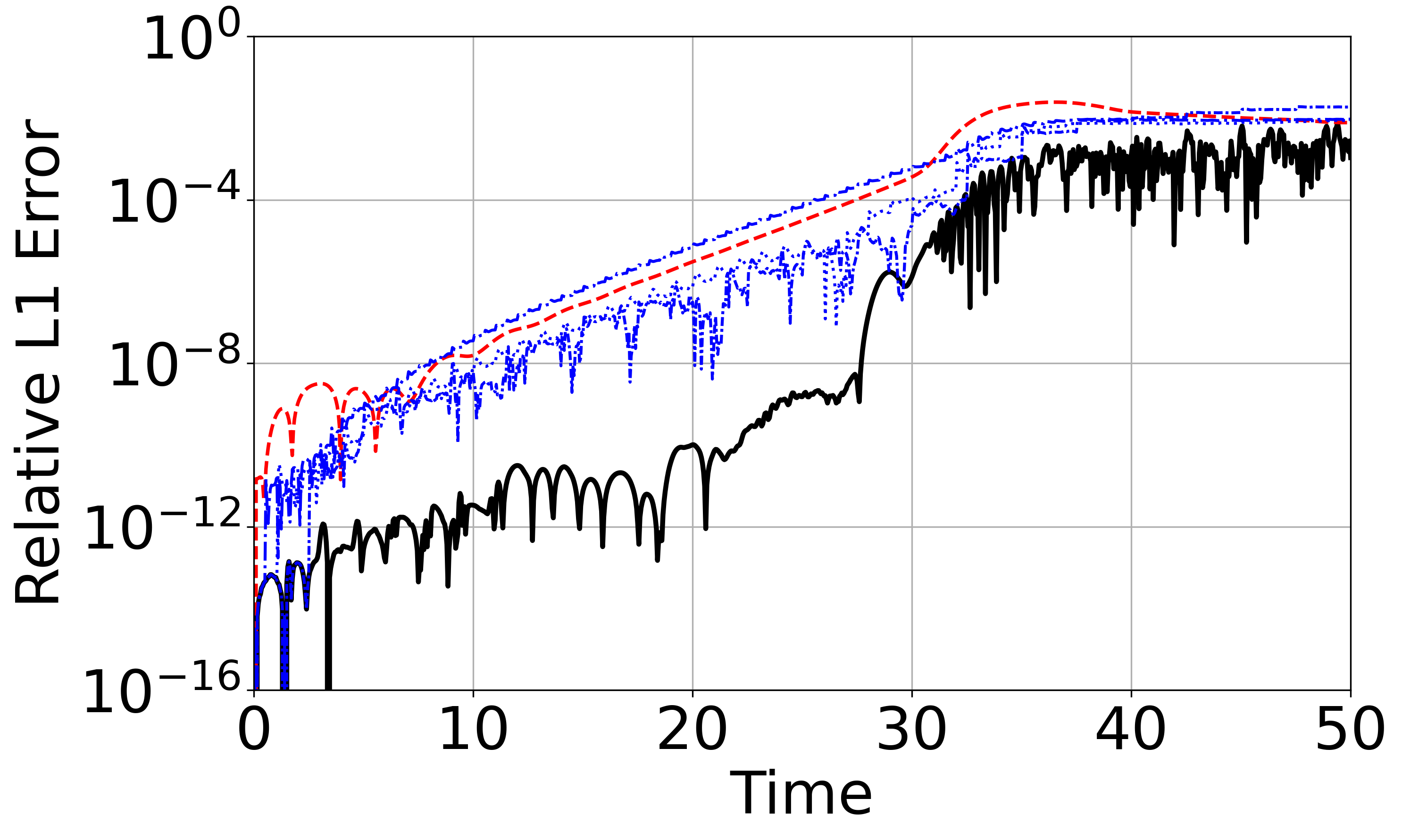}
\subcaption{\label{fig:restarted_l1_norm_conserve_filamentation_instab}
$L^1$-norm
}
\end{subfigure}
\begin{subfigure}{0.49\textwidth}
\includegraphics[width=0.99\textwidth]{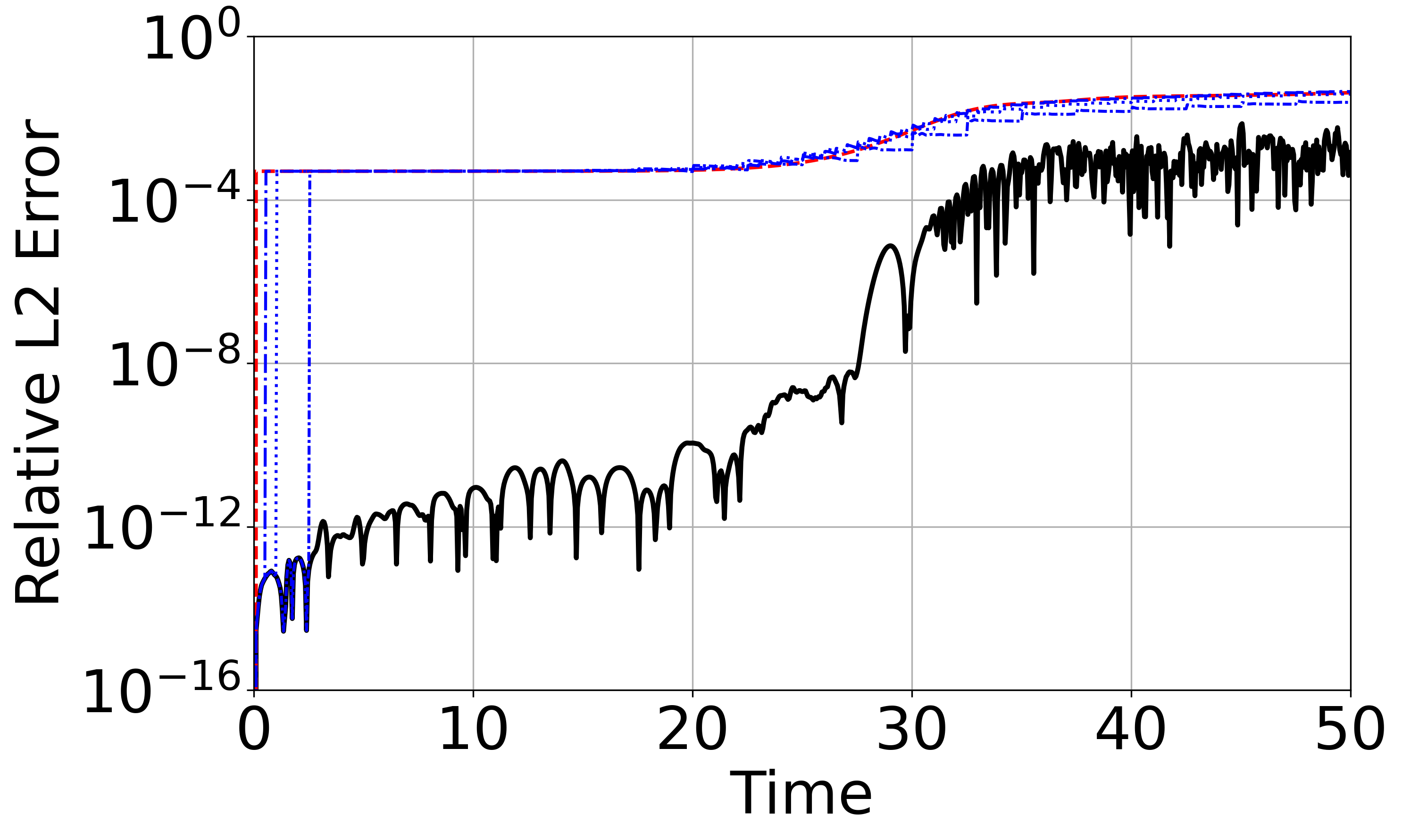}
\subcaption{\label{fig:restarted_l2_norm_conserve_filamentation_instab}
$L^2$-norm
}
\end{subfigure}
\begin{subfigure}{0.99\textwidth}
\centering
\includegraphics[width=0.59\textwidth]{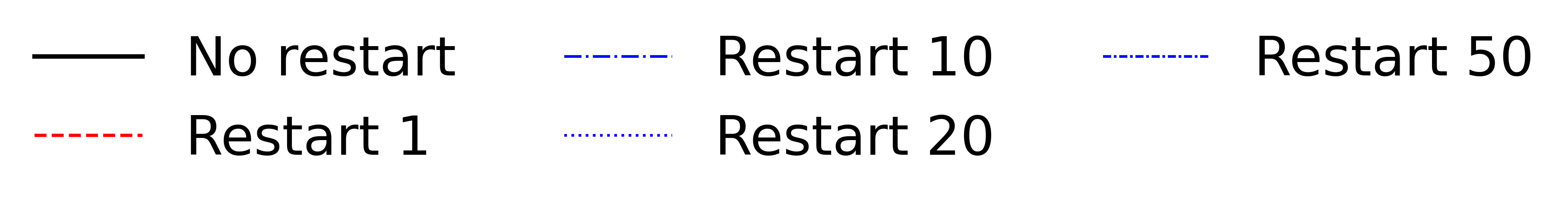}
\end{subfigure}
\caption{\label{fig:restarted_conservation_properties_filamentation_instab}
Conservation of total energy, entropy as well as $L^1$- and $L^2$-norm for 
the \emph{filamentation instability} test case compared between different 
restart times for NuFI-Ham.
}
\end{figure}

The dynamically more interesting \emph{filamentation instability} case is 
considered in figure \ref{fig:restarted_conservation_properties_filamentation_instab}. Here we 
again fix the phase-space resolution to $32 \times 64 \times 16$ and a 
time-step of $\Delta t = \tfrac{1}{10}$, while varying the restart frequency 
$n_t^r \in \{ 1,10,20,50\}$.

Again we observe that the errors of the restarted NuFI are substantially larger 
than for the pure NuFI. 
When comparing the different restart frequencies we actually observe that 
$n_t^r = 10$ and $20$ seem to be better than $n_t^r=1$. For the total energy 
and entropy $n_t^r = 10$ and $20$ show up to 1 order of magnitude smaller 
errors in the linear-growth phase. 
Towards the end of the simulation the errors of all conservation properties 
saturate at comparable levels independent of the restart frequency, however, 
smaller than the errors for SL-Ham, particularly for the total energy and 
$L^2$-norm.

These observations also suggest that with the current restart strategy $n_t^r=50$ is, in this case, worse than $n_t^r=10$ or $20$. This contradicts
the observations from previous 
publications~\autocite{wilhelm2025restartingnumericalflowiteration} where 
increasing $n_t^r$ always improved accuracy. Potentially here the restart 
frequency negatively interacts with a slower oscillatory dynamic of the test 
case at hand, however, the authors do not have a definitive explanation for 
this phenomena at this time and further investigation is necessary to explain 
this observation.
That being said, the observations suggest that while NuFI as standalone is 
always more accurate than a Semi-Lagrangian scheme, even in cases where pure 
NuFI is too expensive, we can still employ it as subcycling for a 
Semi-Lagrangian scheme, i.\,e., restarted NuFI with $n_t^r > 1$ is still 
more physically accurate than pure Semi-Lagrangian advection.

\section{Discussion and outlook}\label{conclusion}

In this work we presented how the Numerical Flow Iteration can be extended from 
the electro-static to the fully electro-magnetic case based upon Hamiltonian 
splitting. We have shown both in theory and numerically that the 
structure-preserving properties of the NuFI time-stepping scheme carry over 
from the electro-static to the electro-magnetic system. The pure NuFI approach 
allows to resolve fine scale dynamics on sub-grid scales beyond the reach of 
any other state-of-the-art method, in particular, including Semi-Lagrangian 
approaches based upon the same Hamiltonian splitting. In principle this allows
us to zoom into dynamically relevant regions, which makes the approach 
particularly interesting for studying dynamical problems which require accurate 
resolution of phase-space structures. 

A remaining major drawback of the pure NuFI time-stepping approach is its 
quadratic time complexity, which makes it again prohibitively expensive for 
long, high-dimensional production simulations. In its pure form NuFI is mostly 
suitable for low dimensional high-fidelity studies or on very limited time 
scales. 
That being said, the present work shows that NuFI has strong 
potential as structure-preserving subcycling for existing Semi-Lagrangian 
codes, as NuFI does not require the storage of intermediate distribution 
functions and therefore not only reduces the memory cost of subcycling but also 
-- due to the increased operation per byte ratio -- potentially speeds it up 
for moderate number of 
subcycles.~\autocite{kirchhart2023numerical,Wilhelm_2025}

The main focus of the paper was to present the electro-magnetic extension of 
the NuFI approach and thus the employed restart mechanism was kept quite 
simple. Arguably, it is better to use restart techniques which preserve at 
least a number of lower order moments and if possible also use either some form
of grid-adaptivity, sparse representation or low rank compression in a 
\textit{step-and-truncate} fashion. How such restart methods can be implemented
both accurately and efficiently is, however, beyond the scope of this work, and 
will be investigated in future work.
Additionally, in on-going and future work we want to look into simulations of 
fully six-dimensional kinetic plasma dynamics, e.g. of beam-driven 
instabilities~\autocite{Bacchini_2024,Pezzini_2024}. Although the codebase 
already supports six-dimensional multi-species simulations, it has not yet been 
optimized for the very long time scales typical of e.g. astrophysical 
applications. Addressing this will require the previously discussed improved 
restart routines as well as targeted computational performance tuning.

\section*{Acknowledgments}

R.-P.W. wants to thank the organizers of the National High Performance 
Computing (NHR) graduate school and the Federal Ministry of Education and 
Research as well as the state governments for supporting this work as part of 
the joint funding of National High Performance Computing (NHR). R.-P.W. wants 
to thank the DFG, who funded part of the work through the IRTG Modern Inverse 
Problems (333849990/GRK2379). 
This work has received funding from the European High Performance Computing 
Joint Undertaking (JU) and Belgium, Czech Republic, France, Germany, Greece, 
Italy, Norway, and Spain under grant agreement No 101093441. Views and opinions 
expressed are however those of the author(s) only and do not necessarily 
reflect those of the European Union or the European High Performance Computing 
Joint Undertaking (JU) and Belgium, Czech Republic, France, Germany, Greece, 
Italy, Norway, and Spain. Neither the European Union nor the granting authority 
can be held responsible for them. 
F.B. acknowledges support from the FED-tWIN programme (profile Prf-2020-004, 
project ``ENERGY'') issued by BELSPO, and from the FWO Junior Research Project 
G020224N granted by the Research Foundation -- Flanders (FWO). 
The resources and services used in this work were provided by the VSC (Flemish 
Supercomputer Center), funded by the Research Foundation - Flanders (FWO) and 
the Flemish Government.

\appendix
\section{Normalization of the multi-species Vlasov--Maxwell system}
\label{appendix_sec:normalization}

We choose to normalize the Vlasov--Maxwell system to the electron scales, 
i.\,e., we choose as reference the electron charge $q_e = e$, electron mass 
$m_e$ and the electron number density $n_0$.
The electron plasma frequency is
\begin{equation}
\omega_{p,e}^2 = \frac{4 \pi n_0 e^2}{m_e}.
\end{equation}

Hence we choose as a reference scale for time $t_0 = \omega_{p,e}^{-1}$, 
spatial dimension $L_0 = \tfrac{c}{\omega_{p,e}}$ and velocity $V_0 = c$. 
Consequentially we choose for the fields $E_0 = \tfrac{m_e c \, \omega_{p,e}}{e}$, $B_0 = E_0$, $f_0 = \tfrac{n_0}{V_0^3} = \tfrac{n_0}{c^3}$.
Finally the charge and current density are scaled to $\rho_0 = e n_0$ and 
$J_0 = e c n_0 $ respectively.

Then our dimensionless variables used in \eqref{eqn:vlasov} -- \eqref{eqn:j}
are
\begin{equation}
\begin{aligned}
t' &= \frac{t}{t_0}, \quad x' = \frac{x}{L_0}, \quad v' = \frac{v}{V_0}, \\
f'_s &= \frac{f_s}{f_0}, \quad E' = \frac{E}{E_0}, \quad B' = \frac{B}{B_0}, \\
\rho' &= \frac{\rho}{\rho_0}, \quad J' = \frac{J}{J_0}.
\end{aligned}
\end{equation}

For the sake of readability we dropped the primes.

\printbibliography

@article{myers_4th-order_2016,
	title = {A 4th-Order Particle-in-Cell Method with Phase-Space Remapping for the {Vlasov--Poisson} Equation},
	issn = {1095--7197},
	volume = {39},
	number = {9},
	pages = {B467--B485},
	doi = {10.1137/16M105962X},
	journal = {{SIAM} Journal on Scientific Computing},
	author = {Myers, A. and Colella, P. and Straalen, B.},
	year = {2016},
	month = {02}
}

@article{wang_particle--cell_2011,
	title = {A Particle-in-cell Method with Adaptive Phase-space Remapping for Kinetic Plasmas},
	volume = {33},
	number = {6},
	pages = {3509--3537},
	doi = {10.1137/100811805},
	journal = {{SIAM} J. Sci. Comput.},
	author = {Wang, B. and Miller, G. H. and Colella, P.},
	year = {2011},
	month = {12}
}

@article{cottet_particle_1984,
	title = {Particle Methods for the One-Dimensional {Vlasov--Poisson} Equations},
	volume = {21},
	issn = {0036--1429},
	pages = {52--76},
	number = {1},
	journal = {{SIAM} Journal on Numerical Analysis},
	author = {Cottet, G.-H. and Raviart, P.-A.},
	year = {1984},
	month = {06},
	doi = {10.1137/0721003}
}

@article{cottet_raviart_pic,
author = {G. H. Cottet and P. A. Raviart},
title = {On particle-in-cell methods for the Vlasov-Poisson equations},
journal = {Transport Theory and Statistical Physics},
volume = {15},
number = {1-2},
pages = {1-31},
year = {1986},
publisher = {Taylor & Francis},
doi = {10.1080/00411458608210442}
}

@article{besse_semi-lagrangian_2003,
	title = {Semi-{Lagrangian} schemes for the {Vlasov} equation on an unstructured mesh of phase space},
	volume = {191},
	issn = {0021--9991},
	doi = {10.1016/S0021-9991(03)00318-8},
	pages = {341--376},
	number = {2},
	journal = {Journal of Computational Physics},
	author = {Besse, N. and Sonnendrücker, E.},
	year = {2003-11}
}

@book{hockney_computer_2021,
	title = {Computer Simulation Using Particles},
	isbn = {978-1-4398-2205-0},
	publisher = {{CRC} Press},
	author = {Hockney, R.W. and Eastwood, J.W.},
	year = {2021},
	month = {03}
}

@article{cheng_integration_1976,
	title = {The integration of the {Vlasov} equation in configuration space},
	volume = {22},
	issn = {0021--9991},
	doi = {10.1016/0021-9991(76)90053-X},
	pages = {330--351},
	number = {3},
	journal = {Journal of Computational Physics},
	author = {Cheng, C.Z. and Knorr, G.},
	year = {1976},
	month = {11}
}

@article{pfaffelmoser_global_1992,
	title = {Global classical solutions of the {Vlasov--Poisson} system in three dimensions for general initial data},
	volume = {95},
	issn = {0022--0396},
	doi = {10.1016/0022-0396(92)90033-J},
	pages = {281-303},
	number = {2},
	journal = {Journal of Differential Equations},
	author = {Pfaffelmoser, K.},
	year = {1992},
	month = {02}
}

@article{fijalkow_numerical_1999,
	title = {A numerical solution to the {Vlasov} equation},
	volume = {116},
	issn = {0010--4655},
	doi = {10.1016/S0010-4655(98)00146-5},
	pages = {319--328},
	number = {2},
	journal = {Computer Physics Communications},
	author = {Fijalkow, E.},
	year = {1999},
	month = {02}
}

@article{filbet_conservative_2001,
	title = {Conservative Numerical Schemes for the {Vlasov} Equation},
	volume = {172},
	issn = {0021--9991},
	doi = {10.1006/jcph.2001.6818},
	pages = {166--187},
	number = {1},
	journal = {Journal of Computational Physics},
	author = {Filbet, F. and Sonnendrücker, E. and Bertrand, P.},
	year = {2001},
	month = {09}
}

@book{birdsall_plasma_1985,
	title = {Plasma Physics via Computer Simulation},
	author = {Birdsall, C.K. and Langdon, A.B.},
	year = {2004},
	publisher = {Taylor \& Francis},
        isbn = {978-1-4822-6306-0},
        series = {Series in Plasma Physics},
}

@book{intro_plasma_physics_chen,
	title = {Introduction to Plasma Physics and Controlled Fusion },
	isbn = {978-3-319-22309-4},
	publisher = {Springer Cham},
	author = {Chen, F.},
    doi = { 10.1007/978-3-319-22309-4},
	year = {2015}
}

@phdthesis{ameres_stochastic_2018,
	title = {Stochastic and Spectral Particle Methods for Plasma Physics},
	school = {Technische Universität München},
	author = {Ameres, Jakob},
	year = {2018},
}

@article{ARBER2002339,
	title = {A Critical Comparison of {Eulerian}-Grid-Based {Vlasov} Solvers},
	journal = {Journal of Computational Physics},
	volume = {180},
	number = {1},
	pages = {339-357},
	year = {2002},
	issn = {0021--9991},
	doi = {10.1006/jcph.2002.7098},
	author = {Arber, T.D.  and Vann, R.G.L. }
}

@article{HarlowEvans2,
	title = {The Particle-In-Cell method for Hydrodynamic Calculations},
	author = {Evans, M.W. and Harlow, F.H.},
	journal = {Report LA-2139, Los Alamos Scientific laboratory of the university of {California}},
	year = {1957}
}

@article{ROSSMANITH20116203,
title = {A positivity-preserving high-order semi-Lagrangian discontinuous Galerkin scheme for the Vlasov–Poisson equations},
journal = {Journal of Computational Physics},
volume = {230},
number = {16},
pages = {6203--6232},
year = {2011},
issn = {0021--9991},
doi = {doi.org/10.1016/j.jcp.2011.04.018},
author = {James A. Rossmanith and David C. Seal}
}

@article{COTTET2018362,
title = {Semi-Lagrangian particle methods for high-dimensional Vlasov–Poisson systems},
journal = {Journal of Computational Physics},
volume = {365},
pages = {362-375},
year = {2018},
issn = {0021--9991},
doi = {10.1016/j.jcp.2018.03.042},
author = {Georges-Henri Cottet}
}

@article{trefethen_trapezoidal,
author = {Trefethen, Lloyd N. and Weideman, J. A. C.},
title = {The Exponentially Convergent Trapezoidal Rule},
journal = {SIAM Review},
volume = {56},
number = {3},
pages = {385-458},
year = {2014},
doi = {10.1137/130932132},
}

@article{EINKEMMER2019937,
title = {A performance comparison of semi-Lagrangian discontinuous Galerkin and spline based Vlasov solvers in four dimensions},
journal = {Journal of Computational Physics},
volume = {376},
pages = {937-951},
year = {2019},
issn = {0021-9991},
doi = {10.1016/j.jcp.2018.10.012},
author = {Lukas Einkemmer}
}

@article{kormann_tensor_train,
author = {Kormann, Katharina},
title = {A Semi-Lagrangian Vlasov Solver in Tensor Train Format},
journal = {SIAM Journal on Scientific Computing},
volume = {37},
number = {4},
pages = {B613--B632},
year = {2015},
doi = {10.1137/140971270},
}

@article{WILHELM2023111720,
title = {An interpolating particle method for the Vlasov–Poisson equation},
journal = {Journal of Computational Physics},
volume = {473},
pages = {111720},
year = {2023},
issn = {0021-9991},
doi = {10.1016/j.jcp.2022.111720},
author = {R. Paul Wilhelm and Matthias Kirchhart},
}

@article{cottet_etancelin_perignon_picard_2014,
title={High order semi-Lagrangian particle methods for transport equations: numerical analysis and implementation issues},
volume={48},
DOI={10.1051/m2an/2014009},
number={4},
journal={ESAIM: Mathematical Modelling and Numerical Analysis},
publisher={EDP Sciences},
author={Cottet, G.-H. and Etancelin, J.-M. and Perignon, F. and Picard, C.},
year={2014},
pages={1029–1060}
}

@article{CROUSEILLES20101927,
title = {Conservative semi-Lagrangian schemes for Vlasov equations},
journal = {Journal of Computational Physics},
volume = {229},
number = {6},
pages = {1927-1953},
year = {2010},
issn = {0021-9991},
doi = {10.1016/j.jcp.2009.11.007},
author = {Nicolas Crouseilles and Michel Mehrenberger and Eric Sonnendrücker},
}

@book{hairer2006,
    title     = {Geometric Numerical Integration},
    subtitle  = {Structure-Preserving Algorithms for Ordinary Differential
    Equations},
    author    = {Hairer, Ernst and Wanner, Gerhard and Lubich, Christian},
    edition   = {2},
    series    = {Springer Series in Computational Mathematics},
    number    = {31},
    year      = {2006},
    publisher = {Springer},
    isbn      = {3540306633},
    issn      = {0179--3632},
    doi       = {10.1007/3-540-30666-8}
}

@article{kirchhart2023numerical,
author = {Kirchhart, Matthias and Wilhelm, R. Paul},
title = {The Numerical Flow Iteration for the Vlasov–Poisson Equation},
journal = {SIAM Journal on Scientific Computing},
volume = {46},
number = {3},
pages = {A1972-A1997},
year = {2024},
doi = {10.1137/23M154710X},
}

@article{sl_456d_einkemmmer_moriggl,
author = {Lukas Einkemmer and Alexander Moriggl},
title ={Semi-Lagrangian 4d, 5d, and 6d kinetic plasma simulation on large-scale GPU-equipped supercomputers},
journal = {The International Journal of High Performance Computing Applications},
volume = {37},
number = {2},
pages = {180-196},
year = {2023},
doi = {10.1177/10943420221137599},
}

@article{sldg2015,
    title={{High performance computing aspects of a dimension independent semi-Lagrangian discontinuous Galerkin code}},
    author={Einkemmer, L.},
    journal={Comput. Phys. Commun.},
    volume={202},
    pages={326-336},
    year={2016},
    doi = {10.1016/j.cpc.2016.01.012}
}

@article{EINKEMMER2023112060,
title = {A robust and conservative dynamical low-rank algorithm},
journal = {Journal of Computational Physics},
volume = {484},
pages = {112060},
year = {2023},
issn = {0021-9991},
doi = {10.1016/j.jcp.2023.112060},
author = {Lukas Einkemmer and Alexander Ostermann and Carmela Scalone},
}

@InProceedings{sparse_kormann_sonnendruecker,
author="Kormann, Katharina
and Sonnendr{\"u}cker, Eric",
editor="Garcke, Jochen
and Pfl{\"u}ger, Dirk",
title="Sparse Grids for the Vlasov--Poisson Equation",
booktitle="Sparse Grids and Applications - Stuttgart 2014",
year="2016",
publisher="Springer International Publishing",
address="Cham",
pages="163--190",
isbn="978-3-319-28262-6",
doi = {10.1007/978-3-319-28262-6_7}
}

@ARTICLE{Krah2023905,
	author = {Krah, Philipp and Yin, Xi-Yuan and Bergmann, Julius and Nave, Jean-Christophe and Schneider, Kai},
	title = {A Characteristic Mapping Method for Vlasov-Poisson with Extreme Resolution Properties},
	year = {2023},
	journal = {Communications in Computational Physics},
	volume = {35},
	number = {4},
	pages = {905 – 937},
	doi = {10.4208/cicp.OA-2024-0012},
	type = {Article},
}

@ARTICLE{fftw_reference,
  author={Frigo, M. and Johnson, S.G.},
  journal={Proceedings of the IEEE}, 
  title={The Design and Implementation of FFTW3}, 
  year={2005},
  volume={93},
  number={2},
  pages={216-231},
  doi={10.1109/JPROC.2004.840301}
}

@article{case_for_electron_astrophysics,
  title={A Case for Electron-Astrophysics},
  author={Daniel Verscharen and Wicks RT and Alexandrova O and Bruno R 
	and  Burgess D and Chen CHK and D'Amicis R and De Keyser J and de Wit TD
	and Franci L and He J and Henri P and Kasahara S and Khotyaintsev Y
	and Klein KG and Lavraud B and Maruca BA and Maksimovic M and Plaschke F
	and Poedts S and Reynolds CS and Roberts O and Sahraoui F and Saito S
	and Salem CS and Saur J and Servidio S and Stawarz JE and Štverák Š and Told D.},
  journal={Experimental Astronomy},
  volume={54},
  number={2},
  year={2022},
  publisher={Springer Nature},
  doi = {10.1007/s10686-021-09761-5}
}

@article{multi_scale_solar_wind,
  title={The multi-scale nature of the solar wind},
  author={Daniel Verscharen and KG Klein and BA Maruca },
  journal={Living reviews in solar physics},
  volume={16},
  number={1},
  year={2019},
  publisher={Springer Nature},
  doi = {10.1007/s41116-019-0021-0}
}

@article{multi_species_kin_instability_nufi,
  title={Simulation of multi-species kinetic instabilities with the Numerical Flow Iteration},
  author={Rostislav-Paul Wilhelm and Manuel Torrilhon},
  journal={Proceedings of the 33rd International Symposium on Rarefied Gas Dynamics},
  year={2024},
   pages={441-449},
publisher={Springer Nature Switzerland},
doi = {10.1007/978-3-032-00094-1_42}
}

@article{Wilhelm_2025,
doi = {10.1088/1361-6587/ad9fdb},
year = {2025},
month = {jan},
publisher = {IOP Publishing},
volume = {67},
number = {2},
pages = {025011},
author = {Wilhelm, Rostislav-Paul and Eifert, Jan and Torrilhon, Manuel and Orland, Fabian},
title = {High fidelity simulations of the multi-species Vlasov equation in the electro-static, collisional-less limit},
journal = {Plasma Physics and Controlled Fusion}
}

@article{lapenta_mercury,
author = {Lapenta, Giovanni and Schriver, David and Walker, Raymond J. and Berchem, Jean and Echterling, Nicole F. and El Alaoui, Mostafa and Travnicek, Pavel},
title = {Do We Need to Consider Electrons' Kinetic Effects to Properly Model a Planetary Magnetosphere: The Case of Mercury},
journal = {Journal of Geophysical Research: Space Physics},
volume = {127},
number = {4},
pages = {e2021JA030241},
doi = {10.1029/2021JA030241},
note = {e2021JA030241 2021JA030241},
year = {2022}
}

@article{Palmroth2018,
author = { Minna Palmroth and
	  Urs Ganse and
Yann Pfau-Kempf and
Markus Battarbee and
Lucile Turc and
Thiago Brito and
Maxime Grandin and
Sanni Hoilijoki and
Arto Sandroos and
Sebastian von Alfthan},
title = {Vlasov methods in space physics and astrophysics},
journal = {Living Reviews in Computational Astrophysics},
volume = {4},
number = {1},
doi = {10.1007/s41115-018-0003-2},
year = {2018}
}

@article{gene_1,
    author = {Jenko, F. and Dorland, W. and Kotschenreuther, M. and Rogers, B. N.},
    title = {Electron temperature gradient driven turbulence},
    journal = {Physics of Plasmas},
    volume = {7},
    number = {5},
    pages = {1904-1910},
    year = {2000},
    month = {05},
    issn = {1070-664X},
    doi = {10.1063/1.874014},
}

@article{GORLER20117053,
title = {The global version of the gyrokinetic turbulence code GENE},
journal = {Journal of Computational Physics},
volume = {230},
number = {18},
pages = {7053-7071},
year = {2011},
issn = {0021-9991},
doi = {https://doi.org/10.1016/j.jcp.2011.05.034},
author = {T. Görler and X. Lapillonne and S. Brunner and T. Dannert and F. Jenko and F. Merz and D. Told},
}

@article{JUNO2018110,
title = {Discontinuous Galerkin algorithms for fully kinetic plasmas},
journal = {Journal of Computational Physics},
volume = {353},
pages = {110-147},
year = {2018},
issn = {0021-9991},
doi = {10.1016/j.jcp.2017.10.009},
author = {J. Juno and A. Hakim and J. TenBarge and E. Shi and W. Dorland},
}

@article{Mandell_Hakim_Hammett_Francisquez_2020, 
title={Electromagnetic full-$f$ gyrokinetics in the tokamak edge with discontinuous Galerkin methods}, 
volume={86}, 
DOI={10.1017/S0022377820000070}, 
number={1}, 
journal={Journal of Plasma Physics}, 
author={Mandell, N. R. and Hakim, A. and Hammett, G. W. and Francisquez, M.}, 
year={2020}, 
pages={905860109}
}

@InProceedings{review_num_kin_filbet,
author="Filbet, F.
and Sonnendr{\"u}cker, E.",
title="Numerical methods for the Vlasov equation",
booktitle="Numerical Mathematics and Advanced Applications",
year="2003",
publisher="Springer Milan",
address="Milano",
pages="459--468",
isbn="978-88-470-2089-4",
doi={10.1007/978-88-470-2089-4_43}
}

@Article{vlasiator_1,
AUTHOR = {Ganse, U. and Pfau-Kempf, Y. and Zhou, H. and Juusola, L. and Workayehu, A. and Kebede, F. and Papadakis, K. and Grandin, M. and Alho, M. and Battarbee, M. and Dubart, M. and Kotipalo, L. and Lalag\"ue, A. and Suni, J. and Horaites, K. and Palmroth, M.},
TITLE = {The Vlasiator 5.2 ionosphere -- coupling a magnetospheric hybrid-Vlasov simulation with a height-integrated ionosphere model},
JOURNAL = {Geoscientific Model Development},
VOLUME = {18},
YEAR = {2025},
NUMBER = {2},
PAGES = {511--527},
DOI = {10.5194/gmd-18-511-2025}
}

@article{VONALFTHAN201424,
title = {Vlasiator: First global hybrid-Vlasov simulations of Earth's foreshock and magnetosheath},
journal = {Journal of Atmospheric and Solar-Terrestrial Physics},
volume = {120},
pages = {24-35},
year = {2014},
issn = {1364-6826},
doi = {https://doi.org/10.1016/j.jastp.2014.08.012},
author = {S. {von Alfthan} and D. Pokhotelov and Y. Kempf and S. Hoilijoki and I. Honkonen and A. Sandroos and M. Palmroth},
}

@article{muphy_I,
    author = {Schmitz, H. and Grauer, R.},
    title = {Kinetic Vlasov simulations of collisionless magnetic reconnection},
    journal = {Physics of Plasmas},
    volume = {13},
    number = {9},
    pages = {092309},
    year = {2006},
    month = {09},
    issn = {1070-664X},
    doi = {10.1063/1.2347101},
}

@article{ALLMANNRAHN2024109064,
title = {The muphyII code: Multiphysics plasma simulation on large HPC systems},
journal = {Computer Physics Communications},
volume = {296},
pages = {109064},
year = {2024},
issn = {0010-4655},
doi = {10.1016/j.cpc.2023.109064},
author = {F. Allmann-Rahn and S. Lautenbach and M. Deisenhofer and R. Grauer},
}

@article{kormann_massively_parallel,
author = {Katharina Kormann and Klaus Reuter and Markus Rampp},
title ={A massively parallel semi-Lagrangian solver for the six-dimensional Vlasov–Poisson equation},
journal = {The International Journal of High Performance Computing Applications},
volume = {33},
number = {5},
pages = {924-947},
year = {2019},
doi = {10.1177/1094342019834644},
}

@article{Kraus_Kormann_Morrison_Sonnendrücker_2017, 
title={GEMPIC: geometric electromagnetic particle-in-cell methods}, 
volume={83}, 
DOI={10.1017/S002237781700040X}, 
number={4}, 
journal={Journal of Plasma Physics},
author={Kraus, Michael and Kormann, Katharina and Morrison, Philip J. and Sonnendrücker, Eric}, 
year={2017}, 
pages={905830401}
}

@ARTICLE{pic_on_gpu,
  author={Burau, Heiko and Widera, Renée and Hönig, Wolfgang and Juckeland, Guido and Debus, Alexander and Kluge, Thomas and Schramm, Ulrich and Cowan, Tomas E. and Sauerbrey, Roland and Bussmann, Michael},
  journal={IEEE Transactions on Plasma Science}, 
  title={PIConGPU: A Fully Relativistic Particle-in-Cell Code for a GPU Cluster}, 
  year={2010},
  volume={38},
  number={10},
  pages={2831-2839},
  doi={10.1109/TPS.2010.2064310}
}

@article{piclas,
    author = {Fasoulas, S. and Munz, C.-D. and Pfeiffer, M. and Beyer, J. and Binder, T. and Copplestone, S. and Mirza, A. and Nizenkov, P. and Ortwein, P. and Reschke, W.},
    title = {Combining particle-in-cell and direct simulation Monte Carlo for the simulation of reactive plasma flows},
    journal = {Physics of Fluids},
    volume = {31},
    number = {7},
    pages = {072006},
    year = {2019},
    month = {07},
    issn = {1070-6631},
    doi = {10.1063/1.5097638},
}

@article{DEROUILLAT2018351,
title = {Smilei : A collaborative, open-source, multi-purpose particle-in-cell code for plasma simulation},
journal = {Computer Physics Communications},
volume = {222},
pages = {351-373},
year = {2018},
issn = {0010-4655},
doi = {10.1016/j.cpc.2017.09.024},
author = {J. Derouillat and A. Beck and F. Pérez and T. Vinci and M. Chiaramello and A. Grassi and M. Flé and G. Bouchard and I. Plotnikov and N. Aunai and J. Dargent and C. Riconda and M. Grech},
}

@INPROCEEDINGS{10046112,
  author={Fedeli, Luca and Huebl, Axel and Boillod-Cerneux, France and Clark, Thomas and Gott, Kevin and Hillairet, Conrad and Jaure, Stephan and Leblanc, Adrien and Lehe, Rémi and Myers, Andrew and Piechurski, Christelle and Sato, Mitsuhisa and Zaim, Neïl and Zhang, Weiqun and Vay, Jean-Luc and Vincenti, Henri},
  booktitle={SC22: International Conference for High Performance Computing, Networking, Storage and Analysis}, 
  title={Pushing the Frontier in the Design of Laser-Based Electron Accelerators with Groundbreaking Mesh-Refined Particle-In-Cell Simulations on Exascale-Class Supercomputers}, 
  year={2022},
  volume={},
  number={},
  pages={1-12},
  doi={10.1109/SC41404.2022.00008}
}

@article{PARODI2025102244,
title = {Pantera: A PIC-MCC-DSMC software for the simulation of rarefied gases and plasmas},
journal = {SoftwareX},
volume = {31},
pages = {102244},
year = {2025},
issn = {2352-7110},
doi = {10.1016/j.softx.2025.102244},
author = {P. Parodi and S. Boccelli and F. Bariselli and T.E. Magin},
}

@article{MARKIDIS20101509,
title = {Multi-scale simulations of plasma with iPIC3D},
journal = {Mathematics and Computers in Simulation},
volume = {80},
number = {7},
pages = {1509-1519},
year = {2010},
note = {Multiscale modeling of moving interfaces in materials},
issn = {0378-4754},
doi = {10.1016/j.matcom.2009.08.038},
author = {Stefano Markidis and Giovanni Lapenta and  Rizwan-uddin},
}

@article{MASON1981233,
title = {Implicit moment particle simulation of plasmas},
journal = {Journal of Computational Physics},
volume = {41},
number = {2},
pages = {233-244},
year = {1981},
issn = {0021-9991},
doi = {10.1016/0021-9991(81)90094-2},
author = {Rodney J Mason},
}

@article{BRACKBILL1982271,
title = {An implicit method for electromagnetic plasma simulation in two dimensions},
journal = {Journal of Computational Physics},
volume = {46},
number = {2},
pages = {271-308},
year = {1982},
issn = {0021-9991},
doi = {10.1016/0021-9991(82)90016-X},
author = {J.U Brackbill and D.W Forslund},
}

@ARTICLE{Lapenta2006,
	author = {Lapenta, Giovanni and Brackbill, J.U. and Ricci, Paolo},
	title = {Kinetic approach to microscopic-macroscopic coupling in space and laboratory plasmas},
	year = {2006},
	journal = {Physics of Plasmas},
	volume = {13},
	number = {5},
	doi = {10.1063/1.2173623},
}

@article{LAPENTA2017349,
title = {Exactly energy conserving semi-implicit particle in cell formulation},
journal = {Journal of Computational Physics},
volume = {334},
pages = {349-366},
year = {2017},
issn = {0021-9991},
doi = {10.1016/j.jcp.2017.01.002},
author = {Giovanni Lapenta},
}

@article{bacchini_relsim,
title = {RelSIM: A Relativistic Semi-implicit Method for Particle-in-cell Simulations},
journal = {The Astrophysical Journal Supplement Series},
volume = {268},
number = {2},
year = {2023},
doi = {10.3847/1538-4365/acefba},
author = {Fabio Bacchini},
}

@article{ARSHAD2025109806,
title = {Adaptive mesh refinement in semi-implicit particle-in-cell method},
journal = {Computer Physics Communications},
volume = {316},
pages = {109806},
year = {2025},
issn = {0010-4655},
doi = {10.1016/j.cpc.2025.109806},
author = {Talha Arshad and Yuxi Chen and Gábor Tóth},
}

@article{MARSDEN1982394,
title = {The Hamiltonian structure of the Maxwell-Vlasov equations},
journal = {Physica D: Nonlinear Phenomena},
volume = {4},
number = {3},
pages = {394-406},
year = {1982},
issn = {0167-2789},
doi = {10.1016/0167-2789(82)90043-4},
author = {Jerrold E. Marsden and Alan Weinstein},
}

@article{MORRISON1980383,
title = {The Maxwell-Vlasov equations as a continuous hamiltonian system},
journal = {Physics Letters A},
volume = {80},
number = {5},
pages = {383-386},
year = {1980},
issn = {0375-9601},
doi = {10.1016/0375-9601(80)90776-8},
author = {Philip J. Morrison},
}

@article{CROUSEILLES2015224,
title = {Hamiltonian splitting for the Vlasov–Maxwell equations},
journal = {Journal of Computational Physics},
volume = {283},
pages = {224-240},
year = {2015},
issn = {0021-9991},
doi = {10.1016/j.jcp.2014.11.029},
author = {Nicolas Crouseilles and Lukas Einkemmer and Erwan Faou}
}

@article { Riishøjgaard1998,
      author = "L. P. Riishøjgaard and S. E. Cohn and Y. Li and R. Ménard",
      title = "The Use of Spline Interpolation in Semi-Lagrangian Transport Models",
      journal = "Monthly Weather Review",
      year = "1998",
      publisher = "American Meteorological Society",
      address = "Boston MA, USA",
      volume = "126",
      number = "7",
      doi = "10.1175/1520-0493(1998)126<2008:TUOSII>2.0.CO;2",
      pages=      "2008 - 2016",
}

@ARTICLE{Pfau_Kempf_2018,
AUTHOR={Pfau-Kempf, Yann  and Battarbee, Markus  and Ganse, Urs  and Hoilijoki, Sanni  and Turc, Lucile  and von Alfthan, Sebastian  and Vainio, Rami  and Palmroth, Minna },
TITLE={On the Importance of Spatial and Velocity Resolution in the Hybrid-Vlasov Modeling of Collisionless Shocks},
JOURNAL={Frontiers in Physics},
VOLUME={Volume 6 - 2018},
YEAR={2018},
DOI={10.3389/fphy.2018.00044},
ISSN={2296-424X}
}

@article{Cheng_2014,
author = {Cheng, Yingda and Gamba, Irene M. and Li, Fengyan and Morrison, Philip J.},
title = {Discontinuous Galerkin Methods for the Vlasov--Maxwell Equations},
journal = {SIAM Journal on Numerical Analysis},
volume = {52},
number = {2},
pages = {1017-1049},
year = {2014},
doi = {10.1137/130915091}
}

@article{KORMANN2025118290,
title = {A structure-preserving finite element framework for the Vlasov–Maxwell system},
journal = {Computer Methods in Applied Mechanics and Engineering},
volume = {446},
pages = {118290},
year = {2025},
issn = {0045-7825},
doi = {10.1016/j.cma.2025.118290},
author = {Katharina Kormann and Murtazo Nazarov and Junjie Wen},
}

@misc{wilhelm2025restartingnumericalflowiteration,
      title={Restarting the Numerical Flow Iteration through low rank tensor approximations}, 
      author={Rostislav-Paul Wilhelm and Katharina Kormann},
      year={2025},
      eprint={2509.08474},
      archivePrefix={arXiv},
      primaryClass={math.NA},
      url={https://arxiv.org/abs/2509.08474}, 
      doi = {10.48550/arXiv.2509.08474},
}

@Article{ScRaEi2024,
  author  = {Nils Schild and Mario Räth and Sebastian Eibl and Klaus Hallatschek and Katharina Kormann},
  journal = {Comput. Phys. Commun.},
  title   = {A performance portable implementation of the semi-Lagrangian algorithm in six dimensions},
  year    = {2024},
  pages   = {108973},
  volume  = {295},
  doi     = {https://doi.org/10.1016/j.cpc.2023.108973},
}

@article{Wei_Cheng_2016,
author = {Guo, Wei and Cheng, Yingda},
title = {A Sparse Grid Discontinuous Galerkin Method for High-Dimensional Transport Equations 
	and Its Application to Kinetic Simulations},
journal = {SIAM Journal on Scientific Computing},
volume = {38},
number = {6},
pages = {A3381-A3409},
year = {2016},
doi = {10.1137/16M1060017},
}

@article{WEN2025114079,
title = {An anisotropic nonlinear stabilization for finite element approximation of Vlasov–Poisson equations},
journal = {Journal of Computational Physics},
volume = {536},
pages = {114079},
year = {2025},
issn = {0021-9991},
doi = {10.1016/j.jcp.2025.114079},
author = {Junjie Wen and Murtazo Nazarov},
}

@article{KIECHLE2025113693,
title = {A positivity-preserving Active Flux method for the Vlasov-Poisson system},
journal = {Journal of Computational Physics},
volume = {524},
pages = {113693},
year = {2025},
issn = {0021-9991},
doi = {10.1016/j.jcp.2024.113693},
author = {Yanick Kiechle and Erik Chudzik and Christiane Helzel},
}

@misc{cmm_nufi,
      title={A Hybrid semi-Lagrangian Flow Mapping Approach for Vlasov Systems: Combining Iterative and Compositional Flow Maps}, 
      author={Philipp Krah and Zetao Lin and R. -Paul Wilhelm and Fabio Bacchini and Jean-Christophe Nave and Virginie Grandgirard and Kai Schneider},
      year={2026},
      eprint={2601.21668},
      archivePrefix={arXiv},
      primaryClass={math.NA},
    doi = {10.48550/arXiv.2601.21668}
}

@article{Bacchini_2024,
    author = {Bacchini, Fabio and Philippov, Alexander A},
    title = {Fundamental, harmonic, and third-harmonic plasma emission from beam–plasma instabilities: a first-principles precursor for astrophysical radio bursts},
    journal = {Monthly Notices of the Royal Astronomical Society},
    volume = {529},
    number = {1},
    pages = {169-177},
    year = {2024},
    month = {02},
    issn = {0035-8711},
    doi = {10.1093/mnras/stae521},
}

@article{Pezzini_2024,
doi = {10.3847/1538-4357/ad7465},
year = {2024},
month = {oct},
publisher = {The American Astronomical Society},
volume = {975},
number = {1},
pages = {37},
author = {Pezzini, L. and Zhukov, A. N. and Bacchini, F. and Arrò, G. and López, R. A. and Micera, A. and Innocenti, M. E. and Lapenta, G.},
title = {Fully Kinetic Simulations of Proton-beam-driven Instabilities from Parker Solar Probe Observations},
journal = {The Astrophysical Journal},
}

@article{Bacchini_2019,
doi = {10.1088/1742-6596/1225/1/012011},
year = {2019},
month = {may},
publisher = {IOP Publishing},
volume = {1225},
number = {1},
pages = {012011},
author = {Bacchini, Fabio and Amaya, Jorge and Lapenta, Giovanni},
title = {The relativistic implicit Particle-in-Cell method},
journal = {Journal of Physics: Conference Series},
}

@article{Croonen_2024,
doi = {10.3847/1538-4365/ad31a3},
year = {2024},
month = {apr},
publisher = {The American Astronomical Society},
volume = {271},
number = {2},
pages = {63},
author = {Croonen, J. and Pezzini, L. and Bacchini, F. and Lapenta, G.},
title = {An Exactly Energy-conserving Electromagnetic Particle-in-cell Method in Curvilinear Coordinates},
journal = {The Astrophysical Journal Supplement Series},
}

@incollection{CATMULL1974317,
title = {A CLASS OF LOCAL INTERPOLATING SPLINES},
editor = {ROBERT E. BARNHILL and RICHARD F. RIESENFELD},
booktitle = {Computer Aided Geometric Design},
publisher = {Academic Press},
pages = {317-326},
year = {1974},
isbn = {978-0-12-079050-0},
doi = {10.1016/B978-0-12-079050-0.50020-5},
author = {Edwin Catmull and Raphael Rom}
}

@article{morrison2017,
    author = {Morrison, P. J.},
    title = {Structure and structure-preserving algorithms for plasma physics},
    journal = {Physics of Plasmas},
    volume = {24},
    number = {5},
    pages = {055502},
    year = {2017},
    month = {04},
    issn = {1070-664X},
    doi = {10.1063/1.4982054},
}

@article{CHEN20117018,
title = {An energy- and charge-conserving, implicit, electrostatic particle-in-cell algorithm},
journal = {Journal of Computational Physics},
volume = {230},
number = {18},
pages = {7018-7036},
year = {2011},
issn = {0021-9991},
doi = {10.1016/j.jcp.2011.05.031},
author = {G. Chen and L. Chacón and D.C. Barnes},
}

@article{LIU2025113840,
title = {An asymptotic-preserving conservative semi-Lagrangian scheme for the Vlasov-Maxwell system in the quasi-neutral limit},
journal = {Journal of Computational Physics},
volume = {528},
pages = {113840},
year = {2025},
issn = {0021-9991},
doi = {10.1016/j.jcp.2025.113840},
author = {Hongtao Liu and Xiaofeng Cai and Yong Cao and Giovanni Lapenta},
}

@article{POLLINGER2023112338,
title = {A stable and mass-conserving sparse grid combination technique with biorthogonal hierarchical basis functions for kinetic simulations},
journal = {Journal of Computational Physics},
volume = {491},
pages = {112338},
year = {2023},
issn = {0021-9991},
doi = {10.1016/j.jcp.2023.112338},
author = {Theresa Pollinger and Johannes Rentrop and Dirk Pflüger and Katharina Kormann},
}

@article{BORRIN2025190,
title = {Regular Lagrangian flow for wavelike vector fields and the Vlasov-Maxwell system},
journal = {Journal of Differential Equations},
volume = {416},
pages = {190-226},
year = {2025},
issn = {0022-0396},
doi = {10.1016/j.jde.2024.09.051},
author = {Henrique Borrin},
}

@article{guo_mei_2024,
title = {A Local Macroscopic Conservative (LoMaC) Low Rank Tensor Method for the Vlasov Dynamics},
journal = {Journal of Scientific Computing},
volume = {102},
year = {2024},
doi = {10.1007/s10915-024-02684-1},
author = {Guo, Wei and Qiu, Jing-Mei},
}

@misc{oblapenko2026,
      title={Sparse and low-rank kinetic distribution estimation}, 
      author={Georgii Oblapenko and Lambert Theisen and Rostislav-Paul Wilhelm and Michael Herty and Manuel Torrilhon},
      year={2026},
      eprint={2606.04878},
      archivePrefix={arXiv},
      primaryClass={physics.comp-ph},
    doi = {10.48550/arXiv.2606.04878}
}

\end{document}